\documentclass[11pt]{article}
\pdfoutput=1
\usepackage{amsmath,amsthm}
\usepackage[noticks]{preprint}
\usepackage[T1]{fontenc}
\usepackage[utf8]{inputenc}
\usepackage{graphicx,booktabs,array,longtable,pdfpages,placeins,needspace}
\AtBeginDocument{\newgeometry{left=18mm,right=18mm,top=18mm,bottom=22mm,footskip=10mm}\pagestyle{plain}}
\usepackage[numbers,sort&compress]{natbib}
\usepackage{hyperref,cleveref,enumitem}
\usepackage{caption,subcaption}
\hypersetup{colorlinks=true,linkcolor=blue!60!black,citecolor=blue!60!black,urlcolor=blue!60!black,
pdftitle={StateSync-GKR: Design and Evaluation of a Reusable Sparse GKR Engine},
pdfauthor={Jinwook Kim},pdfsubject={Controlled engineering evaluation of a reusable GKR engine},
pdfkeywords={GKR, sparse prover, memory, batching, proof serving, tail latency, CPU, reproducibility}}
\newcolumntype{L}[1]{>{\raggedright\arraybackslash}p{#1}}
\newcommand{\figroot}{Figures}
\newcommand{\planfig}[2]{\includegraphics[width=#1]{#2}}
\makeatletter
\newcommand{\atlaslabelonce}[1]{\@ifundefined{atlasdone@#1}{\phantomsection\label{atlas:#1}\global\@namedef{atlasdone@#1}{}}{}}
\makeatother
\newcommand{\atlasfamily}[2]{\includepdf[pages=-,fitpaper=false,pagecommand={\thispagestyle{plain}\atlaslabelonce{#1}}]{#2}}
\title{StateSync-GKR: Design and Evaluation of a\\Reusable Sparse GKR Engine\thanks{This version evaluates StateSync-GKR release 1.2.0 against same-day measurements of release 1.1. Version 1 of this report (arXiv:2610.14849v1) evaluates release 1.1.0 and remains available.}}
\author{Jinwook Kim\textsuperscript{1,2}~\href{https://orcid.org/0009-0004-4993-8005}{\mbox{\scalerel*{\begin{tikzpicture}[yscale=-1,transform shape]\pic{orcidlogo};\end{tikzpicture}}{|}}}\\(for the Oraclizer Core Team)\\
\textsuperscript{1}Oraclizer Labs, Inc., Delaware, USA\\
\textsuperscript{2}Oraclizer Labs Korea, Seoul, Korea\\
\texttt{jay@oraclizer.io}}
\date{October 2026}
\begin{document}
\maketitle
\thispagestyle{plain}
\begin{abstract}
Authenticated state transitions need proof engines that conserve memory, avoid repeated setup, keep every result bound to its request under parallel execution and return complete proofs at service rates. We present StateSync-GKR, a reusable Rust sumcheck/GKR core with a sparse-Merkle frontend, and evaluate release 1.2.0 against release 1.1 measured on the same day on the same instance types, request corpus and latency budgets. Sparse layer reduction keeps additional requested allocation 3,242-fold and peak process memory about 89-fold below the engine's dense reference at width 4,096. Release 1.2 changes two parts of the software while proof bytes and verifier acceptance stay fixed: reorganized round arithmetic raises prepared proving throughput 2.4 to 2.7-fold on a 192-core bare-metal host, and a new serving path starts each job as soon as a worker is free, bounds queued work and buffered replies, and returns proofs in completion order bound to request identifiers. On a 192-core VM at 1,000 requests/s, 99th-percentile latency to verified acceptance falls from 440 to 34 ms. One connection carries 3,300 requests/s with every request inside one second, confirmed in three 60-second runs and a 30-minute run of 5.94 million requests without a latency or queue trend; at 3,400 requests/s reply traffic reaches the provider's documented 5-Gbit/s single-flow limit, and two connections passed 5,000 requests/s in 15-second screens. In single 15-second screens on 48 cores, the highest passing rate rose from 500 to 2,000/s. Offered 1.5 times its highest screened one-connection rate under a memory limit, with excess offers held back by the client, the worker held its resident memory near a fifth of that limit and left no submitted request unresolved. For an integrating service, fewer servers deliver the same proof rate with bounded tail latency, and existing verifiers accept the GKR proofs unchanged.
\end{abstract}
\keywords{GKR protocol \and sparse reduction \and memory \and batching \and CPU evaluation \and proof serving \and tail latency \and reproducibility}

\section{Introduction}
Authenticated registries need more than a valid proof algorithm. Their proof engine must fit the computation into memory, avoid rebuilding an unchanged circuit for every request, verify the claimed state transition, and keep independent results attached to the correct requests under parallel execution. A local proving rate answers only one of these questions. Application-facing value also depends on preparation, proof size, record semantics, result handling and the acceptance endpoint.

StateSync-GKR is a reusable sumcheck/GKR engine with a sparse-Merkle application frontend. It implements the published GKR and structured-reduction constructions~\cite{gkr2015,thaler2013,libra2019}, using Plonky3 primitives~\cite{plonky3}. The signed software evaluated here is release 1.2.0~\cite{software}. Release 1.1.0~\cite{softwarev11}, the subject of the first version of this report, is measured again on the same day as the baseline. Its formal companion studies compiler-to-protocol model composition and an explicit partial connection to Rust~\cite{fvpaper}. This paper studies the implemented engine and the engineering consequences of its choices; it adds no cryptographic reduction or machine-checked theorem.

Nine mechanisms organize the design and evaluation: sparse prover reduction, derived wiring, cube/affine fusion, parallel path constraints, reusable preparation, independent proof jobs, three-state records, original-request binding, and separate reusable core/model boundaries. Some admit a controlled numerical contrast. Others are supported by implementation structure, semantic controls, or explicit scope rather than an invented speed ratio. The paper makes each evidence type visible.

Release 1.2 changes two parts of the software and leaves the proof untouched. Its serving path admits requests through a bounded queue, refuses excess work with an explicit reason, and lets each worker start the oldest admitted job as soon as it finishes the previous one; replies return in completion order and stay bound to their requests by identifier. Inside the production sparse oracle, the round arithmetic is reorganized while the two phases, the proof bytes and the verifier stay the same. Both changes use properties the engine already had: proof jobs are independent and bound to their original requests, and the sparse booking fixes what each round must compute. Proof format, verifier acceptance conditions and transcript order are those of release 1.1.

The contributions are a connected account of this reusable artifact, controlled comparisons of memory and repeated work with proof-result preservation, deployment measurements ending at original-request encoded-proof acceptance, and a before-and-after evaluation of release 1.2 against release 1.1 measured on the same day with the same instance types, corpus and budgets. Full distributions, process repeats, failed attempts and complete drain remain part of the evidence. The internal Dense reference identifies a representation cost; it does not stand for every optimized external GKR engine. The release 1.1 records of 6 and 7 October 2026 and the release 1.2 campaign of 25 October 2026 remain separate cohorts, connected by same-day release 1.1 runs. \Cref{tab:headline} collects the main results with their exact comparisons.

\begin{table}[t]
\centering\footnotesize
\caption{Main results. Every row names its comparison; ratios apply only to their stated references and workloads. Operation triples are membership/non-membership/update.}\label{tab:headline}
\begin{tabular}{@{}L{.14\linewidth}L{.26\linewidth}L{.23\linewidth}L{.19\linewidth}L{.08\linewidth}@{}}
\toprule Result & Comparison and scope & Release 1.2 & Release 1.1 & Where\\\midrule
Prover memory & \texttt{SparseLayerOracle} vs test-only \texttt{DenseLayerOracle}; same driver, field, transcript and mixed two-layer circuit, width 4,096 & requested allocation 3,242.41$\times$ lower; process RSS 88.87$\times$ lower & 3,242.41$\times$; 89.46$\times$ (7 Oct record) & Fig.~\ref{fig:memory}\\
Scale within 3 GiB & same pair, cgroup cap, swap disabled & Dense OOM at width 8,192; Sparse completes 65,536 (largest tested) & same outcomes & \S\ref{sec:memory}\\
Allocations per proof & instrumented mixed circuit, width 4,096, same day & 85 & 1,282 & \S\ref{sec:memory}\\
Prover time & same pair as above, width 4,096 & 6,470$\times$ faster than Dense (1.80 ms vs 11.63 s) & 2,507$\times$ (4.63 ms vs 11.61 s; 7 Oct record) & \S\ref{sec:memory}\\
Prepared proving & depth 24, batch 768, 192 workers, bare metal, same day & 7,713.71/\allowbreak 7,220.36/\allowbreak 5,390.07 proofs/s & 3,195.50/\allowbreak 2,947.08/\allowbreak 2,102.02 (same day) & Fig.~\ref{fig:compact}\\
Prepared requests & one thread, depth 24, p50; fresh/prepared ratio in the same run & 16.95/18.22/24.62 ms; 7.66/7.19/10.13$\times$ & 42.10/46.10/65.03 ms (same day); 3.69/3.46/4.45$\times$ (6 Oct record) & Fig.~\ref{fig:preparation}\\
Structured wiring verification & Derived vs Table wiring, same proof, prepared verification call, nine profiles & unchanged in release 1.2 & 10.80--23.68$\times$ & Fig.~\ref{fig:verification}\\
Parallel output policy & complete local batch, depth 24, batch 768, 192 workers & 14.10--14.64$\times$ (higher because proving is faster) & 9.71--10.71$\times$ (6 Oct record) & Fig.~\ref{fig:verification}\\
Compact proofs & depths 24--32, nine profiles & unchanged in release 1.2 & 118 layers; +2.93\% membership bytes & Fig.~\ref{fig:compact}\\
Deterministic proof bytes & release 1.2 vs release 1.1, native and default builds, every serving mode & 13,824 hash comparisons equal; every accepted reply in 187 replay-audited processes equals one of 576 reference proofs & reference & \S\ref{sec:compute}, \S\ref{sec:serving}\\
Tail latency & 1,000 offered requests/s, same 192-core VM and day, three 60-s runs & p99 33.7--33.9 ms (continuous) & p99 433.8--475.1 ms & Tab.~\ref{tab:fivecontrol}\\
One 192-core VM & every timed request within 1 s; one connection: three 60-s runs; two connections: 15-s screens & 3,300/s on one connection, also for 30 min (5.94 million); $\geq$5,000/s on two connections (frontend-limited lower bound) & 1,000/s (7 Oct record); 1,150/s same day (one 60-s run), highest rate measured & Fig.~\ref{fig:capacity}\\
One 48-core VM & same budget, one 15-s screen per rate, same day & 2,000/s, bound by worker CPU & 500/s; 1,000/s fails & Tab.~\ref{tab:hostsize}\\
Overload memory & 5,100 offered requests/s (1.5$\times$ the highest one-connection screen pass), 2,760-MiB limit & worker RSS at most 548,064 KiB sampled (402,948 KiB by GNU time); return budget full, no server refusal & no admission limit in the library & \S\ref{sec:overload}\\
Reusable core & serving crate built on the public prover API only & provided & n/a & \S\ref{sec:servingparts}\\\bottomrule
\end{tabular}
\end{table}

\section{Engine and application boundaries}
\label{sec:engine}
\begin{figure}[t]\centering
\planfig{174mm}{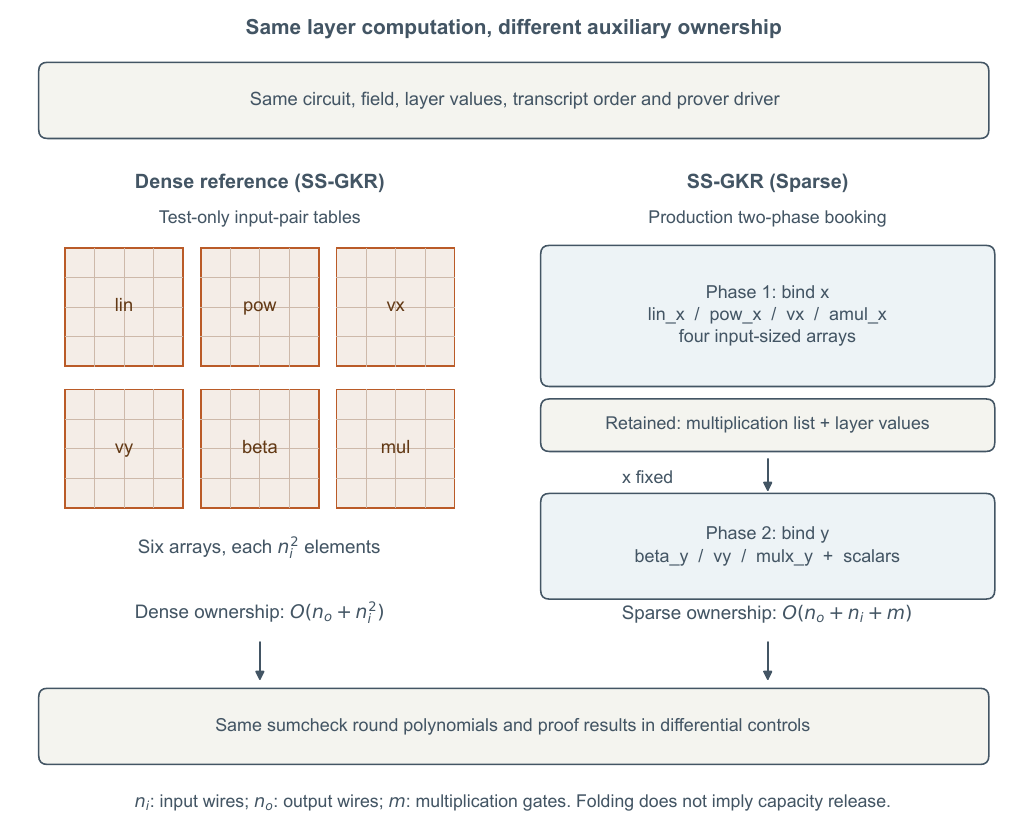}
\caption{SS-GKR layer reduction: six dense input-pair tables versus two-phase sparse booking. The representation explains the memory contrast; differential controls preserve round polynomials and proof results. The structure is unchanged in release 1.2.}\label{fig:architecture}
\end{figure}

\Cref{fig:architecture} contrasts the actual layer-oracle representations; \Cref{fig:lifecycle} then separates shared preparation from request-specific proof state and original-request acceptance.

\subsection{Layer reduction and structured wiring}
Linear-time and structured GKR techniques already establish the algorithmic foundation~\cite{thaler2013,libra2019}. The production oracle uses a Libra-style two-phase sparse booking construction. Weighted linear and cube terms depend on the first input block. Multiplication supplies a sparse list of weighted input pairs. Summing the second block first produces a booking table; binding the first block then produces the remaining second-block coefficients. This implementation avoids materializing the full input-pair hypercube.

For a layer with $n_i$ input wires, $n_o$ output wires and $m$ multiplication gates, the sparse auxiliary ownership is $O(n_o+n_i+m)$. The comparison baseline is StateSync-GKR's own \texttt{DenseLayerOracle}, a test-only reference retaining the initial dense design. It constructs six $n_i^2$ factor tables and constructor temporaries. The production \texttt{SparseLayerOracle} implements Libra-style sparse reduction. Both use StateSync-GKR's \texttt{prove\_impl} driver and transcript order. Thus Dense and Sparse name two representations within this implementation.

StateSync-GKR implements the sumcheck and GKR assembly from their published constructions; it uses Plonky3 0.4.3 for field arithmetic, hash and challenger primitives. The Dense reference is neither a Plonky3 GKR implementation nor the independently probed Lambdaworks GKR engine. The measured ratios isolate an internal representation change, not an external product comparison.

Verification has a separate optimization. \texttt{TableWiring} already iterates a sparse gate list. \texttt{DerivedRegularWiring} evaluates regular progression families in closed form and retains an exact sparse residue. Derivation checks each member's progression, kind and coefficient; unsupported gates stay on the sparse path. Audit tests re-expand the representation and compare weighted gate multisets. This reduces repeated wiring work without changing the proof object. It is distinct from replacing the quadratic prover oracle.

Release 1.2 reorganizes the arithmetic of the production oracle inside the same two phases. Output-wire weights come from scaled equality tables built once per layer instead of a product chain per output wire. Each phase computes the node values of a round polynomial in one pass over its table, taking two nodes from the table halves and the others by difference additions. Round polynomials are interpolated with a constant inverse-Vandermonde matrix built once per layer, which removes the per-round inversions and small allocations of Lagrange interpolation. Each change computes the same field values by an algebraically equivalent route, and field arithmetic is exact, so round polynomials and proofs are byte-identical to release 1.1: the committed test vectors match, the Dense and Sparse oracles still agree, and all 6,912 reference proof hashes of the benchmark corpus match in both the native and the default build (\Cref{sec:compute}). Additional requested allocation at width 4,096 stays at 579,584 bytes. Verification, the wiring oracles and the Dense reference are unchanged in release 1.2.

\subsection{Cube gates and path constraints}
The circuit intermediate representation supports weighted linear, product and cube gates. A \texttt{Pow3} gate is an algebraic operation, not a processor instruction. The builder combines cube/linear taps and constants into one round layer. The Poseidon2 cube and matrix step are thereby represented without separate layers for every arithmetic subexpression~\cite{poseidon2}. The shipped primitives use the pinned Plonky3 0.4.3 registry dependencies~\cite{plonky3}.

The witness computes authentication-path accumulators in order. Compilation checks each compression transition against those supplied intermediate values as parallel residual constraints. More tree levels therefore increase width and work while preserving the hash-template round schedule in the measured profiles. This is a structural benefit; there is no measured unfused-compiler ablation or claim of constant total cost with path depth.

\subsection{Preparation and independent requests}
\texttt{PreparedSync} owns the operation-kind circuit, full circuit commitment and derived wiring. Reuse requires the same kind, sparse Merkle tree (SMT) parameters, strategy/configuration and circuit. Every request keeps its own public inputs, witness, transcript and proof. Preparation reuse saves commitment work as well as compilation; it does not share challenge-dependent proof state.

The indexed parallel batch API returns one proof per input job in input order. Witness construction and output encoding/verification can also be parallelized by the caller. This is independent job processing, not proof aggregation. Remote replies can complete out of order; explicit request IDs bind them to the corresponding original requests.

The continuous serving mode of release 1.2 relies on this property: replies leave in completion order, and the frontend, the client process that sends requests and verifies the returned proofs, matches each reply to its original request by job identifier. Each proof transcript starts from its own job's public inputs, so proof bytes do not depend on serving mode, thread or order. The reference server checks this at startup, before it listens: the sequential and parallel batch APIs must return the same proofs in input order, the per-job completion function must reproduce their encoded bytes, and a tampered control proof must be rejected.

\subsection{Semantics, transport and reuse}
\begin{figure}[t]\centering
\planfig{174mm}{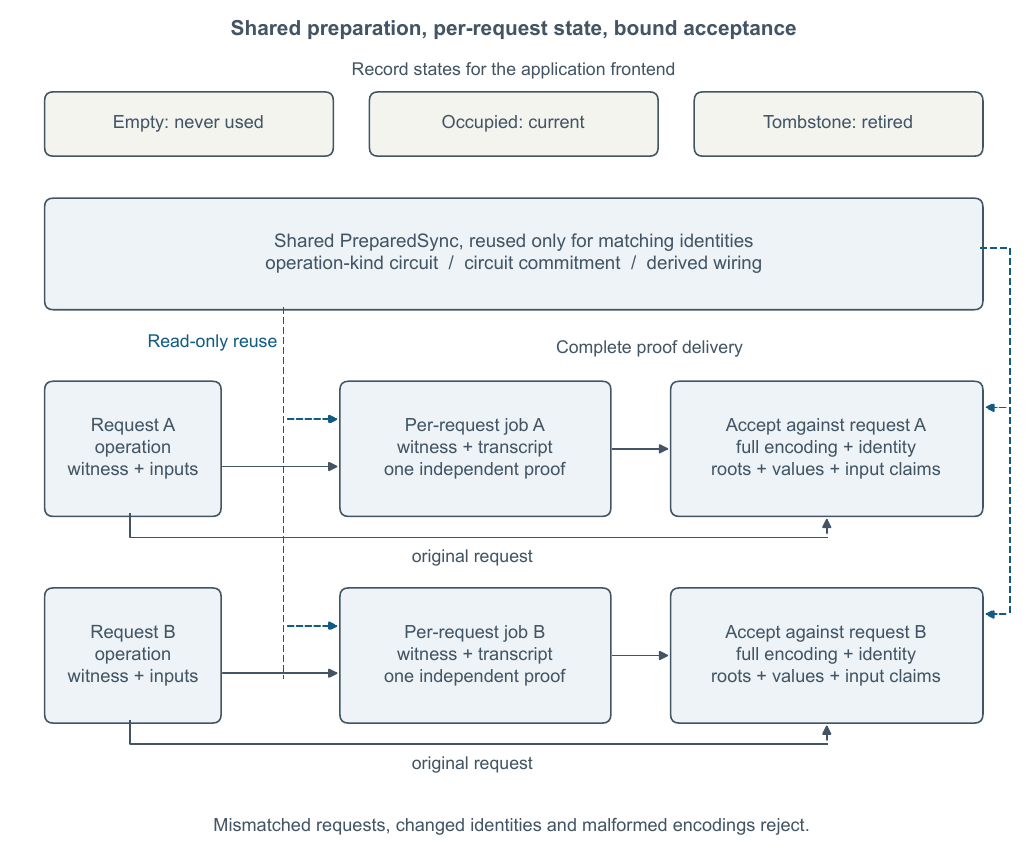}
\caption{Shared preparation serves per-request proof jobs. Each complete encoding is checked against its own original request; the three record states retain distinct application meanings. Unchanged in release 1.2.}\label{fig:lifecycle}
\end{figure}

\Cref{fig:lifecycle} shows what preparation reuse shares and what remains private, including the original request checked at acceptance.

The SMT frontend distinguishes \texttt{Empty}, \texttt{Occupied(payload)} and \texttt{Tombstone}. A never-used and a deleted slot remain semantically different, while either can satisfy non-membership. The general single-leaf update accepts old and new leaf states under the same sibling path and respective root conditions. Insertion, replacement, deletion and restoration are application uses of that operation, rather than an exclusive four-transition policy. The earlier input-variant study exercises eight lifecycle and payload/key cases.

The encoded verifier consumes the complete received inner-proof encoding, checks circuit/configuration identity and verifies against the original request. Roots, value digest and input claims remain required. Malformed trailing bytes and mismatched requests reject. The current SMT verifier also recomputes native leaf/path hashes from the private witness; faster wiring evaluation does not offload that hashing. Release 1.2 enforces in every build two misuse checks that release 1.1 performed only in debug builds: a request whose operation kind differs from the prepared circuit returns an error, and a leaf-hash input of the wrong width stops execution.

The sumcheck/protocol crates have no dependency on the SMT compiler. Another arithmetic frontend supplies its own circuit, witness, output claim and transcript conventions and must consume both residual input multilinear-extension (MLE) claims. The shipped entry points use KoalaBear values, a degree-four extension and the fixed hash primitives. Arbitrary Rust-program compilation and runtime field/hash selection are not bundled services. The current official distribution is release 1.2.0, source-available under BSL~1.1, which permits non-production research and reproduction. Previously granted MIT or Apache-2.0 rights and required third-party notices remain valid; they do not provide an alternative current official distribution. Data and figure rights are recorded separately. A separate serving crate (\Cref{sec:servingparts}) builds its service components on the public prover API alone, which exercises the reuse boundary from outside the core.

\subsection{Formal results as a design basis}\label{sec:formal}
The companion formal study~\cite{fvpaper} proves that compiler-model acceptance and sparse-Merkle transition validity agree in both directions, then composes that equivalence with a GKR assembly bound in Isabelle/HOL. It also proves a derived/table wiring-polynomial equivalence and an independent-batch preservation contract. These results address semantic preservation: a smaller representation or a shared preparation path must retain the intended computation and each request's outcome.

The measurements complement that basis. The wiring contrast quantifies the cost of the implementation's structured representation; proof-byte controls check independent output handling; lifecycle and request-binding controls exercise the application's checks. The formal statements do not supply empirical speed ratios, and matching observations do not discharge the partial Rust/model value-correspondence or transcript/Fiat--Shamir assumptions. This combination gives the engine an inspectable semantic basis and measured costs at explicit implementation boundaries. Release 1.2 changes proof-generation arithmetic inside the sparse oracle and adds a separate serving crate; the verifier functions, transcript order and formal models stay as in release 1.1, so the model-level results apply as before and the partial Rust connection keeps its stated assumptions.

\subsection{What the application requests mean}
An application can ask whether a named record is present in an authenticated ledger (membership), whether a record is absent from the claimed ledger (non-membership), or whether an old record has been replaced by the requested new record consistently with the stated old and new roots (update). A bank-account register provides an intuitive example: proving that an entry is present, that an unused or retired entry is absent, or that the requested old-to-new change agrees with the resulting register. The experiments use synthetic fixtures rather than bank customers or executed payments. The circuit checks these state and root conditions; an application's authorization and financial execution remain separate. These operation meanings explain why update has more path work than a simple presence check.

\subsection{Serving components and modes}\label{sec:servingparts}
Release 1.2 adds a serving crate that depends on the product only through the public \texttt{statesync-gkr} API and leaves the product workspace, proof format, verifier acceptance conditions and transcript order unchanged. Its library modules give an integrating service the parts that the prover library does not provide: a request and reply envelope bound to run and job identifiers, with frame and metadata limits checked before work is admitted; one cap on admitted jobs that have not started computing, shared by all connections, with explicit rejection reasons; a return budget in responses and encoded bytes that keeps a reply charged until its frame is written, so that a stalled peer becomes backpressure instead of memory growth; per-job completion functions that generate the witness, prove, encode, hash and release one request on one thread, together with a continuous runner; and per-job stage timers that record which thread released each job's memory. A reference server composes these parts into four modes (\Cref{tab:modes}). The original mode reproduces the release 1.1 serving loop inside the same binary. The queued-return mode is implemented and functionally checked but was not measured in this campaign, so we make no performance claim for it. Continuous dispatch follows the principle, used in iteration-level scheduling of generative-model serving, that a finished request should not wait for the rest of its batch~\cite{orca2022}; the queue between admission and computation with explicit refusal follows staged service designs~\cite{seda2001}. We claim neither principle as new.

\begin{table}[t]
\centering\footnotesize
\caption{Serving modes of the release 1.2 reference server. All modes share admission, rejection, framing, return-budget and shutdown behavior.}\label{tab:modes}
\begin{tabular}{@{}L{.12\linewidth}L{.31\linewidth}L{.26\linewidth}L{.10\linewidth}L{.11\linewidth}@{}}
\toprule Mode & Scheduling & Hash and release of typed data & Reply order & Measured here\\\midrule
original & one FIFO per operation kind, oldest ready kind first, flush at the batch cap or maximum wait; one compute batch at a time with parallel witness, proof and encoding phases & on the dispatcher, which hashes, writes and logs each result and then releases the typed batch & batch order & release-timer runs\\
queued-return & as original & dispatcher releases the batch; per-connection writers hash and write while the next batch computes & batch order & no\\
batch-fused & as original & each job's witness, proof, encoding, hash and release run as one task on its worker inside the parallel stage; writers only write & batch order & 48-core arrival and fine studies; 192-core one-connection study\\
continuous (default) & one arrival-order FIFO across kinds; up to a set number of workers each start the oldest admitted job when they finish the previous one & on the worker that computed the job & completion order, matched by job identifier & all serving studies\\\bottomrule
\end{tabular}
\end{table}
\FloatBarrier
\section{Experimental design}
\label{sec:methods}
\subsection{Fixed source and separate cohorts}
Release 1.2.0 is the signed tag \texttt{v1.2.0} at commit \texttt{8b2908160f36adc233d7b869a403d2ee4d3502a5} (tree \texttt{be1b62673ea19df806d7276579a997eb012e304b}); its published source archive is 43,366,452 bytes with SHA-256 \texttt{6e859633f7149b45e8c37d0b38ba9ccf1538384f2b4392692054195e263c8d27}. The release 1.1 baseline keeps the immutable kernel commit \texttt{1b8d2f829792172b347dedbfde969016c2c05789}, whose issued source archive is 2,537,304 bytes with SHA-256 \texttt{d22e2200378e21a40c254eab8df4d44dfea81f2806f02f0c9030636300d489e9}; same-day release 1.1 serving runs use the executables preserved from the 7 October retrieval bundle. Same-day release 1.1 CPU and memory builds use commit \texttt{1b47601e34b2bfc0d4566973a0f3d00f8be7f861}, whose prover sources equal those of release 1.1.1, the licensing revision of release 1.1.0. The two archives cover different scopes, so their sizes are not compared. Benchmark-only callers and observers have separate source and binary identities. Tests do not edit the canonical production source; the memory observer is a test child installed in an isolated source copy. Release 1.2 attaches a new observer of the same kind to its sparse oracle and records the observer's source hash.

\Cref{tab:cohorts} identifies the units. Process repeats share a host and fixed corpus and are not independent-machine replications. A repeatedly cycled request is not a new independent application-input sample. Different hosts, profiles and confirmation periods are retained separately.
\begin{table}[htbp]
\centering\small
\caption{Study units and actual sample denominators. Release 1.2 rows include their same-day release 1.1 anchors.}\label{tab:cohorts}
\begin{tabular}{L{.20\linewidth}L{.40\linewidth}L{.32\linewidth}}
\toprule Study & Execution units & Timed and warmup observations\\\midrule
\multicolumn{3}{@{}l}{\emph{Release 1.1 records (6 and 7 October 2026)}}\\
Earlier CPU & 81 conditions, three processes: 243 & 63,540 timed samples; original method retained\\
Earlier supplement & 35 conditions, five processes: 175 & 156,100 timed, 3,925 warmups\\
Generic memory & Four families, eleven widths, two modes, three processes: 264 & 792 timed, 264 warmups\\
Limits/concurrency & 17 attempts: 14 successes, one address-space (AS) abort, two OOMs & 38 successful concurrent jobs; 12 timed and four warmup scale proofs\\
48-core arrivals & 36 screens, one pilot, one confirmation & 730,000 timed, 5,888 warmups\\
Equal resources & 12 frontend and 18 worker processes & 180,000 timed, 3,072 warmups\\
48-core fine study & 19 conditions, three process pairs: 57 & 456,750 timed, 14,592 warmups\\
192-core VM & 47 actual frontend/worker pairs & 1,102,000 timed, 12,032 warmups\\\midrule
\multicolumn{3}{@{}l}{\emph{Release 1.2 campaign (25 October 2026, UTC)}}\\
Bare-metal CPU & 502 cells: 216 release 1.2 and 45 same-day release 1.1 matrix processes, 146 supplement cells, 60 thread-placement probes, 20 prototype cells, nine build-setting processes and six gate or environment cells; 16 probe cells on the serving VM & matrix: 36,540 timed rows for release 1.2 and 9,108 for same-day release 1.1, each one request or one batch as in the earlier study; supplement: 75,100 timed and 2,575 warmups in its 145 timed processes; steady-state proof timings: 954,292 in placement probes and 180,719 on the serving VM; 176,322 prototype timings; builds: 9,000 timed requests\\
Generic memory & 264 processes as above, plus 12 same-day processes of the release 1.1 and 1.2 provers at widths 4,096 and 65,536 & 792 timed, 264 warmups; three timed proofs per same-day process\\
Limits/concurrency & 17 attempts: 14 successes, one AS abort, two OOMs & same structure as the release 1.1 record\\
48-core arrivals & 44 processes: 36 batch-fused and six continuous screens, one pilot, one 900-s run & 775,000 timed, 6,656 warmups\\
Equal resources & 12 frontend/worker sets, continuous mode & 180,000 timed, 256 warmups per process\\
48-core fine study & batch-fused, 19 conditions, three processes: 57; continuous, 16 mapped conditions, three processes: 48 & 456,750 and 389,250 timed, 256 warmups per process\\
Host-size ceilings & 48 cores: 7 processes on the 16-core and 6 on the 48-core frontend; 96 cores: 2 and 5 & 142,500, 270,000, 30,000 and 300,000 timed\\
192-core VM & 196 processes in 27 phases on two VMs: 187 replay-audited processes in 24 phases, a 30-minute run, a send-limited overload run and a seven-process phase with one run on a 1,536-fixture corpus outside the 576-proof reference table and six update-only runs & 16,310,000 timed and 50,944 warmups in all 196 processes, of which 9,844,250 and 47,872 in the replay-audited processes; the 30-minute run alone has 5,940,000 timed and 1,024 warmups\\\bottomrule
\end{tabular}
\end{table}

The earlier studies used separate 192-core C8a baremetal hosts. New serving studies used actual 48/96/192-core C8a VMs with advertised 96/192/384 GiB RAM and a separate 16-core, 32-GiB frontend. Each serving VM is called a worker; worker counts in CPU studies count proving threads. The reported CPU model was AMD EPYC 9R45, with one thread per reported core. The memory contrast ran one job on CPU 0 of the 48-core VM. Limits used the isolated development host in the release 1.1 record and the 48-core VM in release 1.2. OS-visible RAM, allowed CPU sets, kernel/toolchain, native compilation, tick frequency and actual telemetry are retained in metadata. Frequency and individual worker placement were not fixed. Coupled SKU CPU/RAM changes do not identify an independent RAM-capacity effect.

The release 1.2 campaign used the same instance types: a \texttt{c8a.metal-48xl} bare-metal host for CPU studies, \texttt{c8a.12xlarge}, \texttt{c8a.24xlarge} and \texttt{c8a.48xlarge} VMs as 48-, 96- and 192-core workers, the 16-core \texttt{c8a.4xlarge} frontend, and a 48-core \texttt{c8a.12xlarge} frontend for rates at which the 16-core frontend would itself become the bound. The operating-system image changed from Ubuntu 22.04 with kernel 6.8 to Amazon Linux 2023 with kernel 6.18. Measurement groups ran in parallel on separate instances, whereas the earlier campaign ran them in sequence; all instances were in one availability zone without a cluster placement group, and measurement traffic used private addresses. A second 192-core VM repeated the one-connection confirmation. Before release 1.1 records are overlaid as a baseline on the same axes, same-day release 1.1 must reproduce them within a preset band: 5\% for the serving compute-interval rate and 3\% for prepared proving throughput. The serving rate was within 1.1\% of the record; the proving rates differed by $-0.4\%$ at 192 workers and $+0.1\%$ at one worker.

\subsection{External comparison: what was and was not established}
The principal memory contrast compares StateSync-GKR's test-only \texttt{DenseLayerOracle} with its production \texttt{SparseLayerOracle} inside the same prover driver. The named external implementation tested separately was the educational Lambdaworks GKR engine~\cite{lambdaworks} at commit \texttt{3a87850dc405e12ab6c4bbaeba68d52815ab1158}. Common small mixed circuits at widths 2 and 4 completed normal, input-binding and codec checks, with one warmup and three timed proofs per case. The width-8 first normal warmup failed verifier acceptance, so it produced no valid timed performance observations. The probe's transcript and adapter encoding differ from this engine's application encoding. Neither its failed run's small RSS nor missing large-case performance is plotted as a zero-cost competitor. The complete probe remains in the raw archive; a fair large-circuit external performance and memory ranking remains unestablished. The probe was not repeated for release 1.2, and no external engine figure is compared with the release 1.2 measurements.

Prior systems make deployment limits part of the research question. DIZK distributes a single large zkSNARK computation to overcome single-machine memory limits~\cite{dizk2018}. Our measured task is sparse representation and delivery of independent GKR proof requests on CPU workers. The controlled observations identify which resource cost changed and which completion endpoint the caller actually receives. They support instance sizing and serving-policy selection for this artifact without transferring the comparison or security claims of another proof system.

\subsection{Same-driver memory contrast}
Four two-layer circuit families cover mixed linear/product, linear, product and cube/affine arithmetic. Widths are powers of two from 4 through 4,096. Each mode has three independent processes with one warmup and three timed proofs. Common inputs and outputs are bound before proving; protocol verification and both final input-MLE checks occur afterwards. Canonical field-basis proof observations match for shared cases where both modes complete. This benchmark JSON representation is separate from the official \texttt{inner-proof-v1} encoding.

The observer records successful requested-allocation ownership, exposed Vec length/capacity and stage lifetimes. It includes Dense constructor temporaries and retained Sparse buffers. Requested bytes exclude allocator metadata/slack, stacks and internal transient realloc overlap. Counter/stage overhead stays inside both prover timers. JSON construction, hashing, verification and output occur outside those timers but inside whole-process RSS. Different families' fixture representations prevent a causal cross-family RSS interpretation.

The release 1.2 observer adds a count of allocation calls per proof. The inline size of the oracle structure, which grows in release 1.2 by the stored interpolation matrix, is recorded separately and is not added to requested allocation or Vec capacity. The Dense reference is measured again on the same day, so that every paired ratio uses same-day Dense and Sparse processes, and a same-day contrast of the release 1.1 and release 1.2 provers at widths 4,096 and 65,536 isolates the prover change.

Address-space and physical-memory experiments are distinct. A 4-GiB address-space guard can produce an allocation abort; a 3-GiB cgroup with swap disabled can produce a native OOM kill. Successful concurrent jobs share a generic circuit/witness while retaining private oracle/proof state. Readiness barriers expose real overlap without requiring layer lockstep. Sparse-only scale successes have no completed Dense proof counterpart.

\subsection{Arrival and acceptance endpoint}
The load caller schedules periodic arrivals independently of responses. It maintains bounded pending and sender queues. The release 1.1 worker, reproduced as the original mode of the release 1.2 binary, has one FIFO per operation kind, flushes the oldest ready batch subject to a cap/wait policy, runs one compute batch at a time, parallelizes witness/proving/encoding, and then hashes, frames, writes and logs complete encoded proofs. Maximum wait is a readiness trigger, not a bound on time waiting behind another batch.

Release 1.2 is measured in two modes (\Cref{tab:modes}). Batch-fused runs the release 1.1 cap/wait grid cell for cell. Continuous runs map each release 1.1 cell to a limit on jobs computing at once, the smaller of the batch cap and the worker count minus two, and fold the wait axis, which has no continuous counterpart. Same-day release 1.1 runs use the preserved release 1.1 executables. To separate the serving change from the prover change, the 192-core study also builds both release 1.2 modes with a prover whose sparse-oracle source is byte-identical to release 1.1, and runs the original mode with release timers to measure how long freeing each batch takes.

The balanced workload cycles 192 fixed fixtures per membership, non-membership and update kind at depth 24. The frontend checks framing, byte hash, reference proof correspondence and the original request, and ends the measured path only after strict encoded verification accepts. Hash equality alone is not the endpoint. Actual full bodies cross TCP between separate machines; loopback is used only by the small functional replay. Rows overlaid on the release 1.1 records use the same 16-core frontend with 16 verification workers. Higher rates use the 48-core frontend with 44 verification workers, and three same-day release 1.1 runs at 1,000/s from each frontend measure the frontend effect alone.

Every offered request has an identity within its phase/process. Submitted requests retain their send/admit/return/receive/complete chain. Explicit pre-send failures have no server chain. Warmups remain in raw records but leave timed distributions. Server refusal, pre-send overflow, late acceptance, deadline-scanner observation and connection EOF are separate outcomes. All timed offers remain in success-rate denominators; accepted-only CDFs do not hide unsuccessful offers. Release 1.2 servers name the reason for each refusal: total queue full, ingress queue full, stopping, run limit reached before computation, or a reply larger than the return budget. The campaign adds one outcome class: a request still unanswered when the frontend stops waiting at its drain limit. It occurred for same-day release 1.1 at 1,000/s on 48 cores (42 requests, with a 10-s drain limit) and in the send-limited overload run, and it counts as a failure in every denominator.

\paragraph{Why offered rates are controlled.}
The local CPU studies run a fixed amount of prepared work and divide completed jobs by the measured batch interval. The service studies instead schedule new requests independently of response completion, at controlled offered rates. A setting of 500/s means 500 new requests each second, not one batch of 500 requests or an engine requirement to know all requests in advance. Rates, batch caps, waits and worker limits are experimental variables, not product constants. The sweeps locate underload, rising queues and explicit admission limits. Measuring from each intended arrival prevents a slow service from reducing the test's demand merely by delaying its replies~\cite{wrk2method}. The caller is separate code; it does not use wrk2.

The corpus is fixed and cycled, and arrivals are periodic. These choices permit repeated comparisons; they do not represent random customer traffic, bursts or every circuit. Average completed throughput and latency distribution answer different questions. The 500-ms, 1-s and 2-s budgets are reporting views of the same original observations. They are not a universal requirement of GKR. A configuration can satisfy a 2-s application budget while missing 1 s, and can finish a finite run while accumulating a queue that warrants a longer confirmation.

Connection count is also an experimental variable. The caller opens one TCP connection per listed worker address. Between instances outside a cluster placement group, the provider limits a single flow to 5 Gbit/s~\cite{awsbandwidth}. Replies average about 182.6 kB on the wire in the balanced corpus, so one connection must carry 4.82 Gbit/s of reply frames at 3,300 requests/s and 4.97 Gbit/s at 3,400 requests/s before TCP/IP headers. When the one-connection setting first failed at or above 3,000 requests/s without a frontend limit, rates from 3,400 to 5,000 requests/s were screened with two connections to the same worker. A point is marked frontend-limited when the frontend uses more than 70\% of its cores or its verification-wait p99 exceeds 50 ms while the worker stays below 90\% CPU. Such a point bounds the frontend: a failure there is not attributed to the worker, and a pass there is reported only as a lower bound.

\subsection{Statistics and clocks}
New arrival quantiles use the observed nearest rank at sorted index $\lceil qN\rceil-1$, with integer nanoseconds and no interpolation. The first CPU study retains its floor-index convention $\lfloor(N-1)q\rfloor$; the five-process supplement retains type-7 linear interpolation at $(N-1)q$. Memory summaries use timed medians within each process. Cross-process medians and full ranges are descriptive, not confidence intervals.

Frontend latency is the original monotonic \texttt{scheduled\_to\_accept\_ns}. Original offered timestamps are retained rather than reconstructed from scheduled time and lateness. Unix timestamps locate events but need not equal differences of monotonic intervals. We do not subtract clocks across hosts to produce one-way network latency. CPU observations use HZ 100 ticks and actual sampling intervals; the requested 20-ms cadence has its observed distribution. Missing telemetry remains missing. GNU-time peaks and sampled \texttt{/proc} RSS remain separate when they differ.

Batch timers are shared and counted once, not multiplied by jobs. Encoding-to-last-write and between-batch intervals include hashing, framing, logging, release and other service work. Their entire sum is not labelled network time. Completion-window rates include drain; accepted count divided by scheduled duration can conceal queue growth.

Release 1.2 records per-job stage times on the computing worker: dispatch, witness, proof, encoding, hash and release, together with the thread that released the job's memory. Batch timers exist only in the batch modes. In continuous runs, a job's compute interval runs from dispatch to the end of encoding; the compute union is the union of these intervals over a run. The mean number of jobs computing at once is the sum of compute intervals divided by the length of their union, and the compute-union rate is the number of jobs divided by that length. These quantities take the place of batch intervals in the continuous panels of the appendix.

\paragraph{Reading the main contrasts.}
Dense and Sparse denote two StateSync-GKR oracle representations, with the same circuit, field, transcript order and proof driver. Figure labels identify SS-GKR (Sparse) and the SS-GKR Dense reference as this internal comparison. The memory curves retain every measured width and process; width-4096 bars separate requested allocation from whole-process RSS. They are not an external-engine ranking. Every series has a legend entry. Release 1.1 records carry their date, StateSync-GKR v1.1 (10-07) for serving and memory and StateSync-GKR v1.1 (10-06) for CPU; StateSync-GKR v1.1 same-day marks release 1.1 runs repeated in the release 1.2 campaign; StateSync-GKR v1.2 batch-fused and StateSync-GKR v1.2 continuous name the two measured serving modes. Memory figures compare SS-GKR (Sparse) v1.2 with the Dense reference measured on the same day; the twelve Dense-only appendix panels keep the 7 Oct Dense measurements.

The direct fresh/prepared comparison uses 1,000 timed depth-24 requests per process and three processes per operation/mode, for release 1.2 and, interleaved in the same blocks, for same-day release 1.1 prepared-request anchors; fresh requests were not repeated for release 1.1, so its fresh/prepared ratios are those of the 6 October record. Ratios pair process summaries by run label, not individual requests; encoding, transport and acceptance are outside that timer. The Table/Derived contrasts time the complete prepared typed verification call on the same proof, request and circuit, changing only the wiring oracle; they are release 1.1 measurements and are unchanged in release 1.2. Compilation, wiring construction, commitment, proof generation and encoding are outside those timers. The serial/parallel-output ratios time the complete prepared local batch: witness generation, proving, encoding and encoded verification. Release 1.2 repeats 145 of the 175 supplement processes because proving lies inside that interval; the 30 processes at depths 28 and 32, which feed only the unchanged wiring comparison, are not repeated. Witness/proving are parallel in both modes; encoding/verification alone change their parallel policy. Preparation, transport and post-timer proof hashing/logging are excluded. Proving-only scaling excludes witness, encoding and transport. Complete encoded proof lengths retain all measured operation/depth profiles.

Serving figures retain separate runs, clock origins and all-offer success denominators. Equal-resource runs last 20 seconds with three process pairs per policy. Fine-study p99 is conditional on acceptance, while refusals remain in the detailed all-offer tables. Input-stop markers identify the end of new demand; falling pending counts afterwards mean that requests are still completing. In the release 1.1 record, the 1,150/s confirmation reaches maximum latency 1,598.489 ms and final acceptance at 61.394 s: 48,553 offers meet 1 s, all 69,000 meet 2 s. These are reporting budgets for the same observations.

\section{Memory, scale and result preservation}
\label{sec:memory}
At mixed width 4,096, the median of three process medians for additional requested allocation is 1,879,247,872 bytes in Dense and 579,584 bytes in Sparse, a ratio of 3,242.41, in both releases. Exposed oracle Vec capacity is 1,610,612,736 versus 508,096 bytes, also in both releases. Whole-process GNU-time RSS has a separate ratio: 89.46 in the release 1.1 record (1,811.54 versus 20.25 MiB) and 88.87 in release 1.2 (1,811.76 versus 20.39 MiB, with the Dense reference measured again on the same day). The observed prover intervals are 11,610,216,436 versus 4,630,590 ns in the release 1.1 record and 11,631,531,215 versus 1,797,814 ns in release 1.2, so the sparse prover is about 2,507 times faster than the dense reference in the release 1.1 record and about 6,470 times faster in release 1.2. Release 1.2 makes 85 allocation calls per Sparse proof at this width, against 1,282 for the same-day release 1.1 prover, while its total requested bytes per proof rise from 1,571,688 to 1,710,696: fewer and larger requests replace many small ones, and the requested peak is the same. The three ownership/timing boundaries must not be substituted for one another.

\begin{figure}[t]\centering
\planfig{174mm}{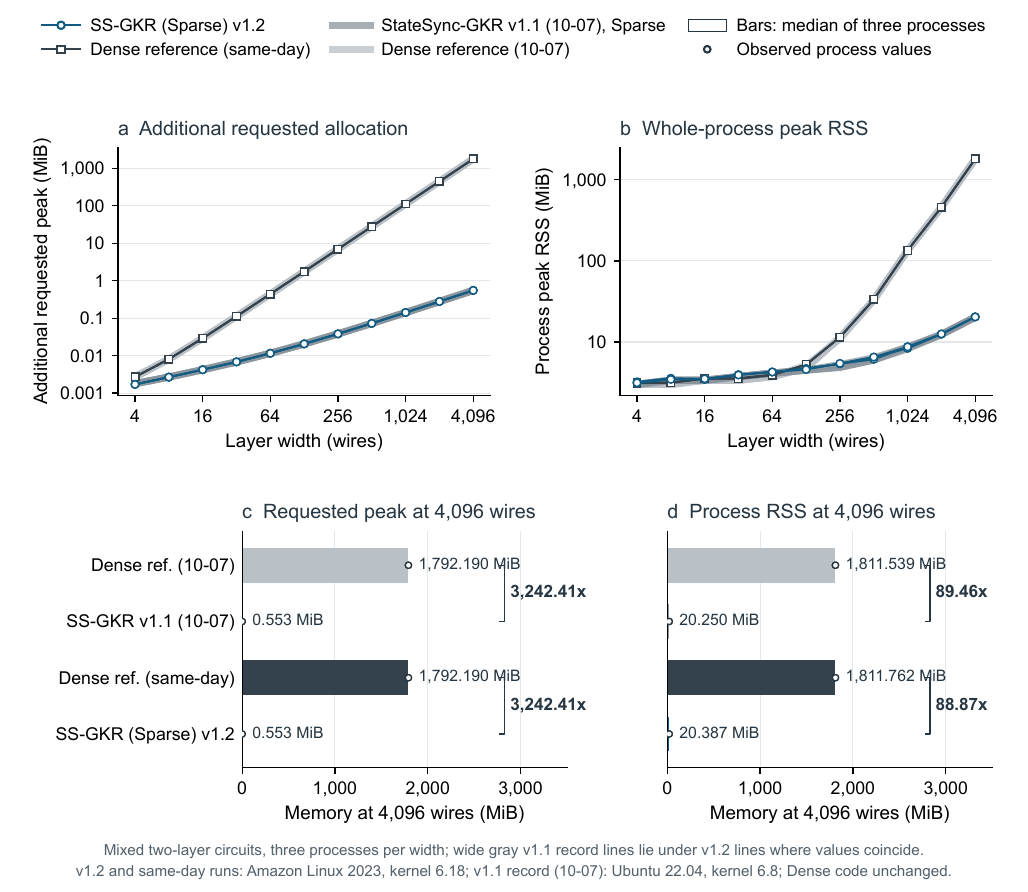}
\caption{SS-GKR (Sparse) versus its internal Dense reference on mixed two-layer circuits. At width 4,096, additional requested allocation is 3,242.41-fold lower in both releases, and process RSS is 88.87-fold lower in release 1.2 (89.46-fold in release 1.1).}\label{fig:memory}
\end{figure}

\Cref{fig:memory} connects the quadratic table construction to the observed growth. This is a controlled internal contrast, with common circuits, field and proof-result checks. It supports avoiding quadratic auxiliary tables when the circuit has a sparse multiplication list. The reduction comes from the sparse representation of release 1.1; release 1.2 retains it and does not claim it anew. It does not measure a production SMT memory reduction of 3,242 times or establish a ranking against every external engine. In the same-day contrast of the two provers, release 1.2 proves in 0.387 times the release 1.1 time at width 4,096 and 0.364 times at width 65,536, with 101 instead of 1,698 allocation calls at the larger width and process RSS within 0.4\%.

At width 2,048, Sparse completes one, two, four and eight concurrent jobs under the 4-GiB AS guard. Dense completes one, two and four, then aborts allocation at eight. Under the distinct 3-GiB cgroup, Sparse four/eight and Dense four succeed; Dense eight is OOM-killed. With active-cap entry records confirming 3 GiB and swap 0, Dense width 8,192 is OOM-killed before returning a proof. Sparse widths 8,192, 16,384, 32,768 and 65,536 each return one warmup and three timed proofs. Release 1.2 repeats all 17 attempts with the same outcome classes. Its eight-job Sparse allocation peak is higher, 2,120,976 bytes under the AS guard and 2,318,912 bytes under the cgroup against 1,481,936 and 1,273,640 bytes in the release 1.1 record. Per-job start and end times point to overlap: in release 1.2 all eight jobs started within 11.5 $\mu$s (5.0 $\mu$s under the cgroup) and ran together, whereas in the release 1.1 record the starts spread over 4.6 ms and at most six jobs overlapped (five under the cgroup). This repeat ran on the 48-core VM, whereas the release 1.1 record used a four-vCPU development host on which the eight jobs shared four cores; the difference in overlap, and with it the higher eight-job peak, is confounded with the host and is not attributed to release 1.2. The one-, two- and four-job peaks are equal or lower. Width 65,536 is the largest tested point, not an established global supported maximum.

The application benefit is additional circuit or independent-job capacity within a memory budget. The comparison preserves unsuccessful attempts and their native dispositions. It does not turn missing failure telemetry into a manufactured peak or interpret a controller's later default limit as the cap active during execution.

Appendix families \hyperref[atlas:A01]{A01}, \hyperref[atlas:A02]{A02}, \hyperref[atlas:A03]{A03}, \hyperref[atlas:A04]{A04}, \hyperref[atlas:A05]{A05}, \hyperref[atlas:A06]{A06} retain all four arithmetic families, stage ownership, distributions and actual telemetry, with release 1.2 overlaid; their twelve Dense-only panels are unchanged in release 1.2. The memory-guard and tested-scale evidence appears in \hyperref[atlas:A07]{A07}, \hyperref[atlas:A08]{A08}, \hyperref[atlas:A09]{A09}, \hyperref[atlas:A10]{A10}, and \hyperref[atlas:A47]{A47} shows the same-day contrast of the two provers.
\section{Completed computation and reusable structure}
\label{sec:compute}
\begin{figure}[t]\centering
\planfig{174mm}{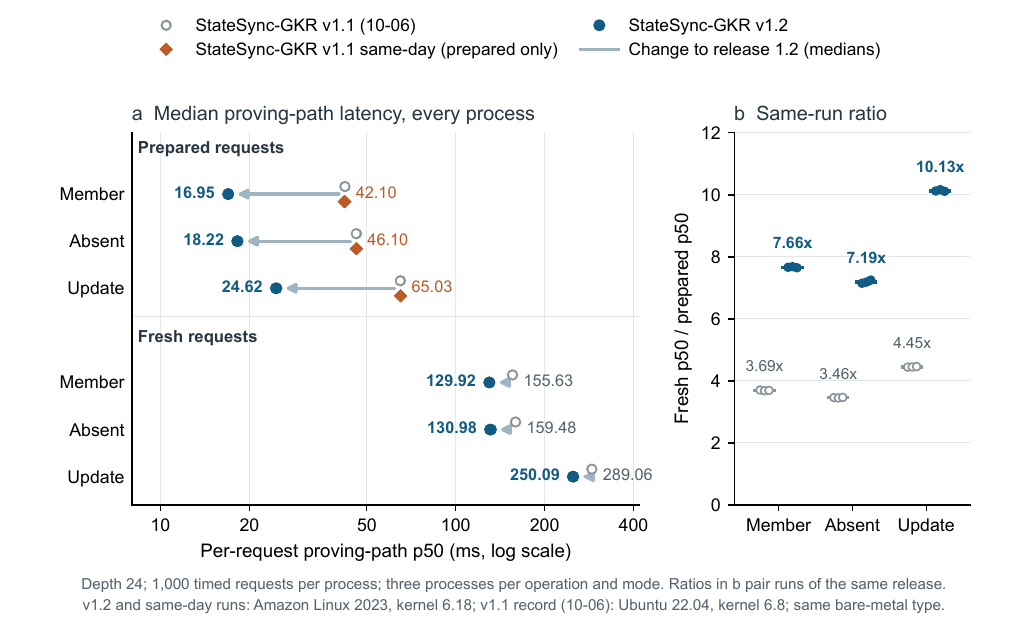}
\caption{Preparation reuse at depth 24. Release 1.2 prepared requests take 16.95 to 24.62 ms at the median against 42.10 to 65.03 ms for same-day release 1.1, so fresh/prepared ratios rise to 7.19--10.13. Points retain three process medians per operation and mode.}\label{fig:preparation}
\end{figure}

\begin{figure}[t]\centering
\planfig{174mm}{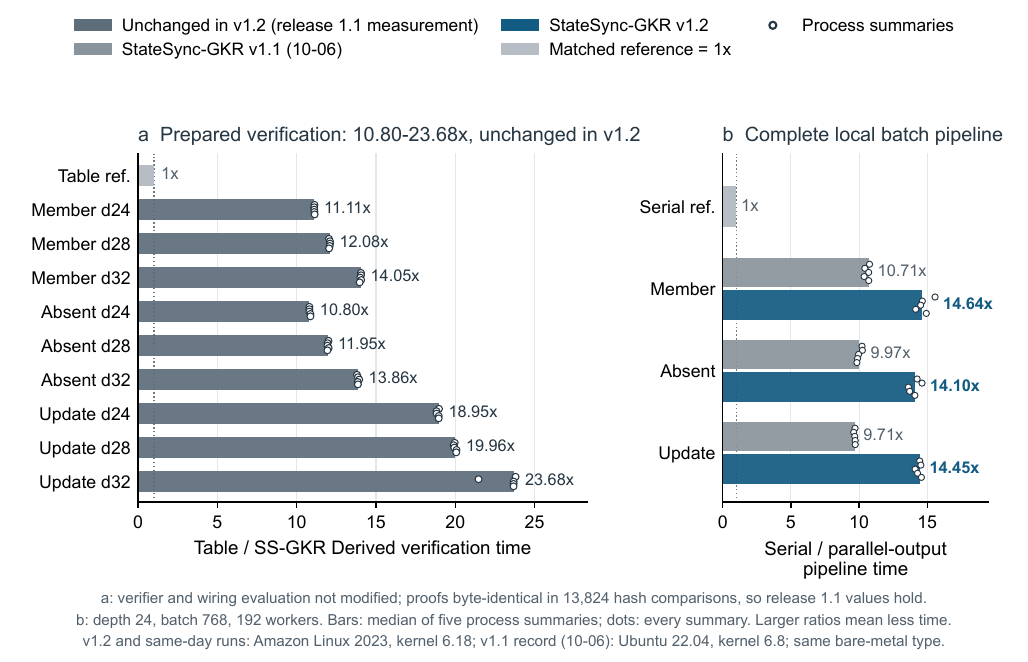}
\caption{Prepared verification and complete local batch speedups in SS-GKR. (a) Table/Derived verifier calls, unchanged in release 1.2; (b) serial/parallel output policies, 14.10--14.64-fold in release 1.2 against 9.71--10.71-fold in release 1.1. References are normalized to 1.}\label{fig:verification}
\end{figure}

\begin{figure}[t]\centering
\planfig{174mm}{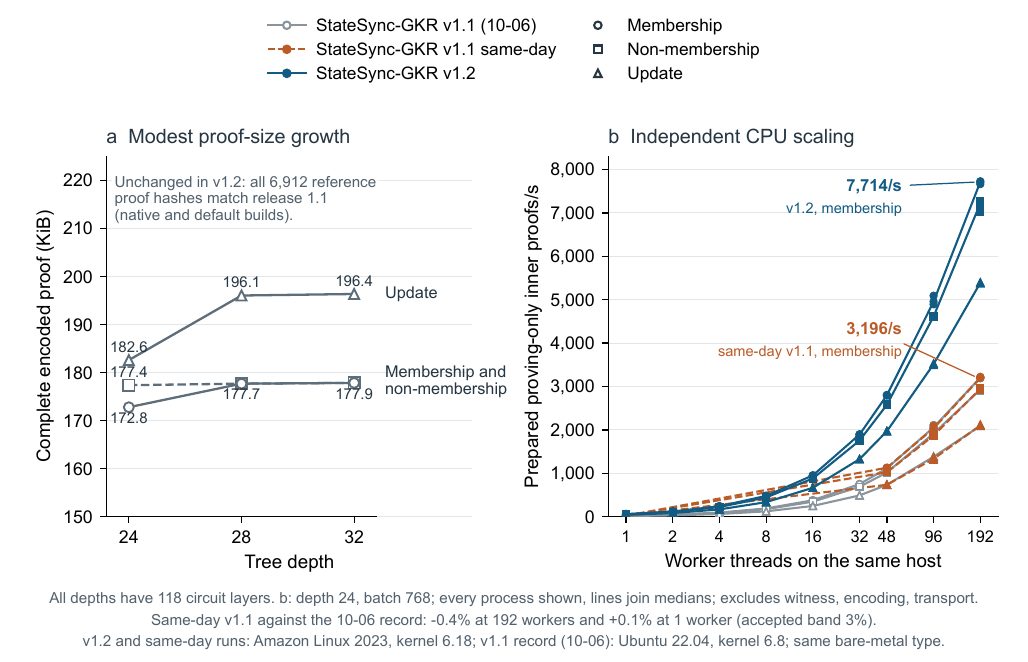}
\caption{Compact proof growth and independent CPU scaling. (a) Complete proof bytes across tree depths, unchanged in release 1.2; (b) prepared proving-only throughput across workers on the same host, where release 1.2 reaches 7,714 membership proofs/s at 192 workers against 3,196/s for same-day release 1.1.}\label{fig:compact}
\end{figure}

\Cref{fig:preparation,fig:verification,fig:compact} separate direct preparation reuse, component reductions and compact proof growth from independent CPU scaling.

On the bare-metal host at depth 24, batch 768 and 192 workers, release 1.2 proves 7,713.71, 7,220.36 and 5,390.07 inner proofs/s for membership, non-membership and update, against 3,195.50, 2,947.08 and 2,102.02/s for same-day release 1.1 and 3,207.74, 2,930.85 and 2,092.03/s in the 6 October record. Relative to same-day release 1.1, the release 1.2 proving rate at 1, 48, 96 and 192 workers is 2.53, 2.48, 2.43 and 2.41 times higher for membership, 2.56, 2.52, 2.47 and 2.45 times for non-membership, and 2.71, 2.68, 2.63 and 2.56 times for update. Parallel witness plus proving yields 6,323.95, 5,899.04 and 4,782.21/s, against 2,881.55, 2,654.03 and 1,984.00/s in the record. The broader internal witness/prove/encode/encoded-acceptance interval yields 2,264.06, 2,107.57 and 1,815.31/s, against 1,593.17, 1,431.66 and 1,167.49/s. Hosts and intervals differ; these are not pooled repeats or serving rates. Every proof hash of the byte-equality check, 768 fixtures for each operation at depths 24, 28 and 32 in the native and the default build, equals the release 1.1 hash: 13,824 comparisons without a mismatch.

The release 1.2 gain is smaller at 192 workers than on one worker. A probe that keeps every thread proving shows why. With one pinned thread as the reference, per-thread proof time at 192 threads is 1.19 times longer for the release 1.1 prover and 1.21 to 1.22 times longer for release 1.2, while the effective clock at 192 threads falls from 4.54 GHz to 3.83 to 3.85 GHz in the release 1.1 runs and 3.75 to 3.77 GHz in the release 1.2 runs. The time ratio divided by the clock ratio is 1.00 to 1.01; five tested thread placements differ by about one percentage point, and the serving VM shows the same shape. A prototype that replaces only the per-round interpolation, removing its small allocations and inversions, recovers 1.6 to 4.2\% of the 192-thread slowdown. The batch driver adds a further 1.20 to 1.25-fold per-thread time over the probe, and its parallel efficiency at 192 workers is 69 to 70\% for release 1.1 and 66 to 67\% for release 1.2. Each of its 192 workers proves four of the 768 jobs and the batch waits for its last job; we read this batch structure as the main source of the extra loss, although per-job times inside these CPU batches were not recorded. \hyperref[atlas:A44]{A44} shows the probe, clock, placement and prototype measurements; the batch ratios use the worker matrix of \Cref{fig:compact}(b).

For depth-24 membership in release 1.2, direct compilation costs 7.816129 ms and circuit commitment 104.2647445 ms; the prover call costs 16.3522 ms, against 41.538730 ms in the release 1.1 profile, where compilation cost 7.902078 ms and commitment 105.027310 ms. Across the three operations proof generation drops to 0.37 to 0.39 of its release 1.1 time, while compilation, commitment and witness generation stay within 0.98 to 1.01 of the record. Preparation reuse therefore removes a substantial commitment as well as compilation, and its relative benefit grows when proving becomes cheaper. \Cref{fig:preparation} measures preparation reuse directly: prepared requests take 16.95, 18.22 and 24.62 ms at the median against 42.10, 46.10 and 65.03 ms for same-day release 1.1, and fresh/prepared ratios rise from approximately 3.69, 3.46 and 4.45 in the release 1.1 record to 7.66, 7.19 and 10.13. Amortization over a chosen reuse count is a cost model using these measurements, not a new observed sample.

Same-proof Table/Derived prepared-verification ratios span 10.8000--23.6757 across nine depth/operation profiles. They are release 1.1 measurements and are unchanged in release 1.2: the verifier and the wiring evaluation are not modified, and the proofs are byte-identical. Each process summarizes paired samples, followed by the five-process median. Table is already a sparse wiring oracle. The timer encloses the complete \texttt{verify\_sync\_op\_with} call, including request validation and native input/path hashing; it is not a wiring-polynomial microbenchmark. Prepared circuit, wiring and commitment construction are excluded. In \Cref{fig:verification}, each matched reference is normalized to 1; the colored bars show the median of five process-level paired-ratio summaries and the points retain all five summaries. Larger ratios mean a smaller component time. Gray reference bars define the normalization rather than add measured observations. The reduction in \Cref{fig:verification}(a) quantifies the verification-call improvement from changing the wiring representation; the companion model-level equivalence~\cite{fvpaper} specifies the semantic relation that the implementation must preserve while leaving the Rust/model correspondence boundary explicit.

Both supplement output modes parallelize witness generation and proving. They differ in output encoding and encoded verification. At depth 24, batch 768 and 192 workers, release 1.2 gives ratios of 14.64, 14.10 and 14.45 for the three operations, each the median over five processes of per-process medians of ten paired serial/parallel interval ratios, against 10.71, 9.97 and 9.71 in the release 1.1 record. The paired interval encloses the complete prepared local batch, from witness and proof generation through encoding and encoded verification, not only its output stages. Proving is a larger share of the parallel batch than of the serial one, so faster proving raises the ratio. Thus the output-policy change lowers whole local batch time while preserving the same computation. It is not the ratio of separately summarized throughput medians and it excludes transport. Smaller batches have their own effects. In the release 1.1 record, a total of 327,360 serial/parallel proof-byte pairs preserve the canonical result; the release 1.2 supplement control cell passed the same check on 4,608 pairs.

\Cref{fig:compact}(a) shows the proof-size consequence of the fixed 118-layer schedule across all nine operation/depth profiles at depths 24, 28 and 32. Membership encoded proof size rises from 176,948 to 182,132 bytes, about 2.93\%. These sizes are unchanged in release 1.2, since all 6,912 reference proof hashes match release 1.1. Fused rounds and parallel path residuals explain the schedule. Width, witness work and total computation still grow; the modest proof growth is not a constant-work claim.

\FloatBarrier
\section{Serving profiles, capacity and configuration choices}
\label{sec:serving}
\begin{figure}[t]\centering
\planfig{174mm}{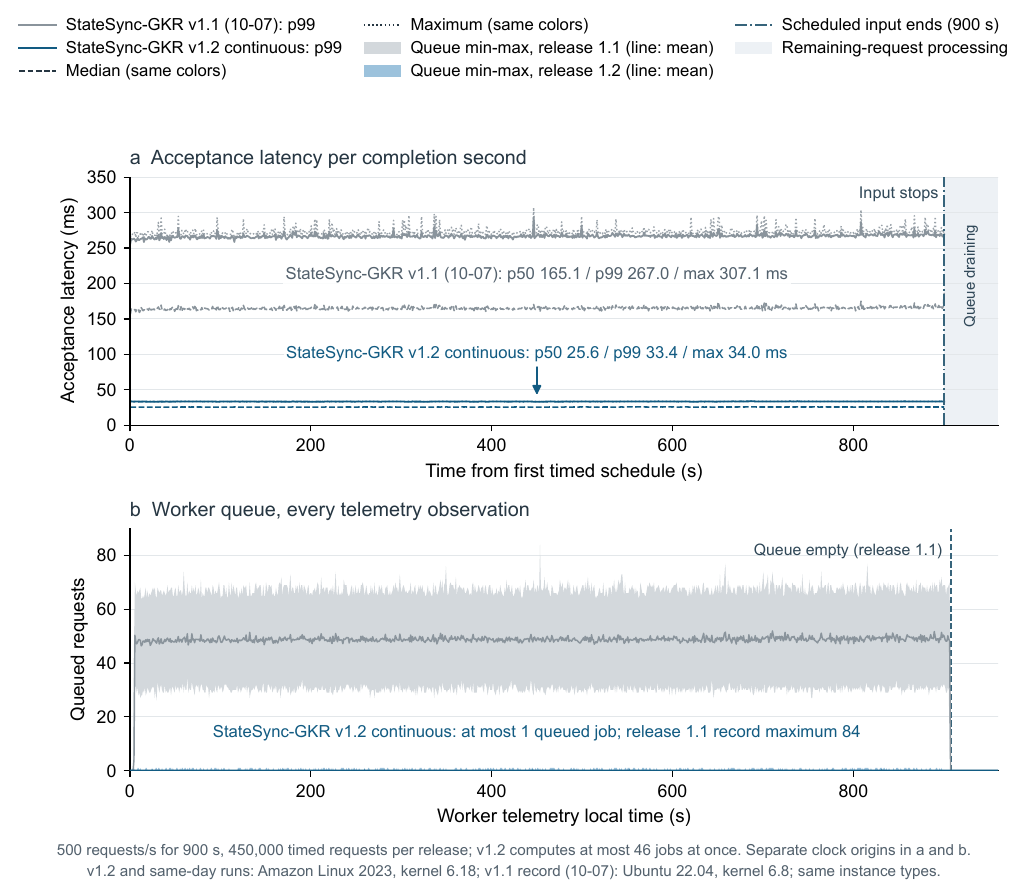}
\caption{One 48-core worker at 500 requests/s for 15 minutes: release 1.2 keeps p99 at 33.4 ms against 267.0 ms in the release 1.1 record. Panels show acceptance latency and worker queue.}\label{fig:stability}
\end{figure}

\begin{figure}[t]\centering
\planfig{174mm}{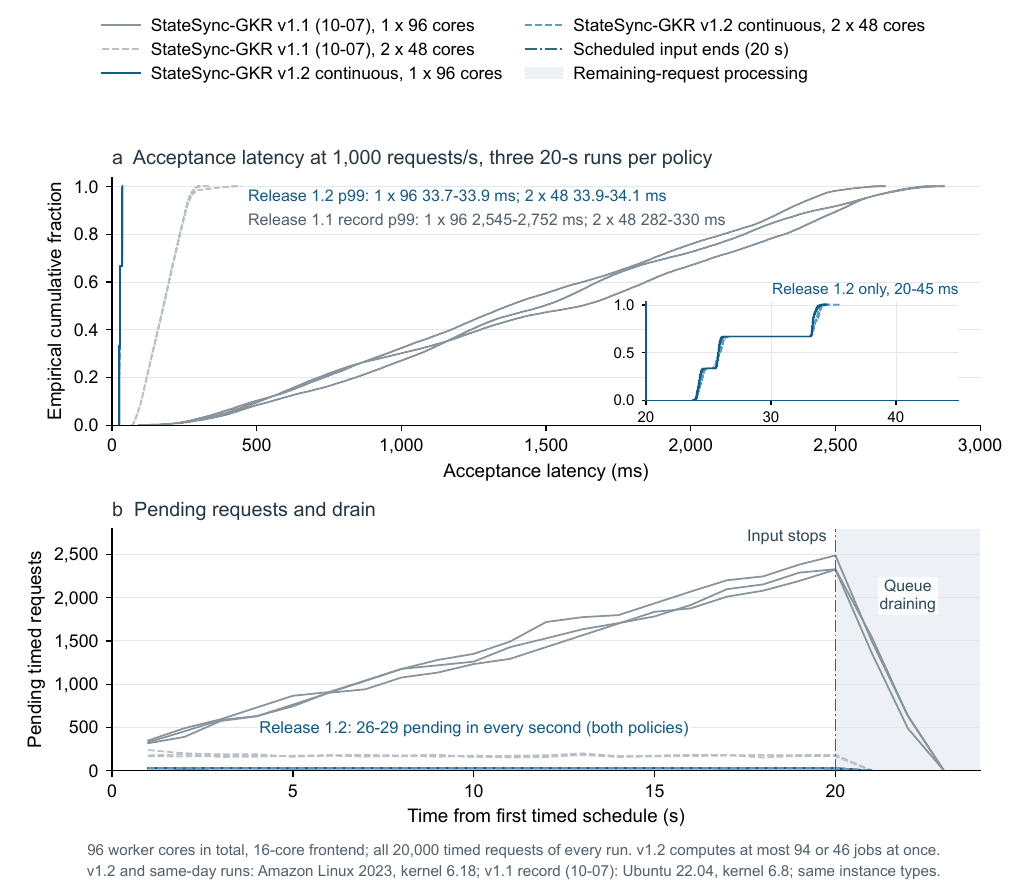}
\caption{Equal-resource SS-GKR deployment policies at 1,000 requests/s. In release 1.2, one 96-core worker and two 48-core workers both keep p99 near 34 ms; in the release 1.1 record only the two-worker policy met one second.}\label{fig:horizontal}
\end{figure}

\begin{figure}[t]\centering
\planfig{174mm}{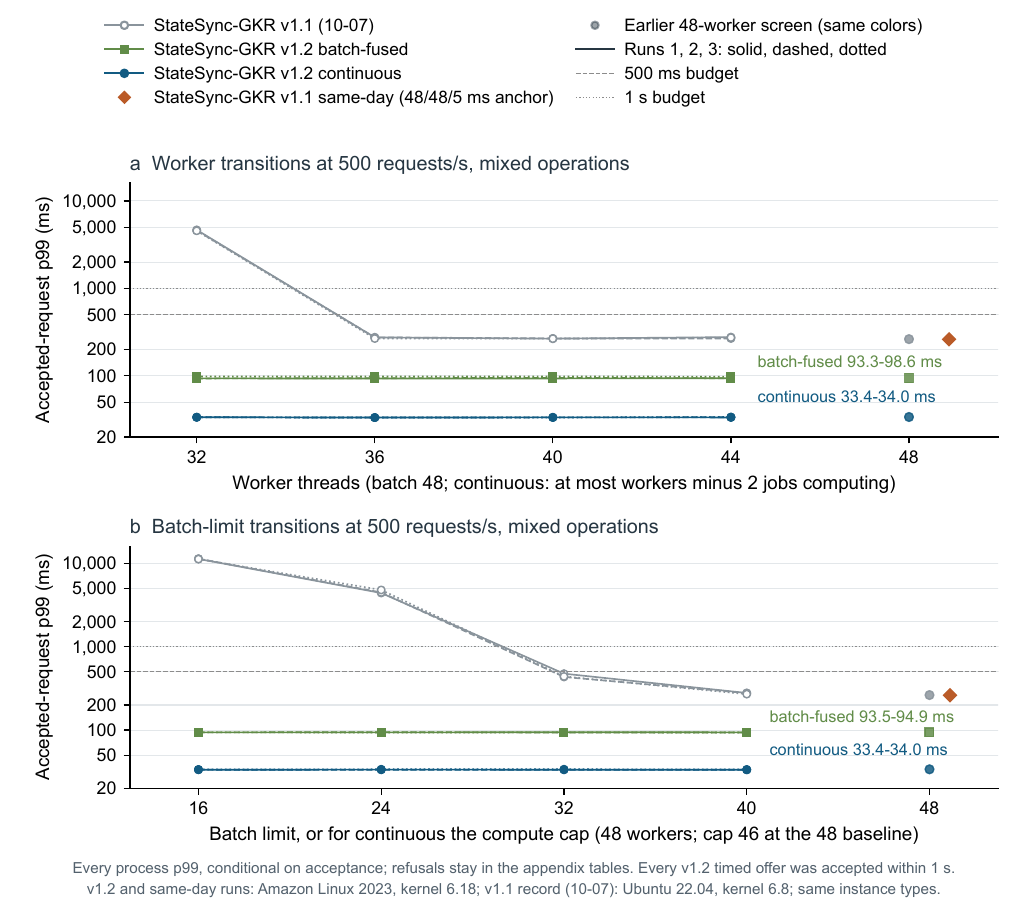}
\caption{Worker and batch settings on one 48-core VM. Release 1.1 tail latency ranges from hundreds of milliseconds to above 10 s, while release 1.2 keeps mixed-workload p99 between 33 and 99 ms. Points retain individual runs.}\label{fig:fine}
\end{figure}

\begin{figure}[t]\centering
\planfig{174mm}{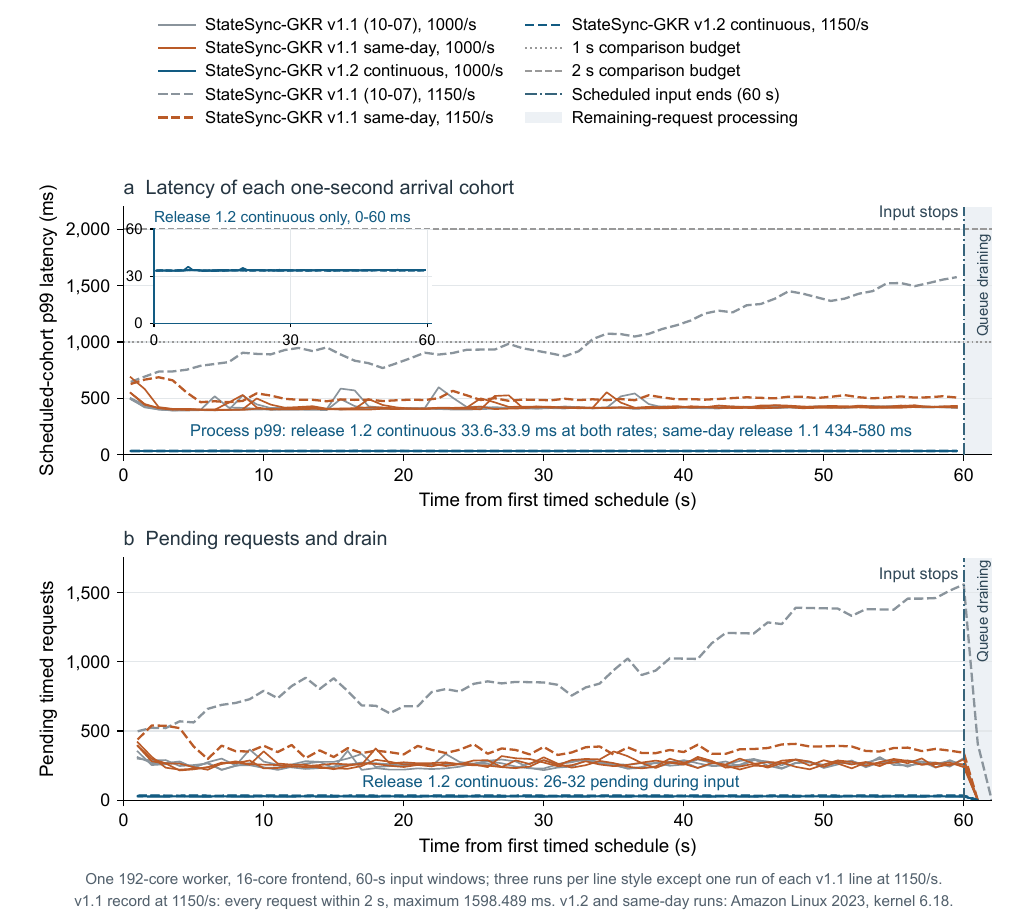}
\caption{SS-GKR delivery at 1,000/s and 1,150/s over 60-second input windows. Release 1.2 keeps p99 at 34 ms at both rates, against 434--580 ms for release 1.1 measured on the same day; input-stop and draining markers preserve completion.}\label{fig:confirmation}
\end{figure}

\Cref{fig:stability,fig:horizontal,fig:fine,fig:confirmation} compare a longer baseline, equal-resource policies, tuning transitions and reporting budgets; \Cref{fig:capacity,fig:timeshare} add the release 1.2 capacity limits and the location of serving time.

\subsection{A confirmed 48-core baseline}
With 48 workers, batch cap 48 and maximum wait 5 ms, the release 1.1 record on an actual 48-core VM accepts 450,000 balanced timed offers at 500/s over 900 seconds. Every acceptance occurs within 500 ms. Exact p50/p95/p99/max are 165.114057/248.254563/267.025246/307.095548 ms; 95.601333\% meet 250 ms. Mean worker/frontend utilization is 27.108/3.479 core equivalents. Queue observations remain bounded in that window, with minute-level mean pending counts around 48--49. Release 1.2 in continuous mode, with 48 workers and at most 46 jobs computing at once, repeats the run on the same VM type: all 450,000 timed offers are accepted within 100 ms, with p50/p95/p99/max of 25.620379/33.312304/33.446495/33.998997 ms, no late acceptance and no refusal. In every 20-ms telemetry sample its worker queue held at most one waiting job, against a maximum of 84 in the release 1.1 record. These are finite observed confirmations, not an unlimited-duration leak or availability guarantee.

Timed proof bodies in the release 1.1 run total 81,837,900,000 bytes, with 82,184,472,577 application-frame bytes; release 1.2 sends the same proof bodies, since the proofs are byte-identical. TCP/IP headers are not included. The result involves real complete proof transport and final frontend verification rather than an acknowledgement standing in for proof reception. The complete short-screen conditions, budgets, batching and 15-minute observations appear in \hyperref[atlas:A11]{A11}, \hyperref[atlas:A12]{A12}, \hyperref[atlas:A13]{A13}, \hyperref[atlas:A14]{A14}, \hyperref[atlas:A15]{A15}, \hyperref[atlas:A16]{A16}; continuous runs form no batches, so A12 gives their per-job stages.

\subsection{Equal-total-resource deployment}
The comparison holds total worker resources at 96 cores/192 GiB and uses the same 16-core frontend. One 96-core worker uses 96 workers/cap 96; two 48-core workers each use 48 workers/cap 48. Maximum wait is 5 ms. Release 1.2 runs the continuous mode with at most 94 and 46 jobs computing at once. Thus it compares deployment policies, not topology alone.

In the release 1.1 record at 1,000/s total, two workers each receive 500/s and all timed offers in three 20-second processes accept within 500 ms; process p99 is 282.108--329.785 ms. The single 96-core policy eventually accepts every submitted request but accumulates queue, with p99 2.545--2.752 s and 27.075--32.31\% acceptance within one second. In release 1.2 both policies accept every timed offer within one second. Process p99 is 33.736--33.885 ms for one 96-core worker and 33.881--34.125 ms for two 48-core workers, with 20.52--20.53 and 20.74--20.82 worker core equivalents; at 500/s the two policies give 33.287--33.451 and 33.685--33.759 ms. The complete proof-body byte total is equal for the same offered workload. Similar CPU usage does not imply similar response latency when batching and the serial remainder of the caller differ, as the release 1.1 record shows.

The release 1.1 one-worker service path includes computation, then hashing/framing/writing/logging/release, plus gaps before the next batch; shared compute timers do not account for that entire path (\Cref{sec:timeshare}). In release 1.2, the choice between one 96-core and two 48-core workers no longer changes latency at these loads, and the two-worker recommendation belongs to the release 1.1 record. Neither study confirms a two-worker 15-minute run. All repeated equal-resource distributions, routing, costs and service phases appear in \hyperref[atlas:A17]{A17}, \hyperref[atlas:A18]{A18}, \hyperref[atlas:A19]{A19}, \hyperref[atlas:A20]{A20}, \hyperref[atlas:A21]{A21}, \hyperref[atlas:A22]{A22}, \hyperref[atlas:A23]{A23}, \hyperref[atlas:A24]{A24}, \hyperref[atlas:A25]{A25}; for continuous runs, A25 divides worker time by the union of per-job stages.

\subsection{Fine transitions}
The fine study tests nineteen conditions with three processes each around worker, cap, wait, rate and operation-mix transitions. In the release 1.1 record, of 456,750 timed offers, 451,429 accept and 5,321 receive explicit server-run-limit refusals. Those refusals occur at cap 16/500/s and cap 48/1000/s when the 30-second run limit leaves jobs uncomputed. They remain in success denominators and outside successful latency CDFs. Release 1.2 repeats the grid in both modes. Batch-fused runs all 19 conditions (57 processes, 456,750 timed offers); continuous runs 16 mapped conditions (48 processes, 389,250 timed offers), since the wait settings collapse into one continuous cell. Every timed offer in both modes is accepted within one second, without a refusal.

At balanced 500/s in the release 1.1 record, 32 workers have p99 around 4.56--4.66 s. Workers 36--40 and cap 40 are short-run candidates with all timed requests below 500 ms in the tested 15-second repetitions. They do not demonstrate an equivalent smaller VM or a long-duration price saving. Caps 16 and 24 lose efficiency; cap 32 and its wait sweep expose tail growth. Maximum wait controls batch readiness, while queue behind an active batch can dominate latency. Update-only 500/s does not meet the one-second objective in every process of the release 1.1 record, so the balanced baseline cannot be assigned to every input mix. Release 1.2 removes this sensitivity at these rates: mixed-workload process p99 spans 93.1--98.6 ms in batch-fused mode and 33.3--34.0 ms in continuous mode across the grid, and update-only 500/s stays at 58.8--63.8 ms and 33.5--33.7 ms. All nineteen tuning conditions, including refusals, accepted distributions and CPU/RSS traces, remain in \hyperref[atlas:A26]{A26}, \hyperref[atlas:A27]{A27}, \hyperref[atlas:A28]{A28}, \hyperref[atlas:A29]{A29}, \hyperref[atlas:A30]{A30}, \hyperref[atlas:A31]{A31}, \hyperref[atlas:A32]{A32}, \hyperref[atlas:A33]{A33}, \hyperref[atlas:A34]{A34}; A28 adds the mean number of continuous jobs computing at once.

\subsection{192-core confirmation at 1,000 and 1,150 requests/s}
The 47 actual VM process pairs of the release 1.1 record comprise one pilot, six 500/1000 matching runs, eighteen high-load tuning runs, eighteen runs spaced 50/s apart around the highest passing rate, one 1,150/s 60-second confirmation and three 1,000/s 60-second confirmations. Total offers are 1,114,032: 1,079,666 submitted/accepted plus 34,366 explicit pre-send \texttt{frontend\_max\_pending} failures. Submitted requests have no server refusal, loss, duplicate or unresolved response. The release 1.2 campaign repeats the short grid on the 16-core frontend in both modes and adds confirmations, a search for the highest passing rate on the 48-core frontend, overload runs, a 30-minute run and a second 192-core VM. Replay audits of 187 processes reproduce the frontend's recorded summaries exactly and match every accepted reply against the 576 reference proofs.

With 192 workers/cap 192/wait 5 ms, the release 1.1 record accepts all 180,000 timed offers within one second in three 60-second runs at 1,000/s. Process p99 is 426.032/438.775/443.471 ms. Acceptance within 500 ms is 99.9183/99.7167/99.6217\%, respectively. On the same day as release 1.2, release 1.1 gives p99 of 433.795/440.236/475.093 ms. \Cref{tab:fivecontrol} separates the serving change from the prover change at this rate. Replacing the release 1.1 executable with the release 1.2 serving binary, still with the release 1.1 prover, lowers p99 to 233.6--234.8 ms in batch-fused mode and 74.9--75.3 ms in continuous mode; the release 1.2 prover then lowers it to 95.6--96.0 and 33.7--33.9 ms. Continuous mode with the release 1.2 prover is the default and the release 1.2 operating point for this caller. These five rows are a controlled contrast of the two release 1.2 changes; they do not attribute shares to the individual arithmetic reorganizations. The difference is present far below either limit: at 100 and 250 offered requests/s, release 1.2 continuous p99 is 33.1--33.4 ms against 233.0--248.9 ms for same-day release 1.1.

\begin{table}[t]
\centering\small
\caption{Latency to verified acceptance at 1,000 offered requests/s on the same 192-core VM and day, three 60-second runs per row, 16-core frontend. Rows separate the serving change from the prover change; rows two to five use the release 1.2 serving binary. Every timed offer was accepted within 1 s.}\label{tab:fivecontrol}
\begin{tabular}{@{}llrrr@{}}
\toprule Serving & Prover & p50 (ms) & p99 per run (ms) & Within 500 ms\\\midrule
release 1.1 loop (release 1.1 executable) & release 1.1 & 265.3--266.4 & 433.8 / 440.2 / 475.1 & 99.28--99.88\%\\
StateSync-GKR v1.2 batch-fused & release 1.1 & 151.5--151.6 & 233.9 / 233.6 / 234.8 & 100\%\\
StateSync-GKR v1.2 continuous & release 1.1 & 53.9 & 75.3 / 75.3 / 74.9 & 100\%\\
StateSync-GKR v1.2 batch-fused & release 1.2 & 61.2--61.4 & 95.6 / 96.0 / 96.0 & 100\%\\
StateSync-GKR v1.2 continuous (default) & release 1.2 & 25.6--25.7 & 33.9 / 33.7 / 33.8 & 100\%\\\bottomrule
\end{tabular}
\par\smallskip{\footnotesize\raggedright The release 1.1 serving loop rebuilt with the release 1.2 prover, run with release timers in the same session, gave p99 of 159.2--162.0 ms in three runs. The same loop with the release 1.1 prover ran only with release timers at 1,150 and 1,500/s; at 1,150/s, a frontend-limited point, it met one second for 73.5\% of offers, so its latency equivalence to the release 1.1 executable is not established.\par}
\end{table}

In the release 1.1 record, three 15-second screens at 1,150/s meet one second for every timed request. The separate 60-second observation meets one second for only 70.3667\%, with 20,447 late timed acceptances and p99 1,471.284 ms. Pending work grows from 562 at five seconds to 1,559 at 60 seconds; all 5,750 offers in the final five-second cohort are late. The first completed one-second-budget miss is preserved, and the other two planned repetitions were not executed. The same run meets a 2-s budget for all offers: its maximum latency is 1,598.489 ms. Its finite-window rate is 1,123.89 accepted requests/s, with the final request accepted at 61.394 s. On the same day as release 1.2, a release 1.1 run at 1,150/s met one second for every timed offer, with p99 580.107 ms and 343 requests pending at the end of the window. We report both release 1.1 results and do not attribute the difference. The same-day check of the compute-interval rate was made at 1,000/s, where it was within 1.1\% of the record; at 1,150/s the same-day rate was 2,290 against 2,128 jobs/s in the record, while full batches computed in a median of 87.2 ms in both runs, and release 1.1 operates near its limit at this rate. Release 1.2 kept p99 at 33.6 ms in continuous mode and 96.1--96.3 ms in batch-fused mode in three runs each, with every offer within one second.

Higher-load observations in the release 1.1 record identify explicit admission limits, queue accumulation and drain, rather than a global kernel ceiling. In the 1,000/s confirmations of that record, mean worker usage is approximately 64.6--65.2 of the 192 available core equivalents. At the same rate, release 1.2 uses 20.6 core equivalents in continuous mode and 23.0 in batch-fused mode, against 65.1 for same-day release 1.1 (medians of three process means over the offering window). The serving caller and proof delivery path therefore matter alongside the internal prover.

The 60-s confirmation deliberately keeps arrivals independent of response completion. It asks whether a setting that passed three short 15-s screens also meets a per-request target during a longer continuous offering window. Sixty seconds is the experiment's input duration; the prover does not stop being usable at that age. In the release 1.1 record at 1,150/s, the 69,000 timed requests all eventually pass cryptographic verification. The one-second comparison budget is met by 48,553 requests (70.3667\%); all 69,000 meet two seconds, with maximum 1,598.489 ms. The remaining 20,447 exceed only the one-second view. No request took 60 seconds to complete. The 256 warmups are excluded from these denominators. At the end of the 60-s window, 67,441 timed requests are complete and 1,559 remain. The final acceptance occurs at 61.393972789 s. The last zero pending point is a one-second-bin endpoint, not an invented exact completion timestamp. The actual-machine comparisons, all offered-load screens and independent confirmations appear in \hyperref[atlas:A35]{A35}, \hyperref[atlas:A36]{A36}, \hyperref[atlas:A37]{A37}, \hyperref[atlas:A38]{A38}, \hyperref[atlas:A39]{A39}, \hyperref[atlas:A40]{A40}, \hyperref[atlas:A41]{A41}, \hyperref[atlas:A42]{A42}, \hyperref[atlas:A43]{A43}; the proof-size panel of A36 is unchanged in release 1.2.

\subsection{Capacity of one 192-core worker and the network bound}\label{sec:capacity}
\begin{figure}[t]\centering
\planfig{174mm}{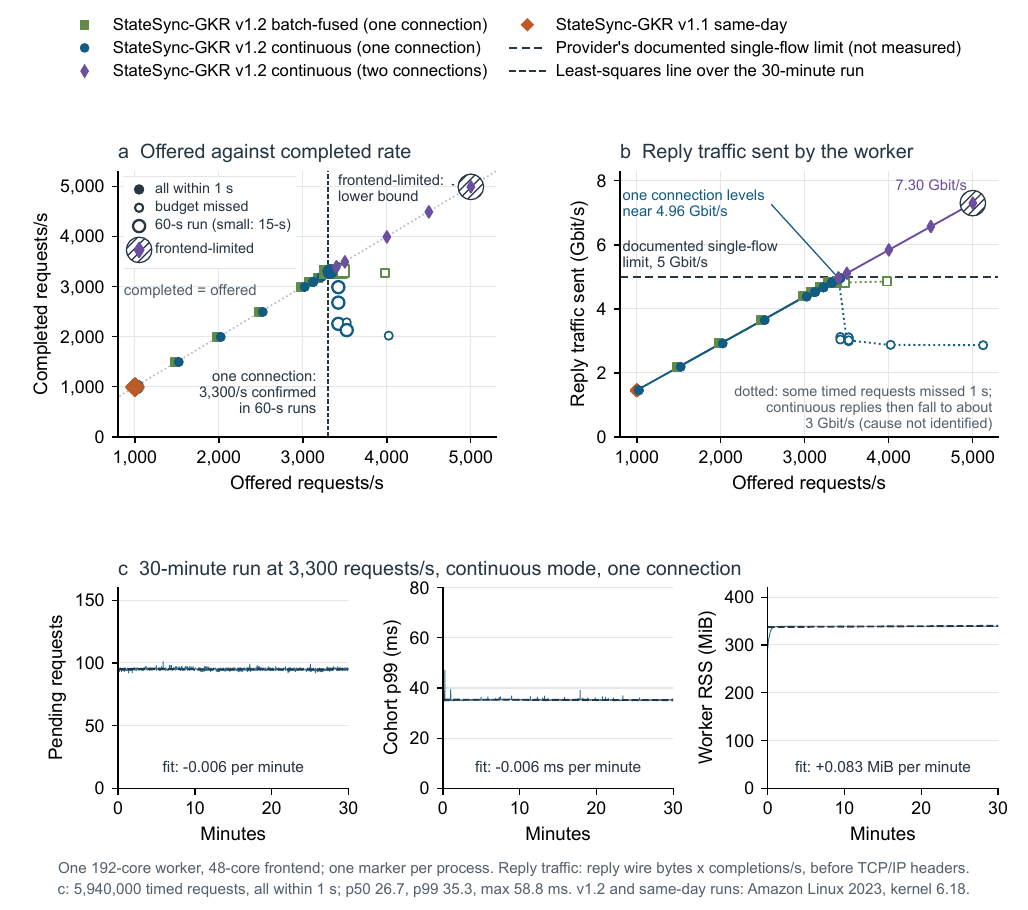}
\caption{Capacity of one 192-core worker in release 1.2. (a) Offered against completed rate and (b) reply traffic: with one connection every request meets one second up to 3,300 requests/s and reply traffic reaches about 5 Gbit/s at 3,400 requests/s; above that rate batch-fused traffic stays near 4.8 to 4.9 Gbit/s while continuous traffic falls to about 3 Gbit/s; two connections pass 5,000 requests/s in 15-second screens. (c) In the 30-minute run at 3,300 requests/s, pending requests and tail latency show no trend, and worker memory rises by 0.08 MiB per minute.}\label{fig:capacity}
\end{figure}

\Cref{fig:capacity} raises the offered rate until the one-second budget fails. On the 48-core frontend with one connection, 15-second screens of both release 1.2 modes met the budget at every rate up to 3,400 requests/s and failed from 3,500/s. Three 60-second confirmations failed at 3,400/s and passed at 3,300/s, with p99 of 35.4--38.5 ms in continuous mode and 225.1--300.3 ms in batch-fused mode. A 30-minute continuous run at 3,300/s accepted all 5,940,000 timed offers within one second, with p50 26.7 ms, p99 35.3 ms and maximum 58.8 ms. Straight-line fits over the run give drifts of $-0.006$ pending requests and $-0.006$ ms of cohort p99 per minute, and $+0.08$ MiB of worker RSS per minute. A second 192-core VM reproduced the confirmation: three 60-second runs at 3,300/s met one second for every timed offer, with p99 of 37.6--46.4 ms.

The one-connection limit is consistent with the provider's single-flow bandwidth limit~\cite{awsbandwidth}. Server records hold no network counters, so reply traffic is computed from the reply wire bytes of completed requests per second. On one connection it follows the offered rate up to 3,300/s (4.82 Gbit/s) and reaches about 4.96 Gbit/s at 3,400/s (4.85 Gbit/s in batch-fused mode) and rises no further, before TCP/IP headers, while the worker used at most 50\% and the frontend at most 52\% of their cores over the offering window. With two connections to the same worker, 15-second screens met the budget at every rate from 3,400 to 5,000 requests/s, and the worker sent 7.30 Gbit/s at 5,000/s. At that rate the frontend used 74\% of its cores and was marked frontend-limited, so 5,000/s is only a lower bound for the worker. Above the one-connection limit the two modes degrade differently: at 3,500 and 4,000 requests/s, continuous completions fell to about 2,000 to 2,200 per second and reply traffic to about 3 Gbit/s, whereas batch-fused held about 3,270 to 3,290 completions per second. We have not identified the cause of this difference.

Same-day release 1.1 met the budget at 1,000 and 1,150/s, the highest rate measured for it in this campaign; its 7 October record met the budget without a frontend limit up to 1,000/s. Changing the frontend from 16 to 48 cores removed the frontend's verification wait for same-day release 1.1 at 1,000/s, with verification-wait p99 falling from 28.7--30.4 ms to below 0.03 ms, while its latency p99 was 457.8--648.9 ms against 433.8--475.1 ms from the 16-core frontend; release 1.2 continuous gave 33.9 ms from the 48-core frontend. The release contrast therefore holds on both frontends (\hyperref[atlas:A46]{A46}). For deployment, these results describe this proof size and topology: above about 3,300 requests/s per connection, a service with replies of this size needs more connections, and its verifying side must grow with them.

\subsection{Overload and memory bounds}\label{sec:overload}
The release 1.1 library has no admission limit, so its overload appears as queue growth. Release 1.2 was offered 5,100 requests/s for 60 seconds, 1.5 times 3,400 requests/s, the highest one-connection rate that passed the 15-second screens, under a 2,760-MiB memory limit derived from the measured server footprint and its in-flight and return budgets. Of 306,000 timed offers, 125,769 (41.1\%) were accepted and 1.15\% met one second. The frontend's bounded pending queue refused 180,231 offers before sending, the worker refused none, and no submitted request was left unresolved. The frontend's pending limit of 8,192 equals the server's queue cap, so the server's admission refusal was not exercised; the return budget reached its 192-reply cap. The worker's maximum RSS was 402,948 KiB by GNU time and 548,064 KiB in the sampled \texttt{/proc} series, 14 to 19\% of the limit, and the worker exited normally. In a second run, the worker's send rate was limited to 400 Mbit/s at 3,400 requests/s, which stands in for a slow reader. With the return budget in place, the worker's maximum RSS was 381,716 KiB by GNU time and 505,232 KiB sampled. That run accepted 20,131 of 204,000 timed offers (9.9\%); 179,756 were refused before sending and 4,113 were still unanswered at the frontend's drain limit, all counted as failures. Both runs end with bounded worker memory and every offer accounted for; acceptance-only distributions would hide these outcomes, so every timed offer stays in the denominators (\hyperref[atlas:A41]{A41}).

\subsection{Host size: 48, 96 and 192 cores}\label{sec:hostsize}
\Cref{tab:hostsize} lists, for each host size, the highest offered rate at which every repeat accepted every timed request within one second, judged only on points where the frontend was not the limit. On 48 and 96 cores each same-day rate was run once for 15 seconds (20 seconds for same-day release 1.1 on 96 cores), so these ceilings are screening results rather than confirmations. Same-day release 1.1 was measured only at 500 and 1,000/s on the smaller hosts, so its ceilings are coarse; the ratio of the highest finite-window completion rates compares overloaded throughput more closely, at 3.31 for 48 cores and 4.10 for 96 cores against same-day release 1.1. The limiting resource differs by host. On 48 cores, release 1.2 continuous met the budget at 2,000/s with p99 147.0 ms and failed at 2,400/s, where 30.0\% met one second; the worker was at 96\% CPU in both cases, so this ceiling is set by computation. On 96 cores, it met the budget at 3,000/s with p99 37.6 ms and the worker at 70\% CPU, and failed at 4,000/s with the worker at 83\% CPU, its return budget full and replies at 4.95 Gbit/s on one connection, the same transfer bound as on 192 cores.

Before the campaign, a queueing model built from release 1.1 batch measurements (compute, return and inter-batch tail intervals) predicted continuous-mode ceilings of 715 to 779/s on 48 cores, 1,349 to 1,470/s on 96 cores and 2,499 to 2,724/s on 192 cores. The measured ceilings on 48 and 96 cores are two to three times higher. The model kept the release 1.1 per-job compute cost, so it omitted the reorganized prover, which proves 2.4 to 2.7 times faster on the same workers (\Cref{sec:compute}). Its per-job costs also came from release 1.1 batches, in which cores wait for the slowest job of each batch; continuous dispatch has no such wait. At 1,000/s on 192 cores, a release 1.2 job takes a median of 18.7 ms from dispatch to completion in either mode. We report the model only to make this difference visible; the ceilings are the measurements (\hyperref[atlas:A45]{A45}).

\begin{table}[t]
\centering\small
\caption{Highest offered rate with every timed request accepted within 1 s. At 48 and 96 cores each same-day rate is one short run (15 s; 20 s for same-day release 1.1 on 96 cores), and the 7 Oct record used three runs per rate; at 192 cores the one-connection values are three 60-s confirmations, and the two-connection value comes from 15-s screens at a frontend-limited point, so it is a lower bound. Release 1.1 grids are coarse.}\label{tab:hostsize}
\begin{tabular}{@{}L{.08\linewidth}L{.12\linewidth}L{.13\linewidth}L{.13\linewidth}L{.19\linewidth}L{.17\linewidth}@{}}
\toprule Host & StateSync-GKR v1.1 (10-07) & StateSync-GKR v1.1 same-day & StateSync-GKR v1.2 batch-fused & StateSync-GKR v1.2 continuous & What bounded release 1.2\\\midrule
48 cores & 500/s & 500/s; 1,000/s fails & $\geq$1,000/s, highest measured & 2,000/s; 2,400/s fails & worker CPU\\
96 cores & 500/s & 500/s; 1,000/s fails & not measured & 3,000/s; 4,000/s fails & reply transfer on one connection\\
192 cores & 1,000/s & $\geq$1,150/s, highest measured & 3,300/s & 3,300/s; on two connections $\geq$5,000/s (frontend-limited lower bound) & one connection, then the frontend\\\bottomrule
\end{tabular}
\end{table}

\subsection{Where the time went}\label{sec:timeshare}
The release 1.1 record located the main observed delay before computation began. Its 1,150/s process has server pre-compute wait p99 1,367.973 ms, compared with 343.967--354.871 ms in the three 1,000/s confirmations; whole-request p99 is 1,471.284 ms versus 426.032--443.471 ms. These are separately summarized distributions: subtracting their p99 values would not isolate a network interval. Mean worker use is 78.09 core equivalents at 1,150/s, against approximately 64.57--65.24 at 1,000/s. Source inspection identified a serial service boundary. One chosen operation-kind batch executes at a time. Its witness/prove/encode work is parallel, but result hashing, response construction, blocking frame writes and logging are performed per request before the next batch begins. In the first one-worker horizontal run, recorded batch intervals include 16.399 s of computation, 3.279 s after encoding to the last write and 3.350 s between batches; explicit socket-write time is only 0.667 s. These worker intervals include warmup and must not be added to a different timed-request denominator.

\begin{figure}[tb]\centering
\planfig{174mm}{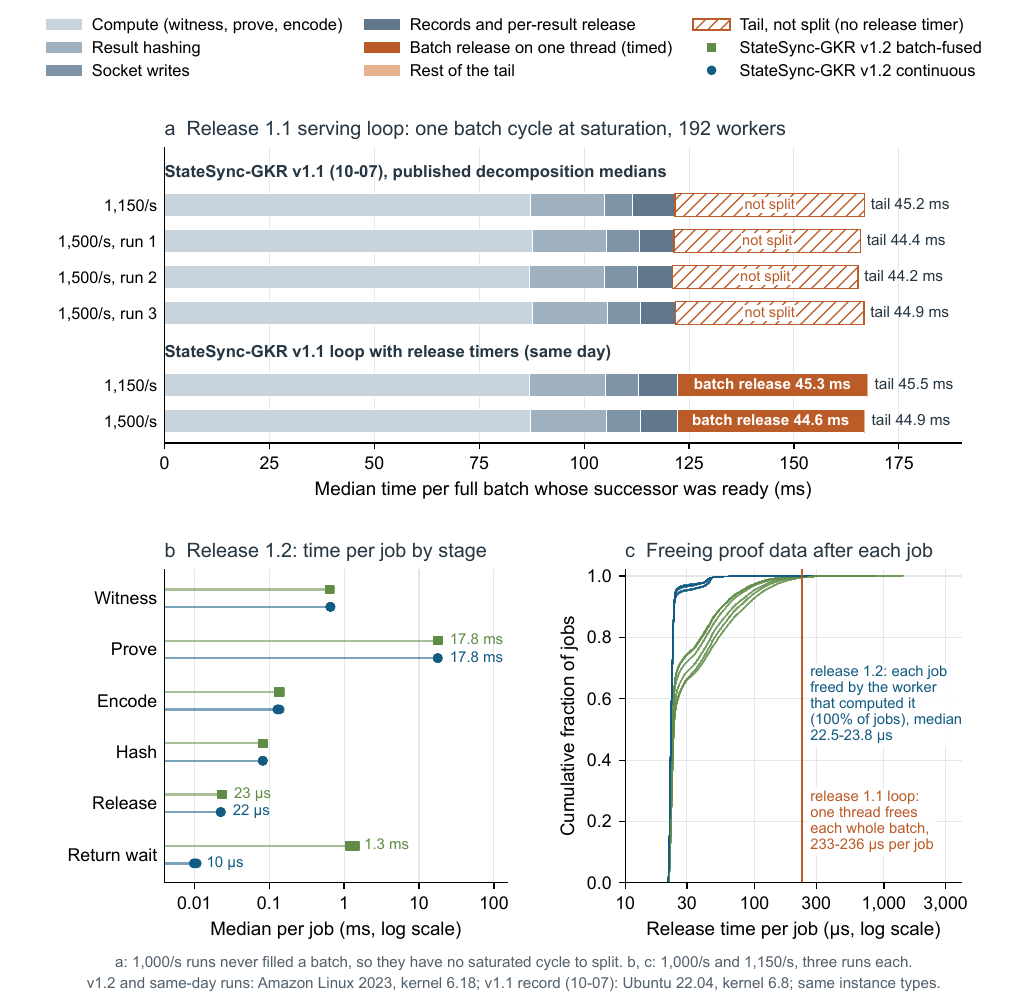}
\caption{Where serving time goes. In the release 1.1 loop, releasing a batch on one thread takes 45.3 ms of the 45.5-ms tail between batches; release 1.2 frees each job on its own worker in about 23 $\mu$s.}\label{fig:timeshare}
\end{figure}

The release 1.1 record proposed passing completed results into a bounded return queue and overlapping their delivery with the next compute batch. Release 1.2 implements that design as the queued-return mode, and its two measured modes go further: batch-fused moves hashing and release into the parallel stage, and continuous removes the batch boundary. Release timers added to the release 1.1 loop show where the inter-batch time went (\Cref{fig:timeshare}). At 1,150/s, full batches whose successor was ready have a median tail of 45.50 ms after the last write, of which releasing the batch's typed witnesses and proofs on the dispatcher thread takes 45.27 ms; at 1,500/s the two values are 44.87 and 44.60 ms. The tail is therefore the release of intermediate data: when one thread frees a whole batch, it takes about 233 to 236 $\mu$s per job with the release 1.1 prover and 221 to 227 $\mu$s with the release 1.2 prover in the same loop. In both release 1.2 modes, the worker that computed a job also frees it, for every job, in a median of 22.5 to 23.8 $\mu$s.

Per-job timers place nearly all remaining service time in proving. At 1,000/s on 192 cores, a job's median proof stage is 17.8 ms of its 18.7 ms from dispatch to completion, and continuous mode finishes writing a reply a median of 0.05 ms after the job completes. Server pre-compute wait p99 at 1,150/s is 471.7 ms for same-day release 1.1, 62.9 ms in batch-fused mode and 0.03 ms in continuous mode. Hashing, logging, scheduling and socket backpressure have not been ablated one by one, so their separate shares remain unknown. \hyperref[atlas:A44]{A44}--\hyperref[atlas:A47]{A47} add the release 1.2 families on CPU contention, host size, frontend size and the same-day prover contrast.

\section{What release 1.2 changes for an integrating service}
\label{sec:product}
\Cref{tab:product} collects the release 1.2 changes that an integrating service sees. Proof bytes, the verifier and the acceptance endpoint of the GKR proofs are those of release 1.1, so a service that already accepts release 1.1 GKR proofs needs no change on the verifying side; the RISC Zero integration of release 1.2 has a new program identity, described under Reproducibility. Costs convert measured rates at the on-demand hourly price of the worker instance, USD 10.34592 for \texttt{c8a.48xlarge} in the us-east-1 region as checked on 25 October 2026, and assume continuous operation at that rate; they estimate worker compute only and exclude the frontend, storage and network charges.

\begin{table}[htbp]
\centering\footnotesize
\caption{Product metrics before and after release 1.2. Operation triples are membership/non-membership/update.}\label{tab:product}
\begin{tabular}{@{}L{.20\linewidth}L{.23\linewidth}L{.29\linewidth}L{.19\linewidth}@{}}
\toprule Product metric & Release 1.1 & Release 1.2 & Condition\\\midrule
Highest offered rate with every request within 1 s & 1,000/s (7 Oct record); 1,150/s same day (one 60-s run), highest rate measured & 3,300/s on one connection; $\geq$5,000/s on two connections (frontend-limited lower bound) & one 192-core VM; one connection: three 60-s runs; two connections: 15-s screens\\
Same, 48-core VM & 500/s & 2,000/s & same day, one 15-s screen per rate\\
Compute cost per million proofs at that rate, worker only & USD 2.87 at 1,000/s; 2.50 at 1,150/s & USD 0.87 at 3,300/s & on-demand price, continuous operation; the same-day release 1.1 value is an upper bound, since its passing rate was not searched further\\
p99 at 1,000/s, 192-core VM & 433.8--475.1 ms (same day) & 33.7--33.9 ms (continuous) & three 60-s runs\\
Long run at the confirmed rate & n/a & 5,940,000 requests in 30 min, all within 1 s, p99 35.3 ms & 3,300/s, one connection\\
Overload at 1.5$\times$ the highest one-connection screen pass & no admission limit in the library & worker RSS at most 548,064 KiB sampled (402,948 KiB by GNU time) under a 2,760-MiB limit; return budget full, no server refusal & 5,100/s, 60 s\\
Slow reader & n/a & worker RSS at most 505,232 KiB sampled (381,716 KiB by GNU time) while the return budget holds replies & send limited to 400 Mbit/s; 3,400/s offered\\
Worker peak RSS & 406.1 MiB (7 Oct record); 381.7 MiB same day & 209.7--223.5 MiB batch-fused; 305.0--307.5 MiB continuous & 1,150/s, 192-core VM; sampled \texttt{/proc} RSS; GNU time maxima on these hosts are 1.3 to 2.2 times lower and are not used as the peak\\
Service parts an integrator writes & request IDs, queue cap and refusal, framing limits, logs and metrics, asynchronous isolation, witness parallelism, output parallelism & library modules for request identity and framing limits, admission and refusal, return budgets, per-job completion and stage timing & serving crate on the public API only\\
Misuse checks in release builds & debug assertions only & operation-kind mismatch rejected; leaf-hash input width enforced & all builds\\
Allocations per proof & 1,282 & 85, with process RSS within 0.4\% & instrumented mixed circuit, width 4,096, same day\\
Prepared request latency, p50 & 42.10/46.10/65.03 ms (same day) & 16.95/18.22/24.62 ms & one thread, depth 24\\
Build configuration & native build recommended; no measured ratio & default release build 1.66--1.71$\times$ native latency; x86-64-v3 build 0.90$\times$ & one process per build\\
Proof format and verifier acceptance & reference & unchanged; proof bytes identical & all builds and modes\\\bottomrule
\end{tabular}
\end{table}

\section{Implications and validity}
\label{sec:discussion}
The nine mechanisms have related but separate payoffs. Sparse reduction gives circuit/job capacity within RAM. Derived wiring lowers prepared verification-call time. Cube/affine fusion and parallel path residuals give a compact schedule. Prepared material removes repeated setup. Independent jobs exploit workers and deployment choices while retaining result identity. Three-state leaves encode application lifecycle. Strict codec/request binding gives a meaningful delivery endpoint. The generic core and formal model provide reusable interfaces and an inspectable assurance boundary. Release 1.2 builds service parts on these mechanisms. In the overload and slow-reader runs, worker memory stayed bounded while the return budget was full and producers waited for it; the frontend's pending limit, equal to the server's queue cap, also bounded outstanding requests, and no run without a return budget was made, so the budget's share of the bound is not isolated. Misuse checks hold in release builds, and the reorganized prover arithmetic lowers the cost of each proof while its bytes stay fixed. These benefits extend independent jobs, request binding and sparse reduction; they add no new mechanism.

The strongest memory claim is the same-driver Dense/Sparse ownership contrast, not the global superiority of this engine. The strongest serving claims name an actual host, mix, batch/wait policy, period and budget. Scale-limit results give tested support points and native failure dispositions. The fixed corpus, common host repeats, frequency/background placement, observation overhead and confirmation lengths restrict generalization. Retaining full data prevents those limitations from being concealed by a representative median. Release 1.2 ran on Amazon Linux 2023 with kernel 6.18, whereas the release 1.1 records used Ubuntu 22.04 with kernel 6.8; same-day release 1.1 runs on the new image separate the software change from the image change, and they reproduced the records within 1.1\% for serving computation at 1,000/s and within 0.4\% for proving; at 1,150/s same-day release 1.1 met one second, which the 60-second run of the record did not. Measurement groups ran in parallel on separate instances, while the earlier campaign ran them in sequence. Network-bound and frontend-bound points describe this measurement topology, and the 15- and 30-minute runs are finite observations.

Cryptographic acceptance here is an inner-proof endpoint. Persistent database writes, sequential adoption of roots, consensus and settlement are not part of this profile. The companion formal composition and preservation results supply the semantic basis for these implementation choices; the Rust connection remains partial under explicit assumptions~\cite{fvpaper}. Neither differential observations nor successful encoded acceptance discharge the online challenge-distribution or multi-round Fiat--Shamir obligations. Release 1.2 leaves this boundary as it was, because its proof bytes and verifier are those of release 1.1.

\clearpage
\subsection{Application-facing meaning of the nine mechanisms}
Figures 1--2 explain implementation structure; Figures 3--6 quantify selected resource and repeated-work effects. Figures 7--12 evaluate deployment behavior and its limits. These twelve figures therefore support a connected account of nine mechanisms, not nine new discoveries or nine independent causal ablations. \Cref{tab:benefits} identifies the measurement, control or formal basis of each application benefit; \Cref{tab:headline} gives the main numbers.
\begin{longtable}{@{}p{.18\linewidth}p{.34\linewidth}p{.42\linewidth}@{}}
\caption{Implementation mechanisms translated into application decisions. Measured ratios apply to their stated internal references and workloads.}\label{tab:benefits}\\
\toprule Mechanism & Application benefit & Evidence and condition\\\midrule\endfirsthead
\toprule Mechanism & Application benefit & Evidence and condition\\\midrule\endhead
Sparse reduction & More room for a circuit or concurrent independent proofs within RAM & Representation in \Cref{fig:architecture}; measured internal contrast in \Cref{fig:memory}, with requested peak 3,242.41-fold lower in both releases and process RSS 89.46-fold (release 1.1) and 88.87-fold (release 1.2) lower at width 4,096.\\
Derived wiring & Less repeated work to check a prepared circuit's connections & Measured same-proof prepared Table/Derived verification call in \Cref{fig:verification}(a), 10.80--23.68-fold, unchanged in release 1.2; model equivalence in the companion formal study~\cite{fvpaper}.\\
Cube/affine fusion & A compact hash-round representation with fewer intermediate layers & Source structure in \Cref{sec:engine}; compact schedule in \Cref{fig:compact}(a), unchanged in release 1.2. No independent unfused-compiler timing ablation.\\
Parallel path constraints & Deeper trees add modest proof-size growth in the tested range & Structure and proof-size consequence in \Cref{fig:compact}(a): 118 layers across nine profiles; membership bytes grow 2.93\%, unchanged in release 1.2. No isolated causal ablation.\\
Prepared reuse & Repeated requests avoid paying identical setup again & Shared/private boundary in \Cref{fig:lifecycle}; direct fresh/prepared contrast in \Cref{fig:preparation}, 3.46--4.45-fold in release 1.1 and 7.19--10.13-fold in release 1.2.\\
Independent jobs & Use parallel workers or multiple servers while keeping each result attached to its request & One result per request in \Cref{fig:lifecycle}; measured worker/batch-pipeline and deployment contrasts in \Cref{fig:compact,fig:verification,fig:horizontal}; release 1.2 continuous serving returns replies in completion order, matched by job identifier (\Cref{sec:servingparts,sec:capacity}).\\
Three-state leaves & Distinguish never-used, active and retired ledger entries & States in \Cref{fig:lifecycle}; eight lifecycle/payload/key controls, unchanged in release 1.2. Semantic evidence, not a database transaction benchmark.\\
Strict request binding & Reject a proof for a different record, root or encoding & Acceptance boundary in \Cref{fig:lifecycle}; byte and negative controls in \Cref{sec:methods}; operation-kind and leaf-width checks enforced in release builds of release 1.2 (\Cref{sec:engine}). No empirical whole-system security claim.\\
Reusable core and models & Adapt the engine to other arithmetic frontends with an inspectable reasoning basis & Source interfaces in \Cref{sec:engine} and formal composition~\cite{fvpaper}; the release 1.2 serving crate uses the public prover API only (\Cref{sec:servingparts}). No additional frontend benchmark or complete Rust refinement claim.\\\bottomrule
\end{longtable}

\paragraph{Reproducibility.} Current source and bounded Linux replay are identified by the BSL~1.1 distribution at \url{https://github.com/Oraclizer/statesync-gkr} and the release 1.2.0 software coordinate \href{https://doi.org/10.5281/zenodo.23643505}{doi:10.5281/zenodo.23643505}. The replay builds separate callers, tests an isolated memory child, checks honest/tampered proofs and CLI rejection, and performs a small local complete-proof transfer. Its source/binary/environment identities are separate from the historical measured binary; \texttt{REPRODUCING.md} separates source-level reproduction from release-asset verification. The serving crate builds with \texttt{cargo build -{}-release -{}-locked -{}-manifest-path benches/serving/Cargo.toml}, and its \texttt{service-checks.py} runs real-TCP functional suites, including byte equality across serving modes; these checks are not capacity evidence. Measurements use native builds (\texttt{-C target-cpu=native}). In one process per build, the default release build took 1.66 to 1.71 times the native latency of a prepared request and an x86-64-v3 build 0.90 times; these build comparisons were not repeated. The repository's capacity reproduction job remains pinned to the release 1.1 measured source, because the archived memory observer accepts only that sparse oracle; the release 1.2 measurement harness, frozen cell plans and raw records are kept in the same companion package (Data availability). The v1.1.0 DOI cited in this paper preserves historical provenance with source files now restricted. Replaying a tiny functional test does not reproduce an archived performance rate. The RISC Zero~\cite{risczero} integration of release 1.2 has its own program identity, image ID beginning \texttt{1a8b55f4}; a proof of that program's fixed depth-24 membership relation was verified through the zkVerify~\cite{zkverify} RISC0 verifier on the public Volta testnet, in submission block 7,775,413 with its aggregation receipt in block 7,775,436. This proof is not attached to the 1.2.0 release, whose attached proof belongs to the release 1.1 program; its bytes are the call argument of that public submission. It adds no measurement to this report. Appendix A displays all 43 detailed scientific figure families with release 1.2 overlays and four new families on uniform A4 portrait pages, in addition to twelve main-text figures. Wide panels are reformatted or divided without dropping values, conditions or not-applicable cells; original exports remain in the archive. The companion archive preserves all observations, detailed tables and figure sources with exact normalization and file contracts. The atlas retains process, family, host, workload and observation boundaries; none of the other experimental conditions is deemed meaningless simply because it shares a chart with several conditions.

\paragraph{Data availability.} The earlier two CPU datasets remain at their public benchmark paths. The release 1.1 raw counterparts and the release 1.2 raw records, kept in the same structure, together with detailed tables, figure sources and replay interfaces, are kept by the authors in a companion package with the same structure; the package has no public deposit or dataset DOI. The first version of this report is arXiv:2610.14849v1.

\paragraph{Assistance and responsibility.} LLM coding and writing tools assisted benchmark/analysis development and manuscript preparation. Recorded native execution, differential and negative controls, frozen source identities and machine-checking records provide inspectable evidence. The human author retains responsibility for the interpretation and final submitted text; tool assistance is not authorship.

\section{Conclusion}
StateSync-GKR connects a reusable GKR/sumcheck core to an authenticated-state frontend with explicit preparation, independent job and encoded-acceptance boundaries. Its nine mechanisms address different application costs: sparse reduction lowers prover ownership; regular wiring reduces repeated verification; fused cube/affine circuits and parallel path constraints keep the schedule and proof growth compact; preparation reuse and indexed jobs lower repeated work and expose CPU parallelism; three-state semantics and original-request binding preserve the meaning of each result; and the separate core/model interfaces support reuse with an inspectable assurance boundary. Release 1.2 shows that the serving path and the arithmetic inside the prover's rounds can change latency and capacity while proof bytes and verifier acceptance stay fixed.

The controlled evidence gives these mechanisms concrete scope. The largest internal memory contrast is a 3,242-fold requested-allocation and about 89-fold process-RSS reduction at mixed width 4,096, retained in release 1.2. Direct fresh/prepared request measurements show 7.19 to 10.13-fold lower proving-path latency in release 1.2, and 3.46 to 4.45-fold in release 1.1; prepared verification calls and complete local batches with parallel output handling show their own substantial reductions. Release 1.2 proves 2.4 to 2.7 times faster than same-day release 1.1 on the same bare-metal workers. On a 192-core VM it keeps p99 at 34 ms at 1,000 offered requests/s, against 440 ms for same-day release 1.1, and meets a one-second budget for every request at 3,300 requests/s on one connection, including a 30-minute run of 5.94 million requests; two connections pass 5,000 requests/s in short screens, where the frontend becomes the bound. On 48 cores, in single 15-second screens, the highest rate meeting that budget rises from 500 to 2,000 requests/s. These are measured offered loads under stated budgets, which differ from both a proving-only kernel rate and an indefinitely stable capacity.

The artifact therefore supplies a memory-efficient and reusable implementation, measured choices for repeated proof work, service components that kept worker memory bounded in the overload runs while the return budget was full, and explicit application acceptance semantics. Full conditions and outcomes preserve the limits of those results. The 15- and 30-minute confirmations are finite observations; larger-scale external ranking and the remaining cryptographic and code-correspondence arguments stay outside the established claims.

\FloatBarrier
\bibliographystyle{plainnat}
\bibliography{references}
\clearpage\appendix
\section{Complete detailed evidence atlas}\label{app:atlas}
The main figures connect implementation mechanisms to the most informative contrasts. This appendix retains all 43 detailed families of the release 1.1 report and all original conditions, overlays release 1.2 on the same axes, and adds four families. Release 1.1 series keep their original values. Panels whose measured code path release 1.2 did not change carry the note ``Unchanged in v1.2'' with its basis; they are the twelve Dense-only memory panels of A01--A03 and the proof-size panel of A36. Continuous runs form no batches, so A12, A25, A28 and A36 add per-job panels for them. Four zero-refusal time-series panels in A31 are replaced by exact condition-specific statements, and release 1.2 also had no refusal in those conditions; the original graphs and CSV remain archived. Each family is an independently addressable figure set: A01--A06 cover generic memory; A07--A10 cover memory guards and scale; A11--A16 cover arrivals and the longer baseline; A17--A25 cover equal-resource deployments; A26--A34 cover fine transitions; A35--A43 cover actual-machine scaling and confirmations; A44--A47 cover CPU contention, host size, frontend size and the same-day prover contrast.

All pages use A4 portrait paper with continuous manuscript page numbers. Many-panel figures continue across pages so their labels remain readable. Common headings, legends and original notes are repeated. Four wide families have their affected panels redrawn or divided at tick boundaries from the same frozen aggregate values; the original PDF/SVG and raw observations remain unchanged in the archive. The routing matrix retains both numeric and not-applicable cells. Release 1.2 measurements are overlaid on the same axes with release 1.1 series unchanged; where a heat map, density or screening surface cannot hold two series (A20, A37, A42), release 1.2 occupies an adjacent panel with the same axes. No smoothed observations or omitted unsuccessful outcomes are introduced. The four A31 replacements concern explicit server refusals only; they do not assert that latency-budget misses or every failure type were zero.

Input-stop markers are derived from actual recorded events in the role's own clock. A falling queue after input ends means that pending requests are being completed. It is not evidence of a crash. Different-duration or unsynchronized-clock cohorts remain separate. Release 1.2 runs used Amazon Linux 2023 with kernel 6.18 and the release 1.1 records Ubuntu 22.04 with kernel 6.8, on the same instance types; same-day release 1.1 anchors appear where measured.

\begin{longtable}{@{}p{.10\linewidth}p{.17\linewidth}p{.54\linewidth}p{.10\linewidth}@{}}
\caption{Detailed family index and first-page links. Each family may span several A4 pages.}\label{tab:atlas}\\
\toprule Family & Study & Figure topic & Pages\\\midrule\endfirsthead
\toprule Family & Study & Figure topic & Pages\\\midrule\endhead
\hyperref[atlas:A01]{A01} & Memory & actual RSS time series & \pageref{atlas:A01}\\
\hyperref[atlas:A02]{A02} & Memory & capacity retention by stage & \pageref{atlas:A02}\\
\hyperref[atlas:A03]{A03} & Memory & empirical time distributions & \pageref{atlas:A03}\\
\hyperref[atlas:A04]{A04} & Memory & memory growth all points & \pageref{atlas:A04}\\
\hyperref[atlas:A05]{A05} & Memory & telemetry coverage and cadence & \pageref{atlas:A05}\\
\hyperref[atlas:A06]{A06} & Memory & time and paired ratios & \pageref{atlas:A06}\\
\hyperref[atlas:A07]{A07} & Limits & independent job memory and time & \pageref{atlas:A07}\\
\hyperref[atlas:A08]{A08} & Limits & independent job observations & \pageref{atlas:A08}\\
\hyperref[atlas:A09]{A09} & Limits & limit policy disposition & \pageref{atlas:A09}\\
\hyperref[atlas:A10]{A10} & Limits & physical cap success observations & \pageref{atlas:A10}\\
\hyperref[atlas:A11]{A11} & Arrival & actual 20-ms zoom and cadence & \pageref{atlas:A11}\\
\hyperref[atlas:A12]{A12} & Arrival & FIFO batches, continuous job stages and stage distributions & \pageref{atlas:A12}\\
\hyperref[atlas:A13]{A13} & Arrival & latency budget success surface & \pageref{atlas:A13}\\
\hyperref[atlas:A14]{A14} & Arrival & screen latency CDF and quantiles & \pageref{atlas:A14}\\
\hyperref[atlas:A15]{A15} & Arrival & screen output and backlog & \pageref{atlas:A15}\\
\hyperref[atlas:A16]{A16} & Arrival & stability full 15-minute envelope & \pageref{atlas:A16}\\
\hyperref[atlas:A17]{A17} & Horizontal & backlog output 1,000 & \pageref{atlas:A17}\\
\hyperref[atlas:A18]{A18} & Horizontal & backlog output 500 & \pageref{atlas:A18}\\
\hyperref[atlas:A19]{A19} & Horizontal & CPU memory traces & \pageref{atlas:A19}\\
\hyperref[atlas:A20]{A20} & Horizontal & endpoint routing counts & \pageref{atlas:A20}\\
\hyperref[atlas:A21]{A21} & Horizontal & latency distributions & \pageref{atlas:A21}\\
\hyperref[atlas:A22]{A22} & Horizontal & process operating points & \pageref{atlas:A22}\\
\hyperref[atlas:A23]{A23} & Horizontal & request latency 1,000 & \pageref{atlas:A23}\\
\hyperref[atlas:A24]{A24} & Horizontal & resources cost network & \pageref{atlas:A24}\\
\hyperref[atlas:A25]{A25} & Horizontal & service phases 1,000, continuous job-stage union & \pageref{atlas:A25}\\
\hyperref[atlas:A26]{A26} & Fine & accepted latency CDFs & \pageref{atlas:A26}\\
\hyperref[atlas:A27]{A27} & Fine & accepted tail timelines & \pageref{atlas:A27}\\
\hyperref[atlas:A28]{A28} & Fine & batch queue resource points, jobs computing at once & \pageref{atlas:A28}\\
\hyperref[atlas:A29]{A29} & Fine & condition deadline heatmaps & \pageref{atlas:A29}\\
\hyperref[atlas:A30]{A30} & Fine & CPU time galleries & \pageref{atlas:A30}\\
\hyperref[atlas:A31]{A31} & Fine & refusal backlog output & \pageref{atlas:A31}\\
\hyperref[atlas:A32]{A32} & Fine & RSS time galleries & \pageref{atlas:A32}\\
\hyperref[atlas:A33]{A33} & Fine & wait operation effects & \pageref{atlas:A33}\\
\hyperref[atlas:A34]{A34} & Fine & worker batch rate transitions & \pageref{atlas:A34}\\
\hyperref[atlas:A35]{A35} & Hardware & accepted latency distributions & \pageref{atlas:A35}\\
\hyperref[atlas:A36]{A36} & Hardware & batch, job and service phases & \pageref{atlas:A36}\\
\hyperref[atlas:A37]{A37} & Hardware & confirmation request density & \pageref{atlas:A37}\\
\hyperref[atlas:A38]{A38} & Hardware & CPU and RSS traces & \pageref{atlas:A38}\\
\hyperref[atlas:A39]{A39} & Hardware & hardware scaling & \pageref{atlas:A39}\\
\hyperref[atlas:A40]{A40} & Hardware & memory price and transport & \pageref{atlas:A40}\\
\hyperref[atlas:A41]{A41} & Hardware & overload and drain, memory and send limits & \pageref{atlas:A41}\\
\hyperref[atlas:A42]{A42} & Hardware & screening surfaces & \pageref{atlas:A42}\\
\hyperref[atlas:A43]{A43} & Hardware & selection and confirmation & \pageref{atlas:A43}\\
\hyperref[atlas:A44]{A44} & CPU & 192-thread contention, clock and placement & \pageref{atlas:A44}\\
\hyperref[atlas:A45]{A45} & Hardware & host-size ceilings at 48, 96 and 192 cores & \pageref{atlas:A45}\\
\hyperref[atlas:A46]{A46} & Hardware & frontend size alone & \pageref{atlas:A46}\\
\hyperref[atlas:A47]{A47} & Memory & same-day release 1.1 and 1.2 provers & \pageref{atlas:A47}\\
\bottomrule\end{longtable}
\atlasfamily{A01}{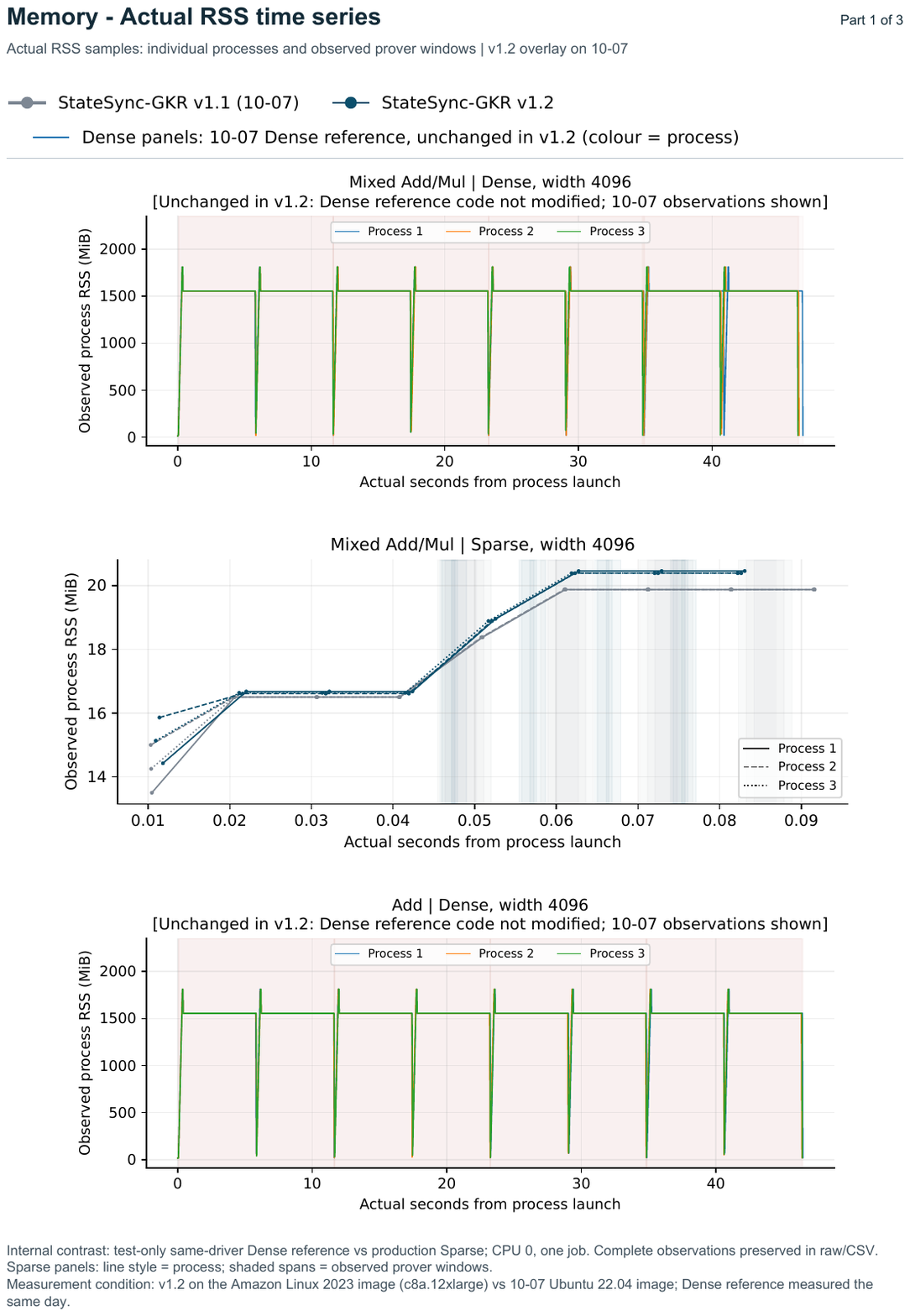}
\atlasfamily{A02}{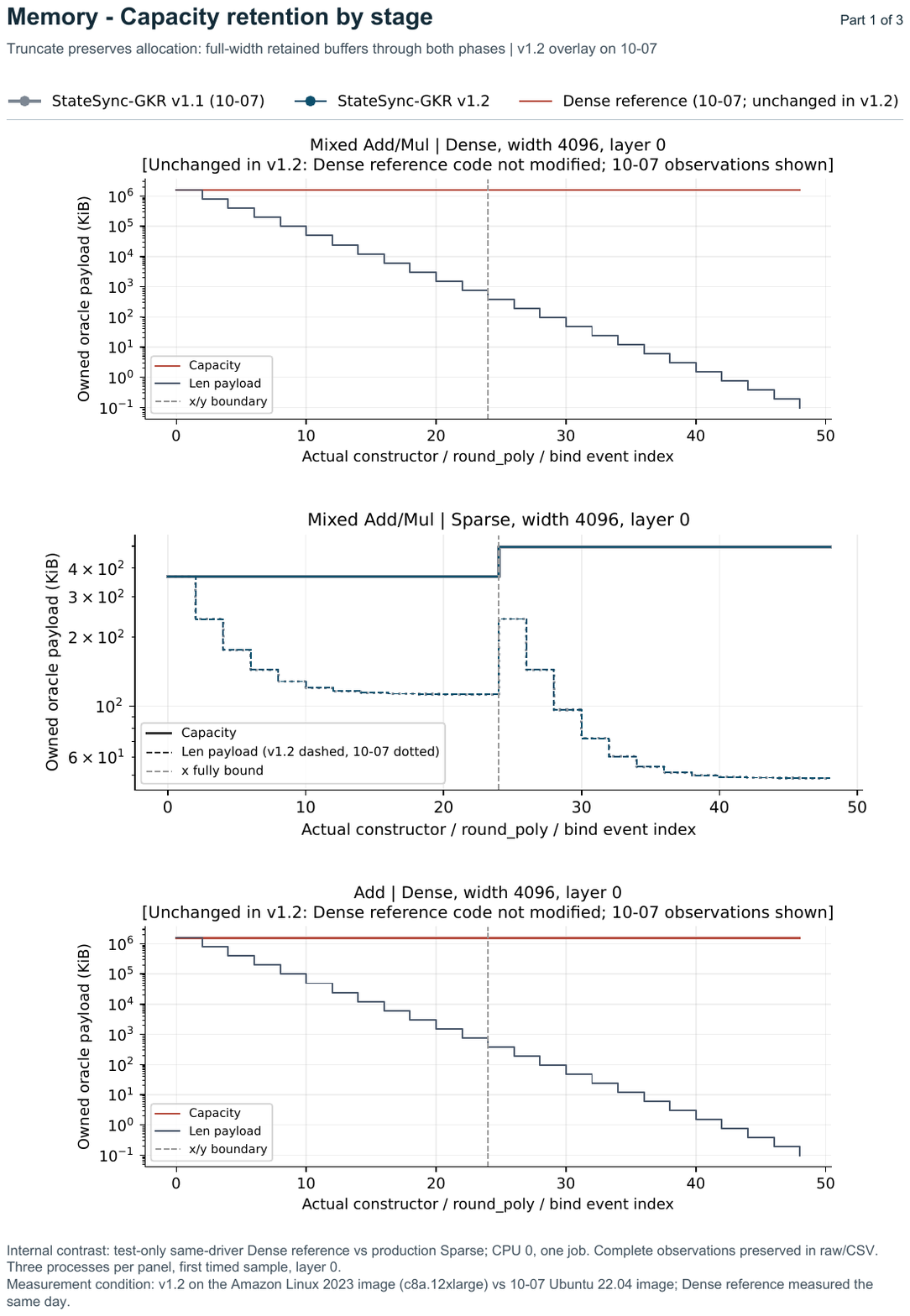}
\atlasfamily{A03}{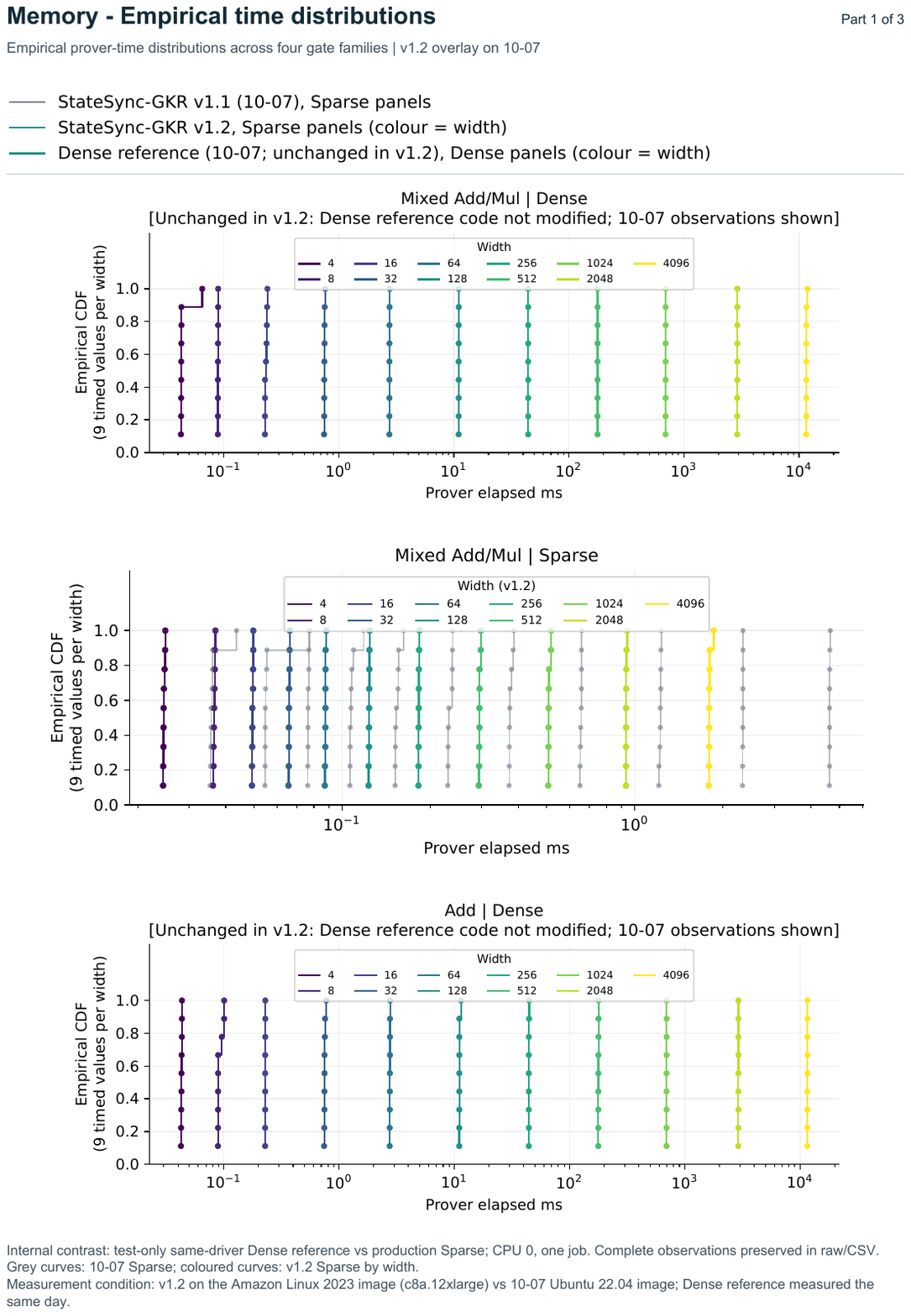}
\atlasfamily{A04}{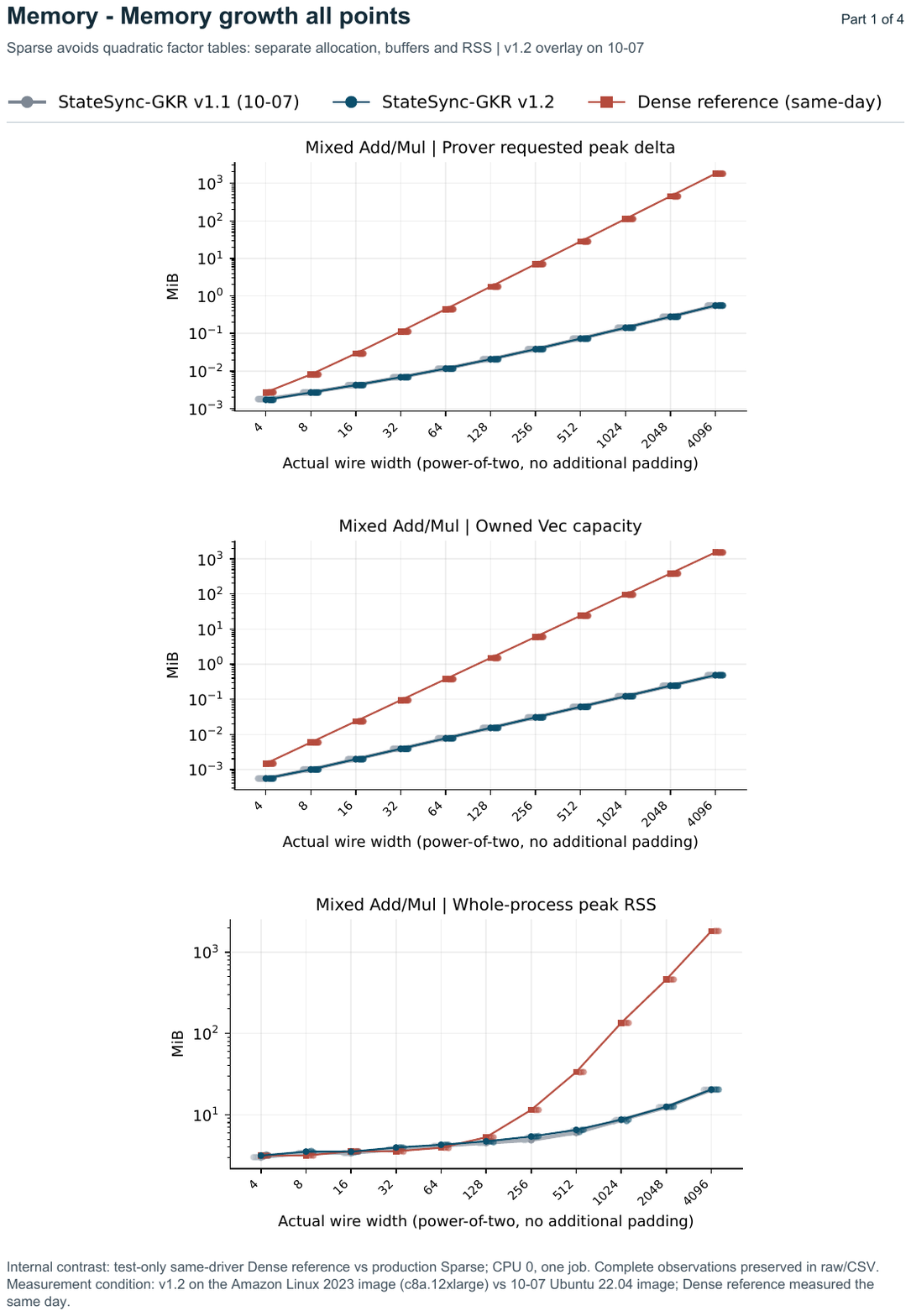}
\atlasfamily{A05}{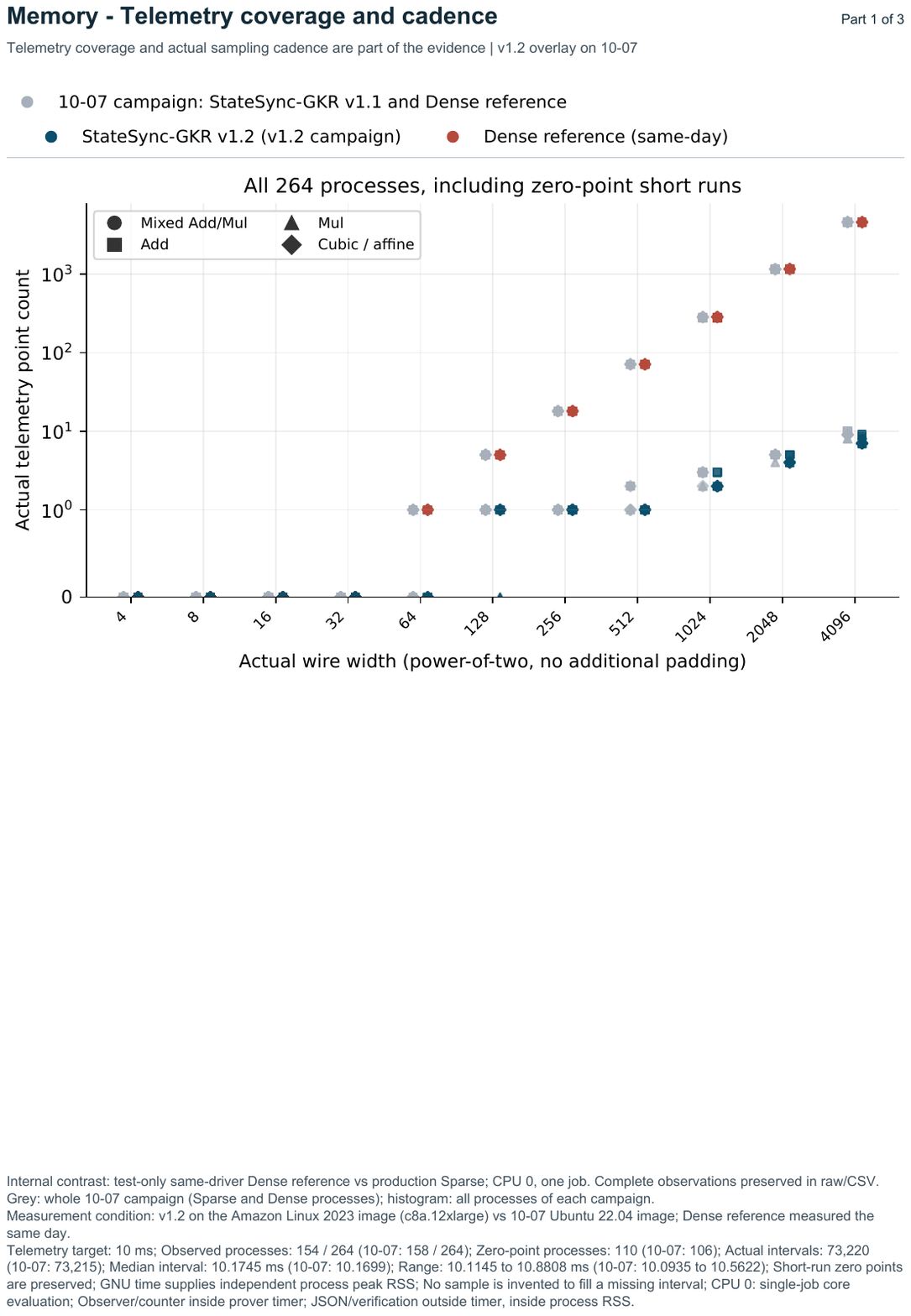}
\atlasfamily{A06}{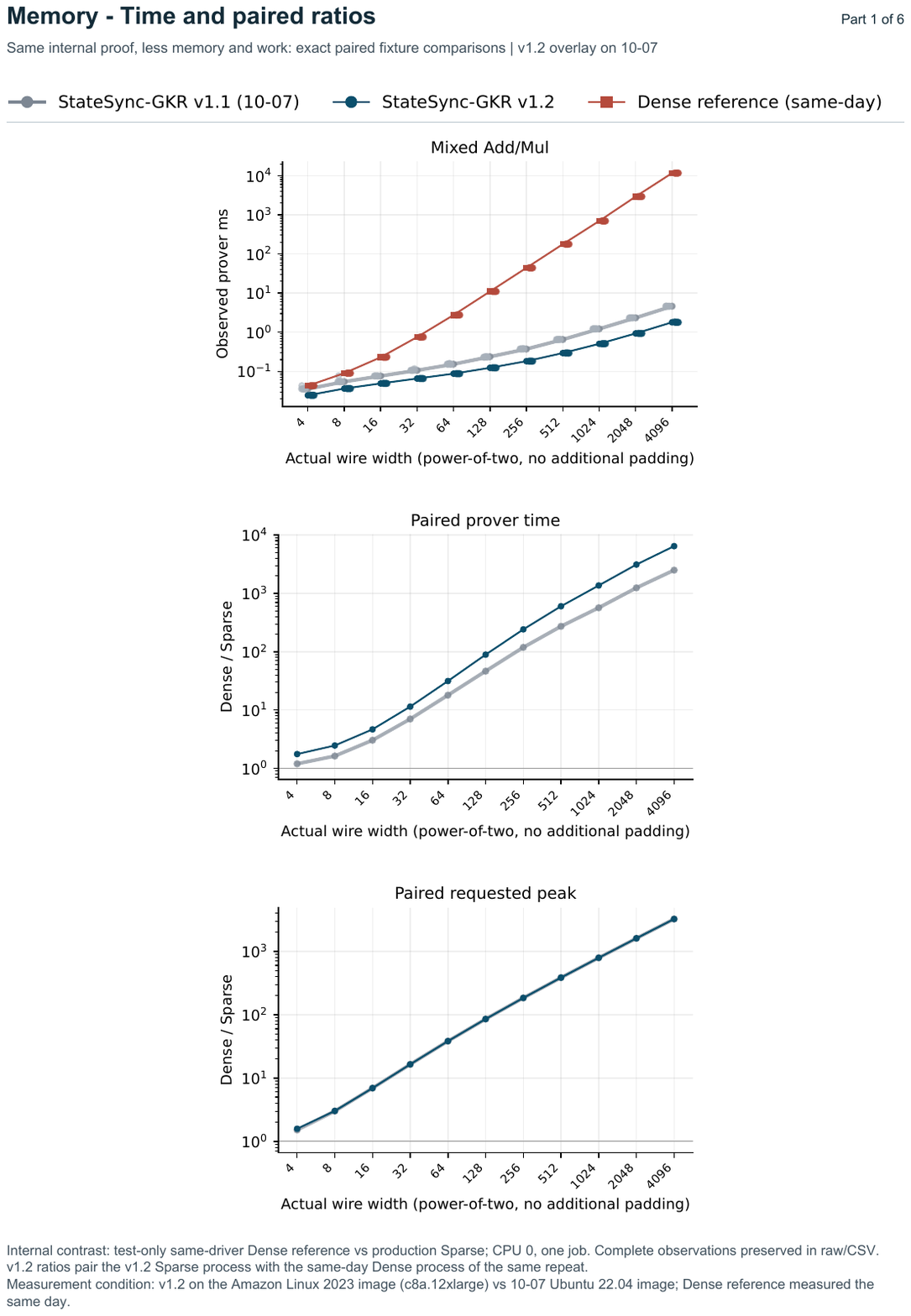}
\atlasfamily{A07}{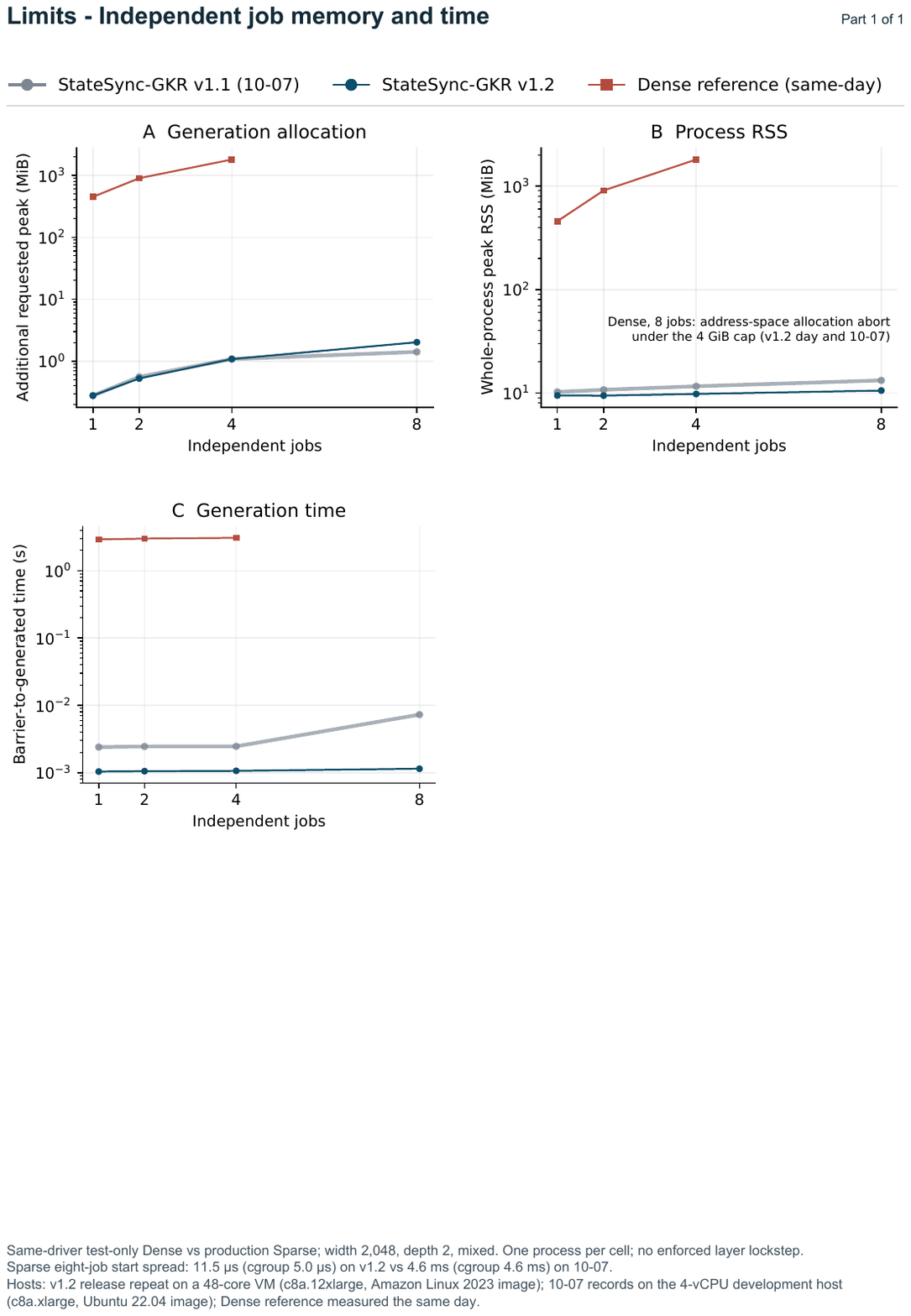}
\atlasfamily{A08}{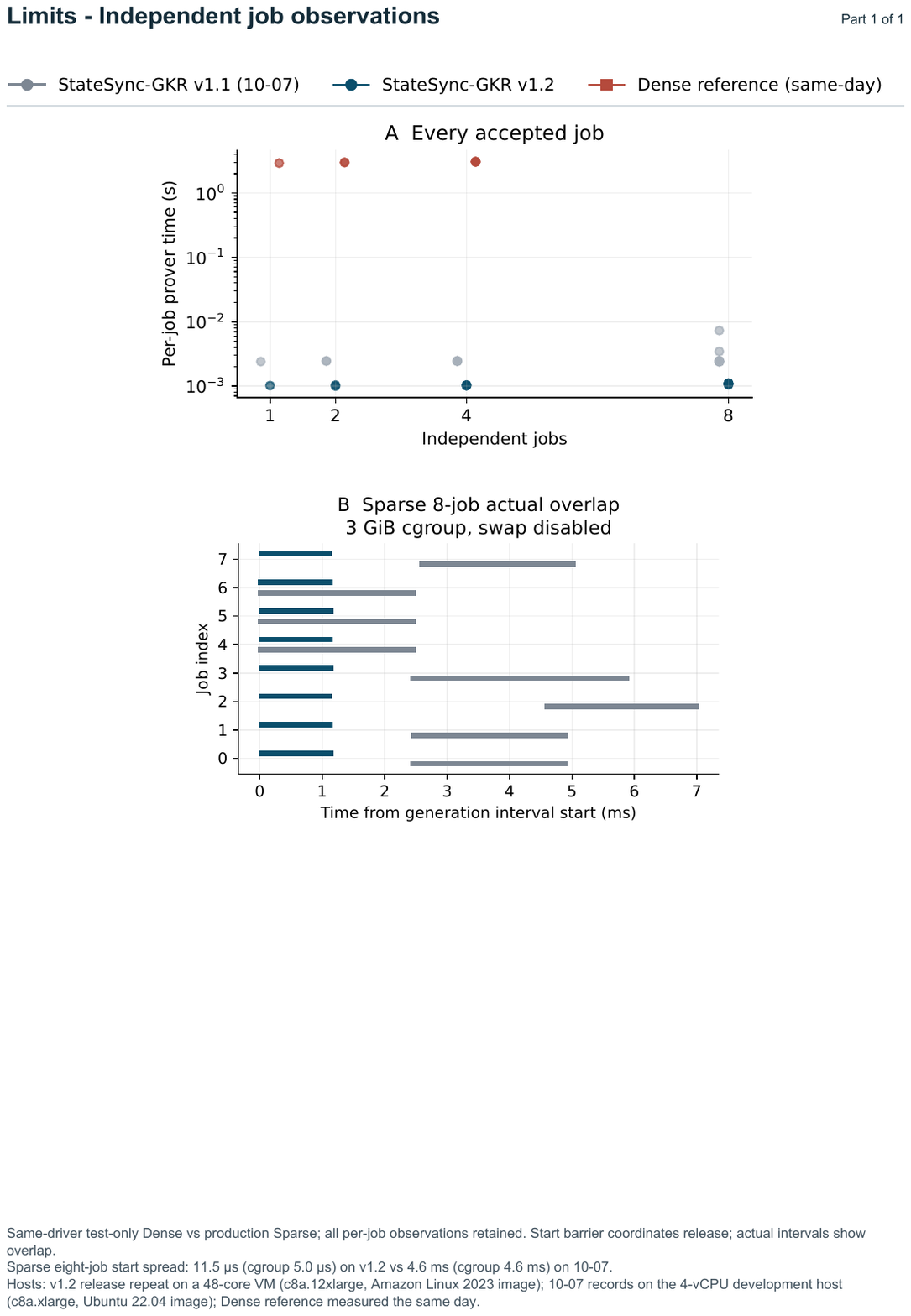}
\atlasfamily{A09}{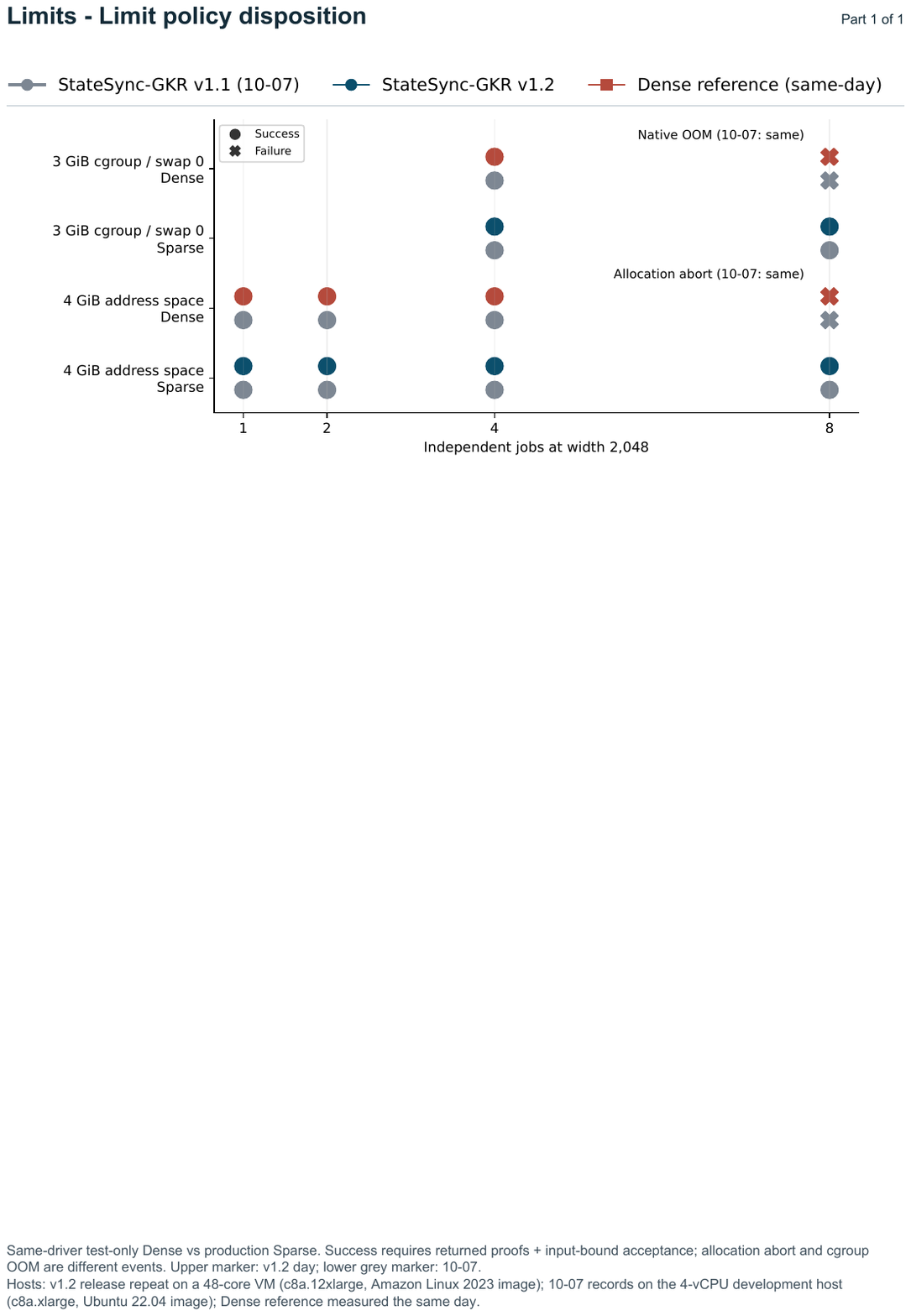}
\atlasfamily{A10}{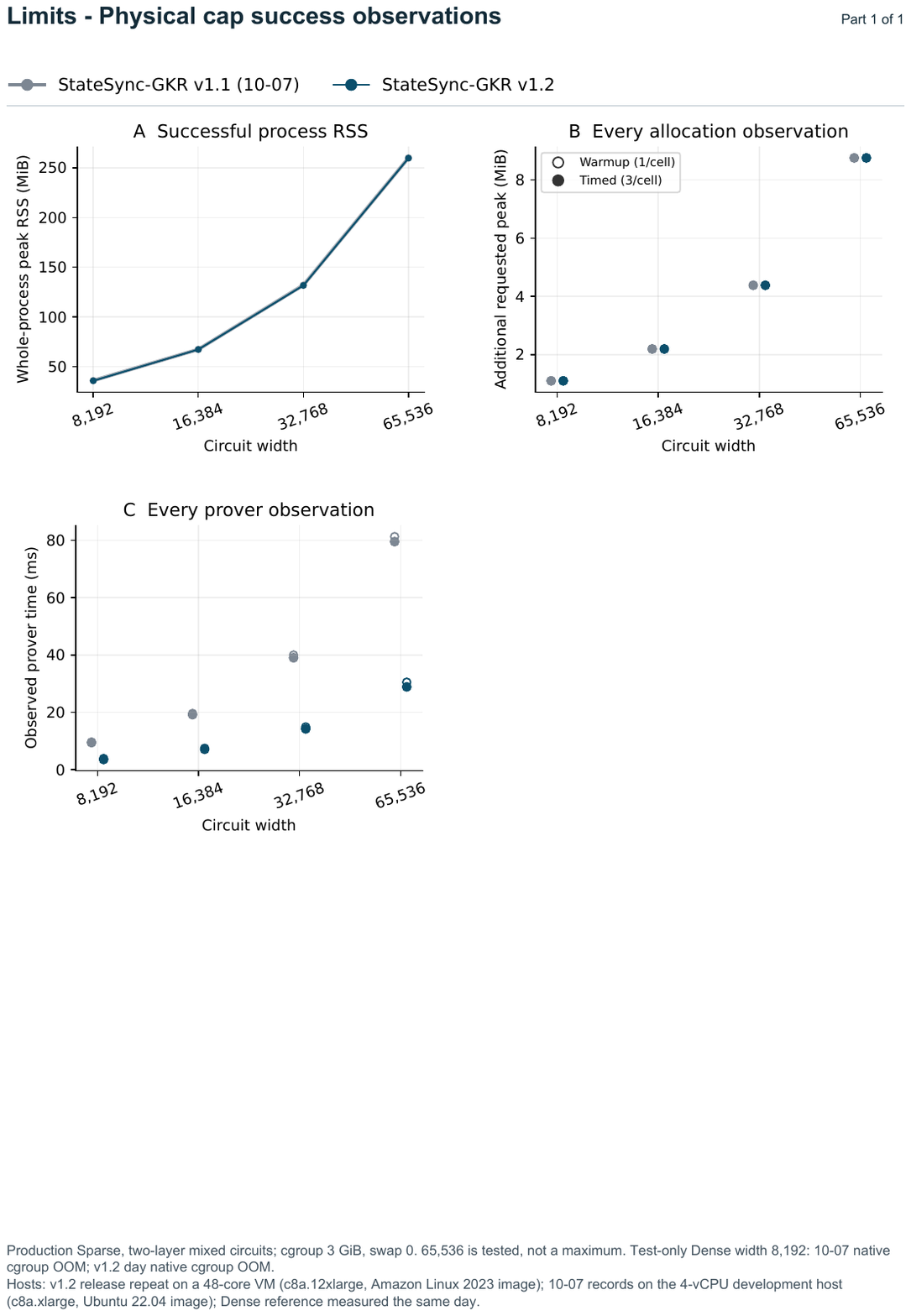}
\atlasfamily{A11}{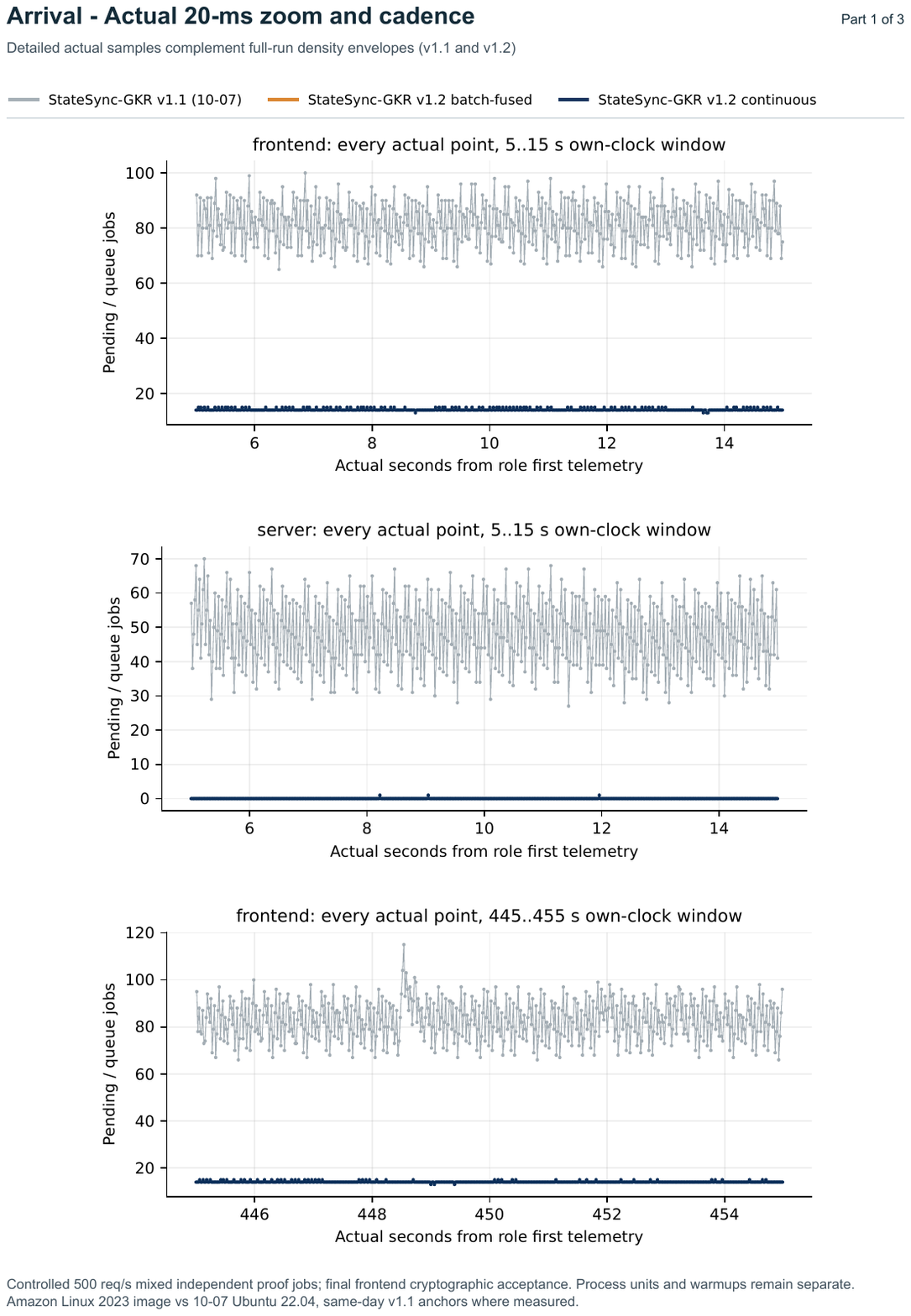}
\atlasfamily{A12}{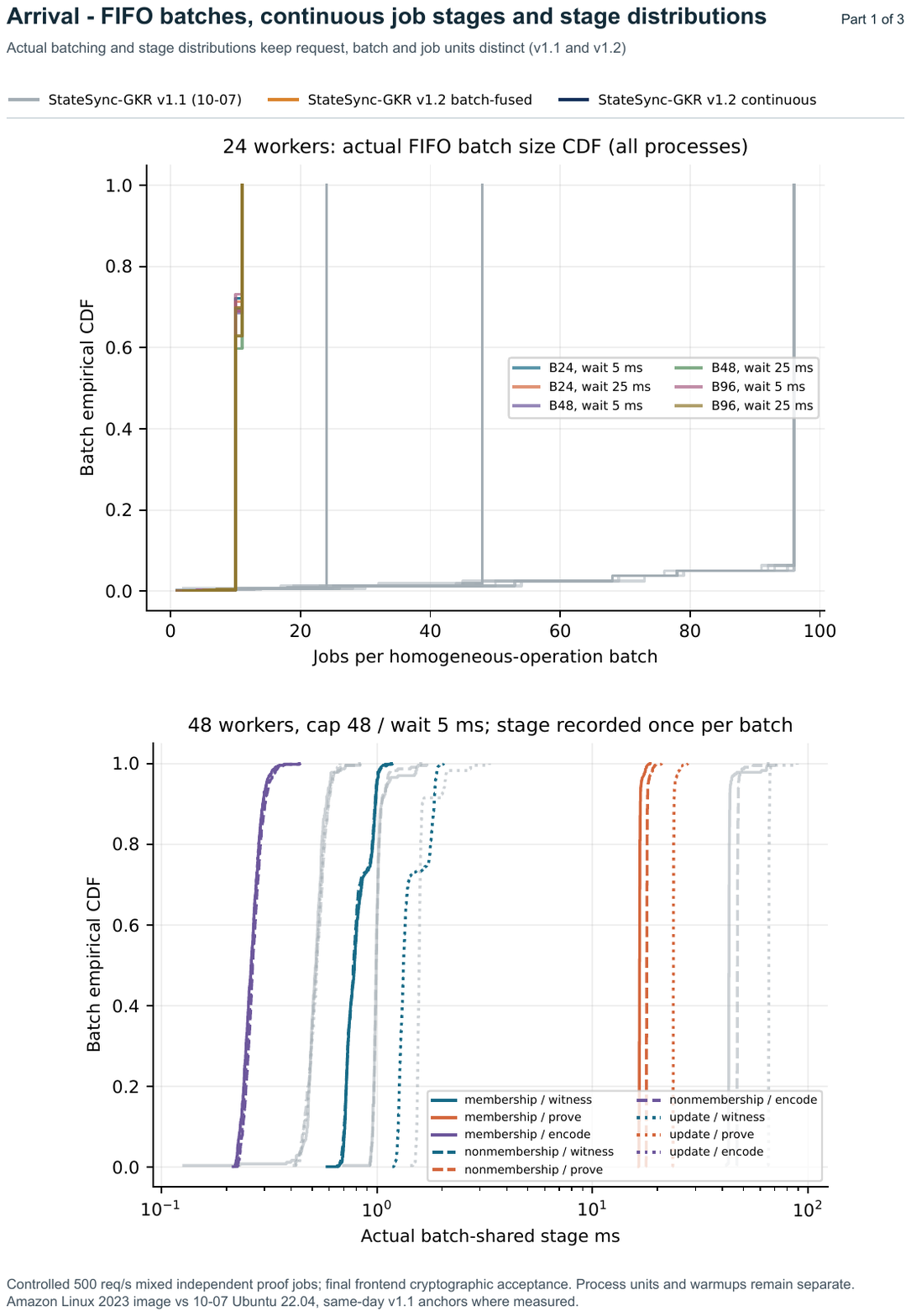}
\atlasfamily{A13}{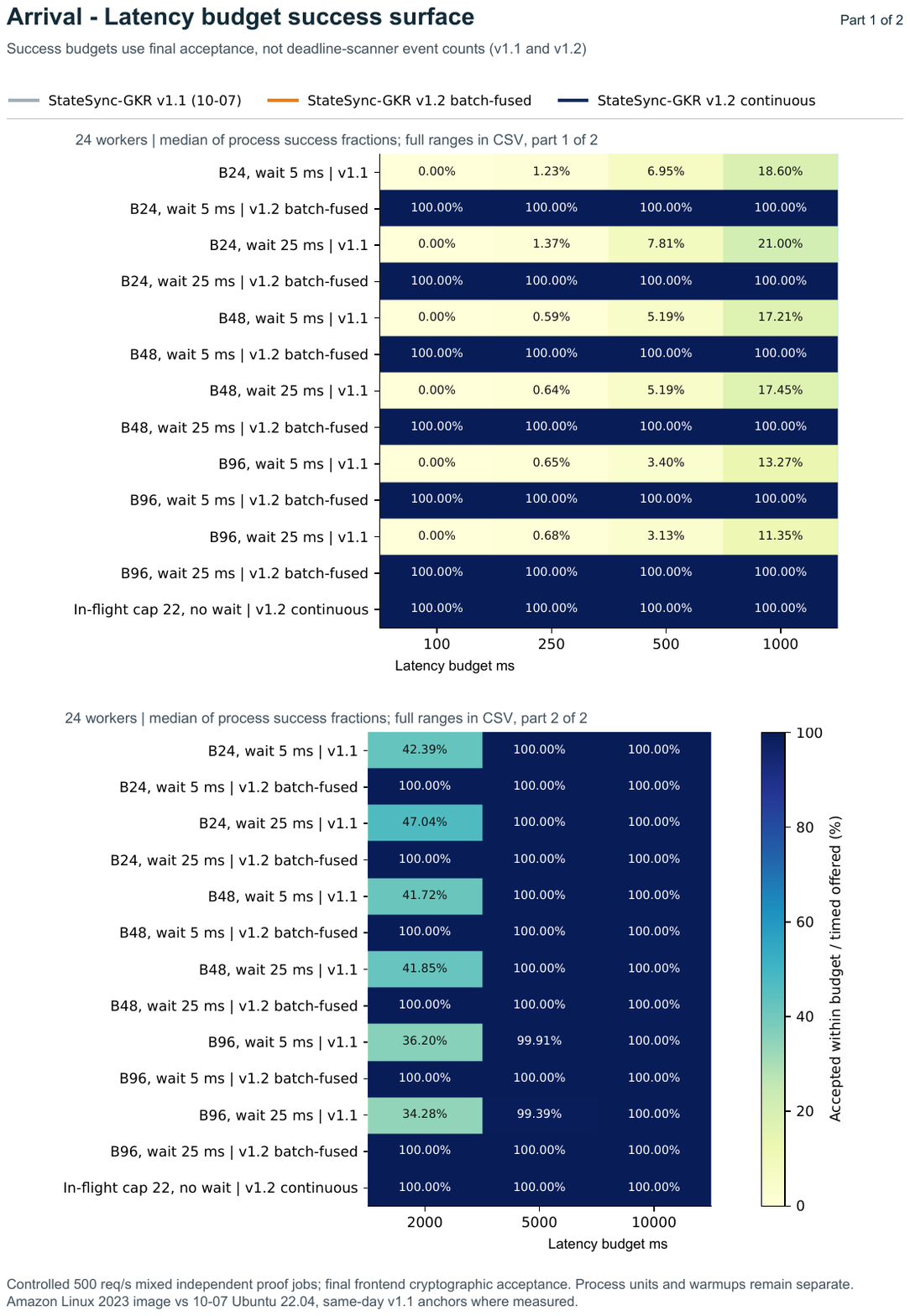}
\atlasfamily{A14}{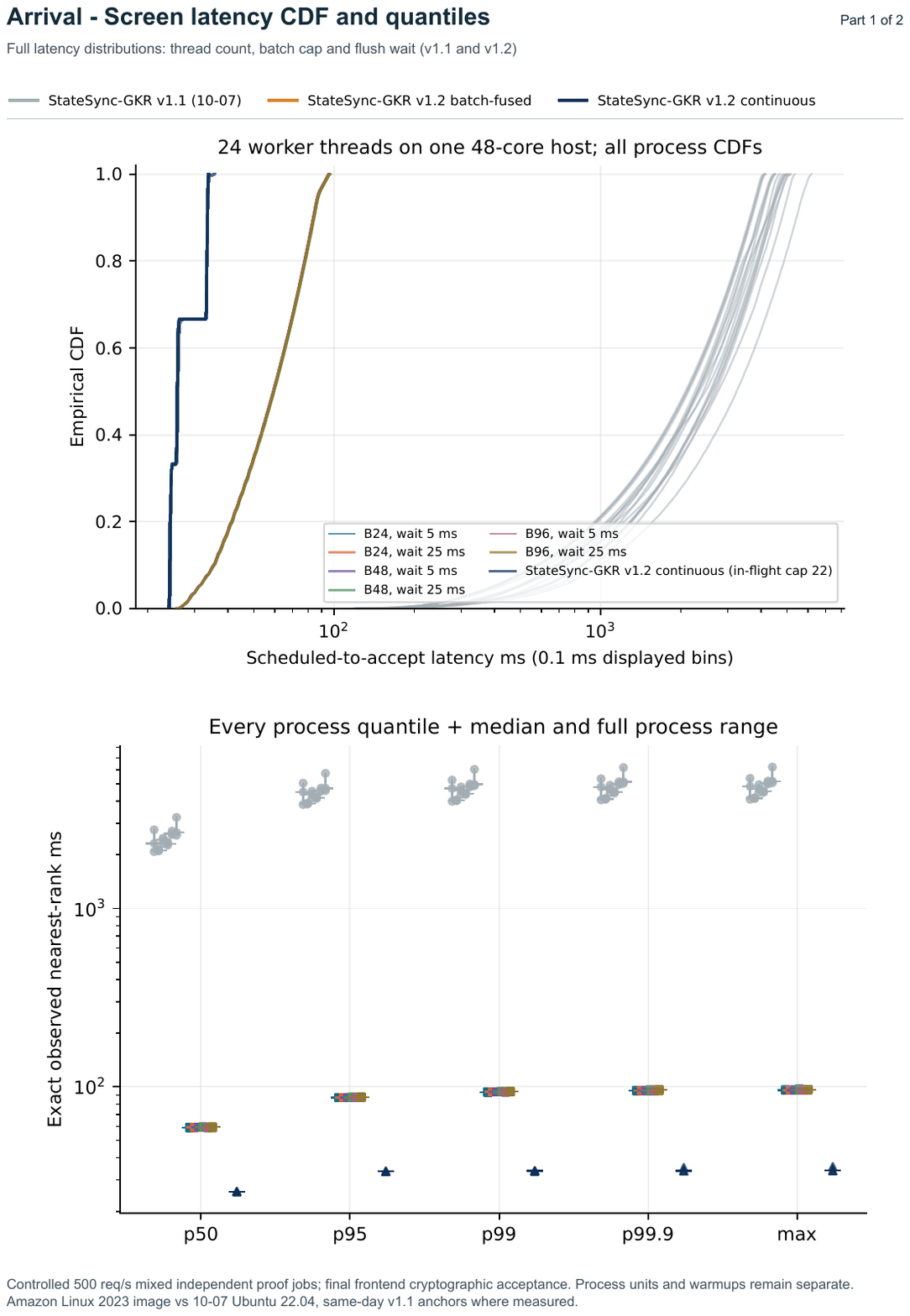}
\atlasfamily{A15}{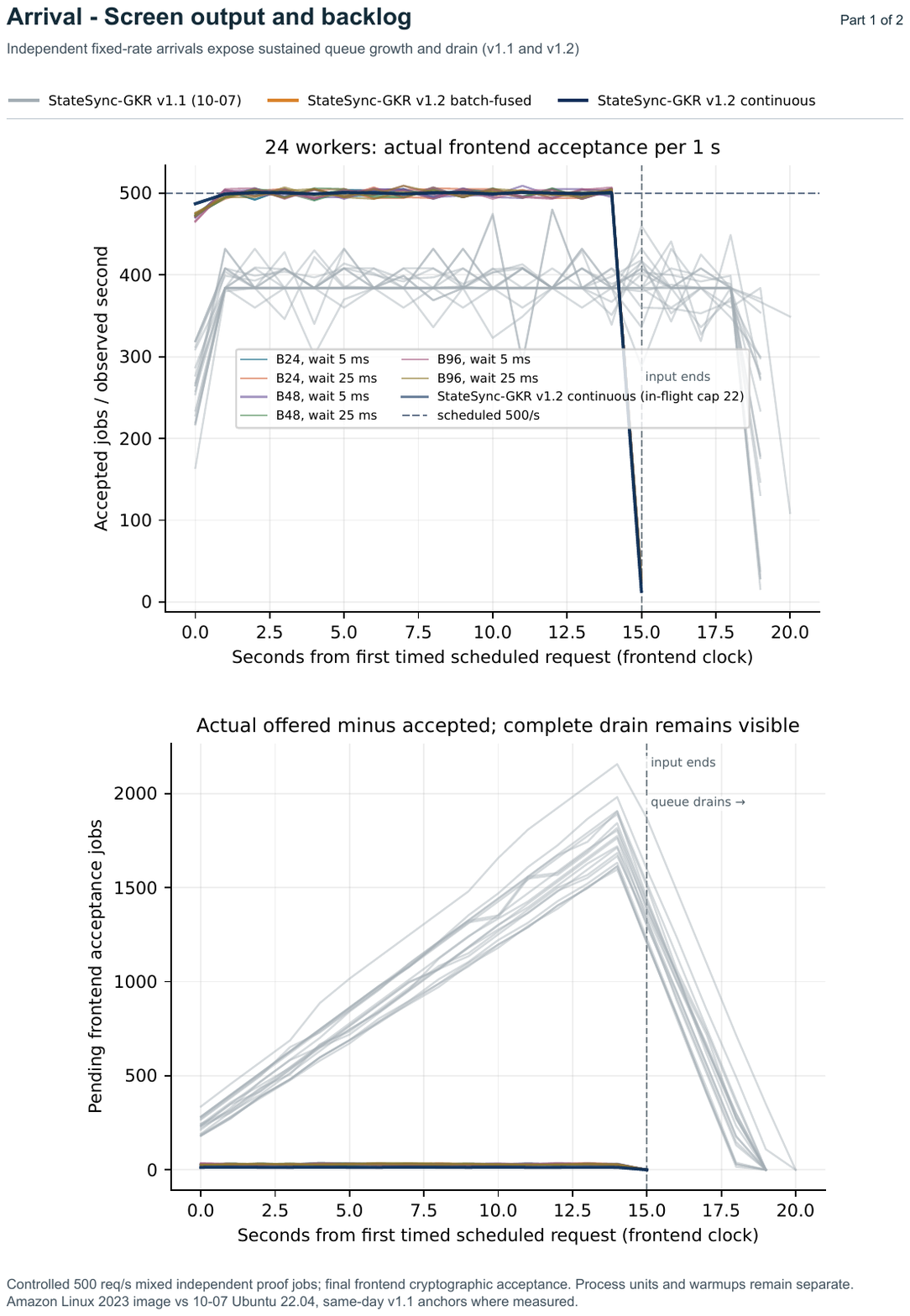}
\atlasfamily{A16}{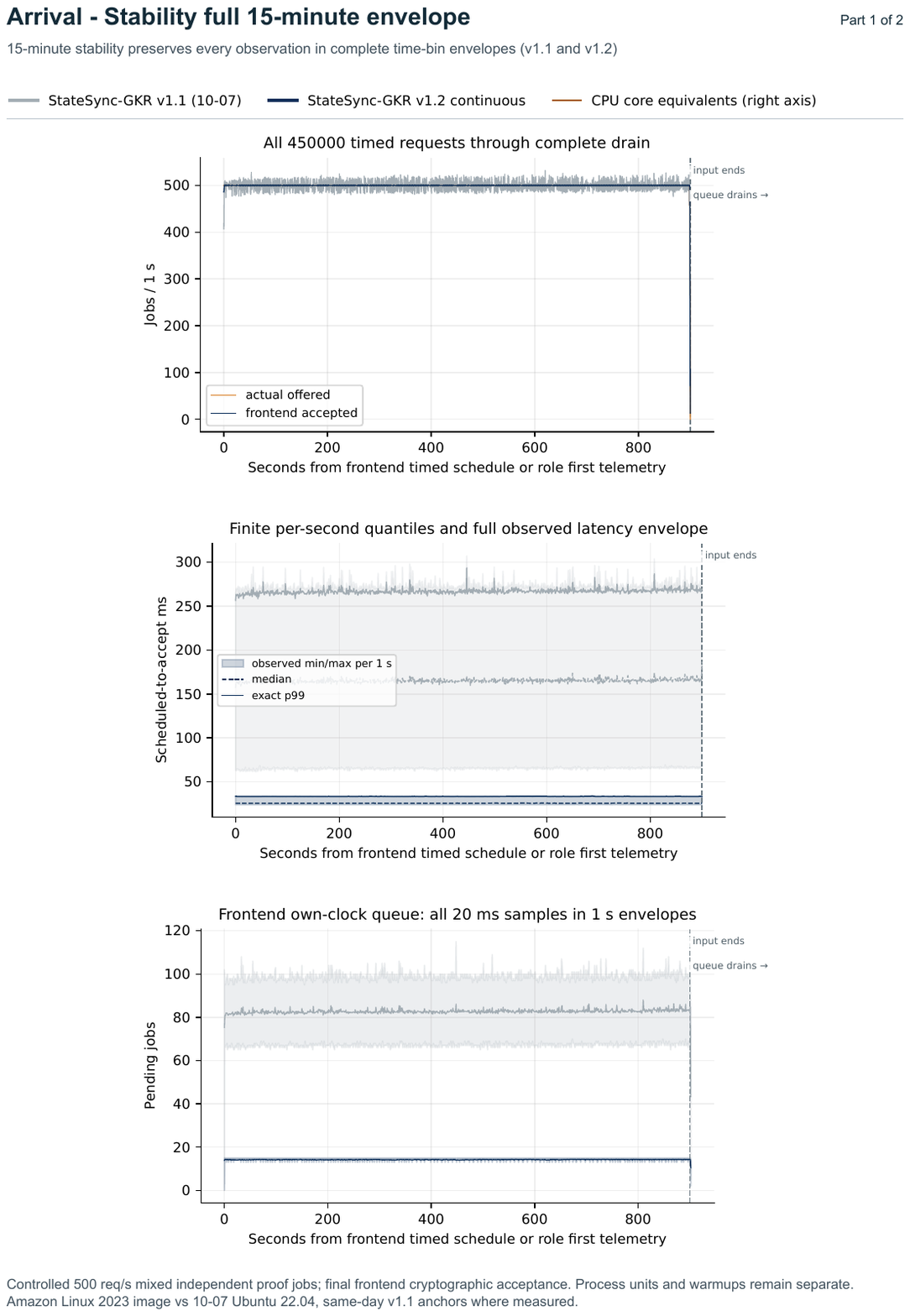}
\atlasfamily{A17}{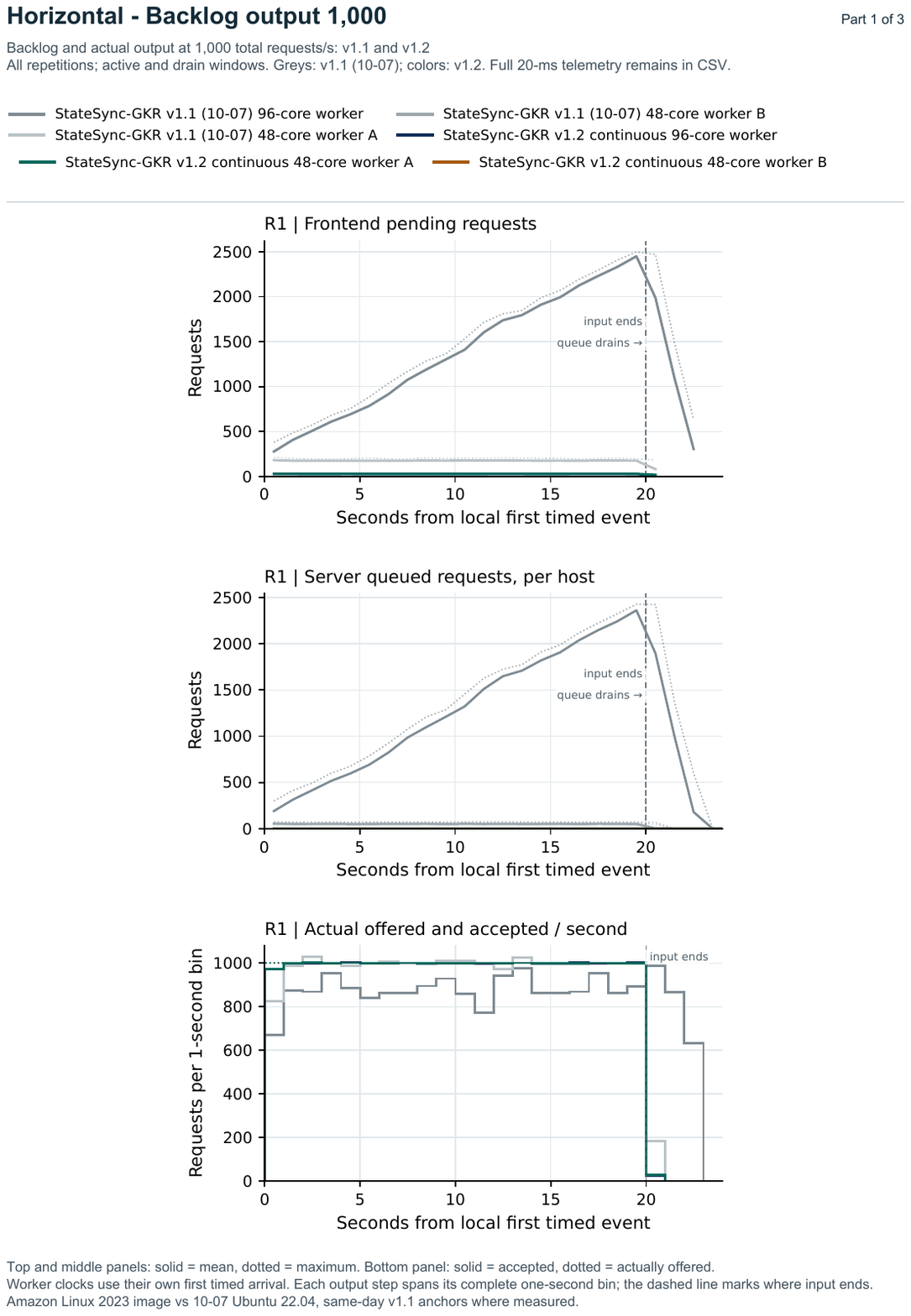}
\atlasfamily{A18}{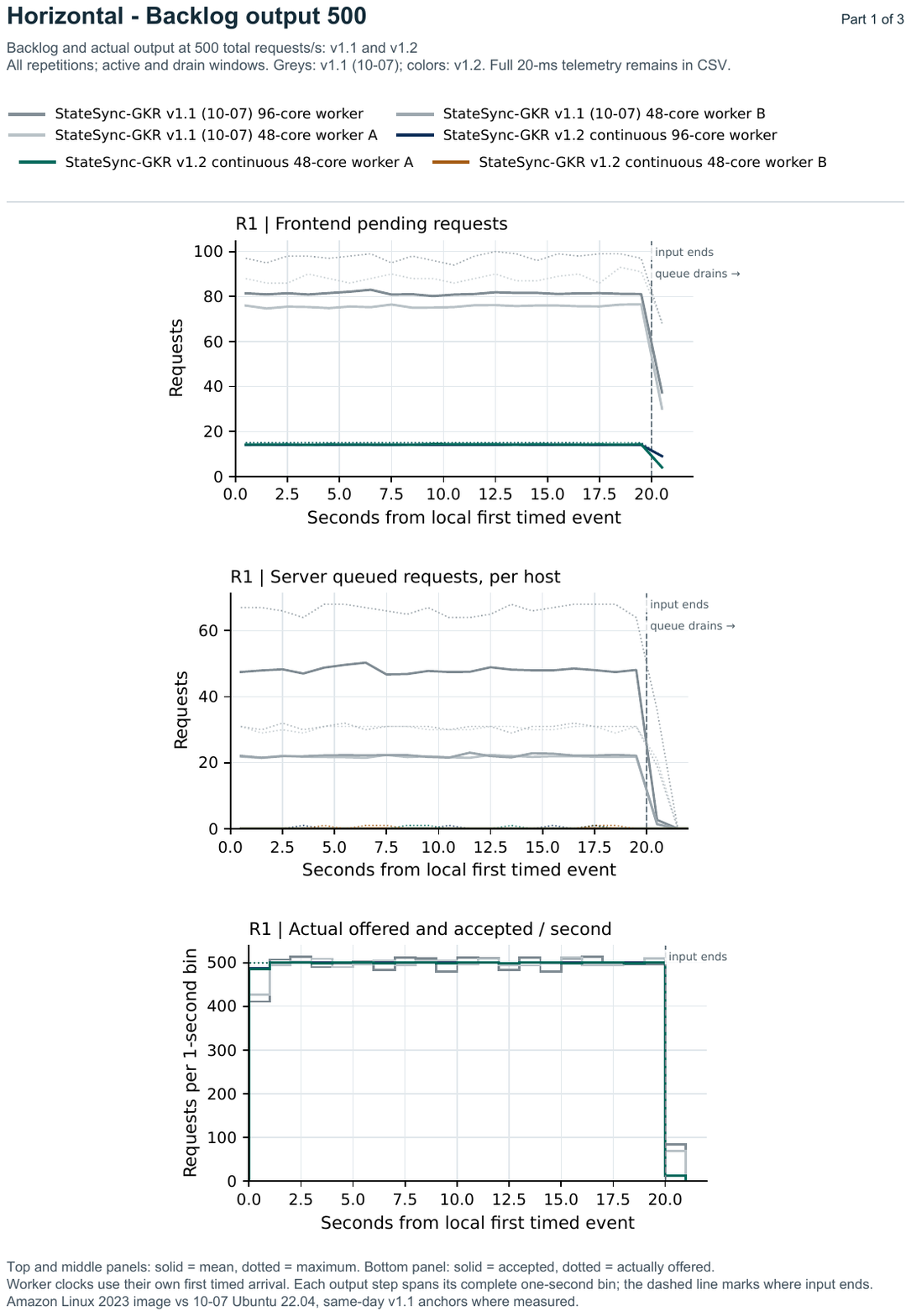}
\atlasfamily{A19}{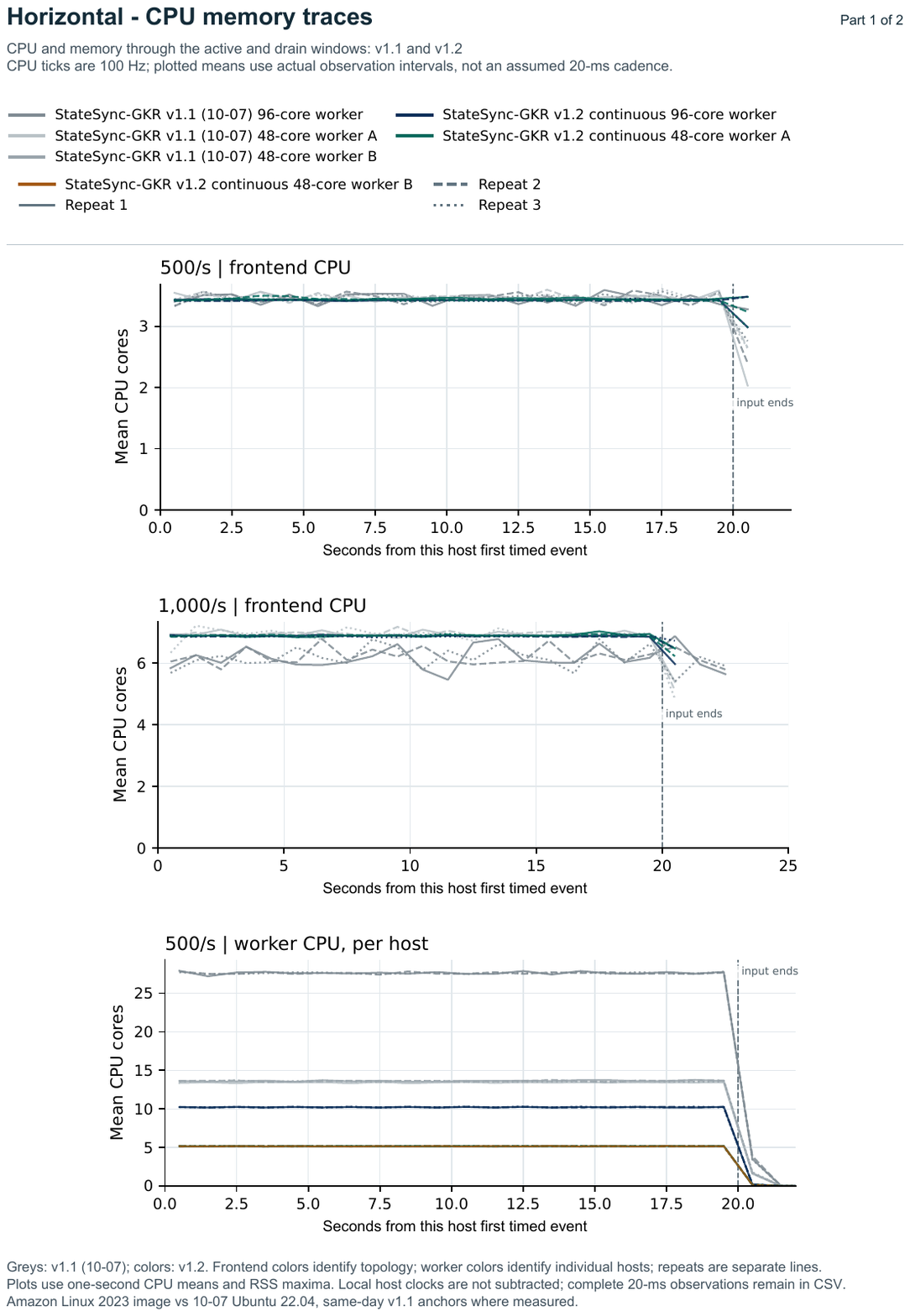}
\atlasfamily{A20}{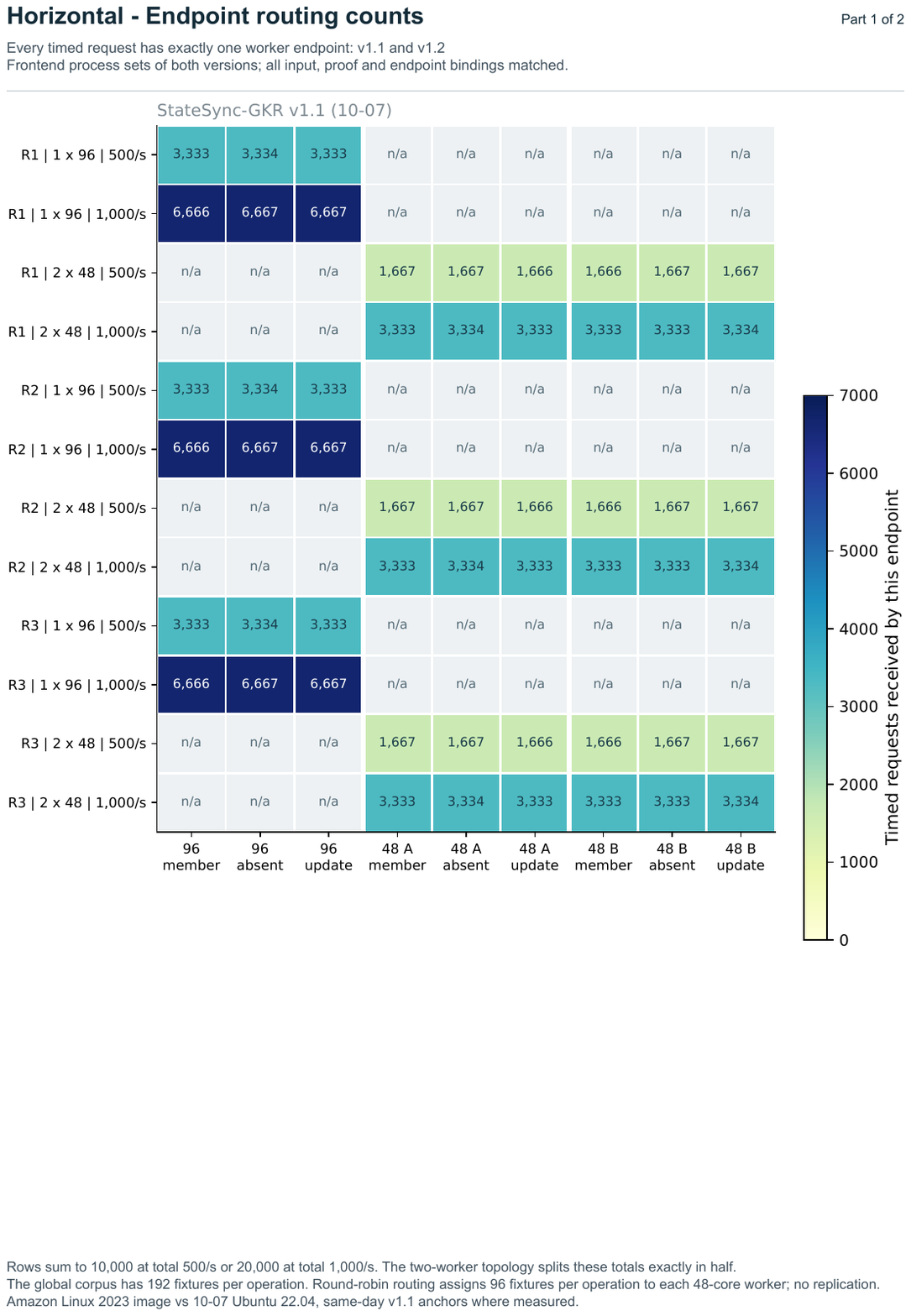}
\atlasfamily{A21}{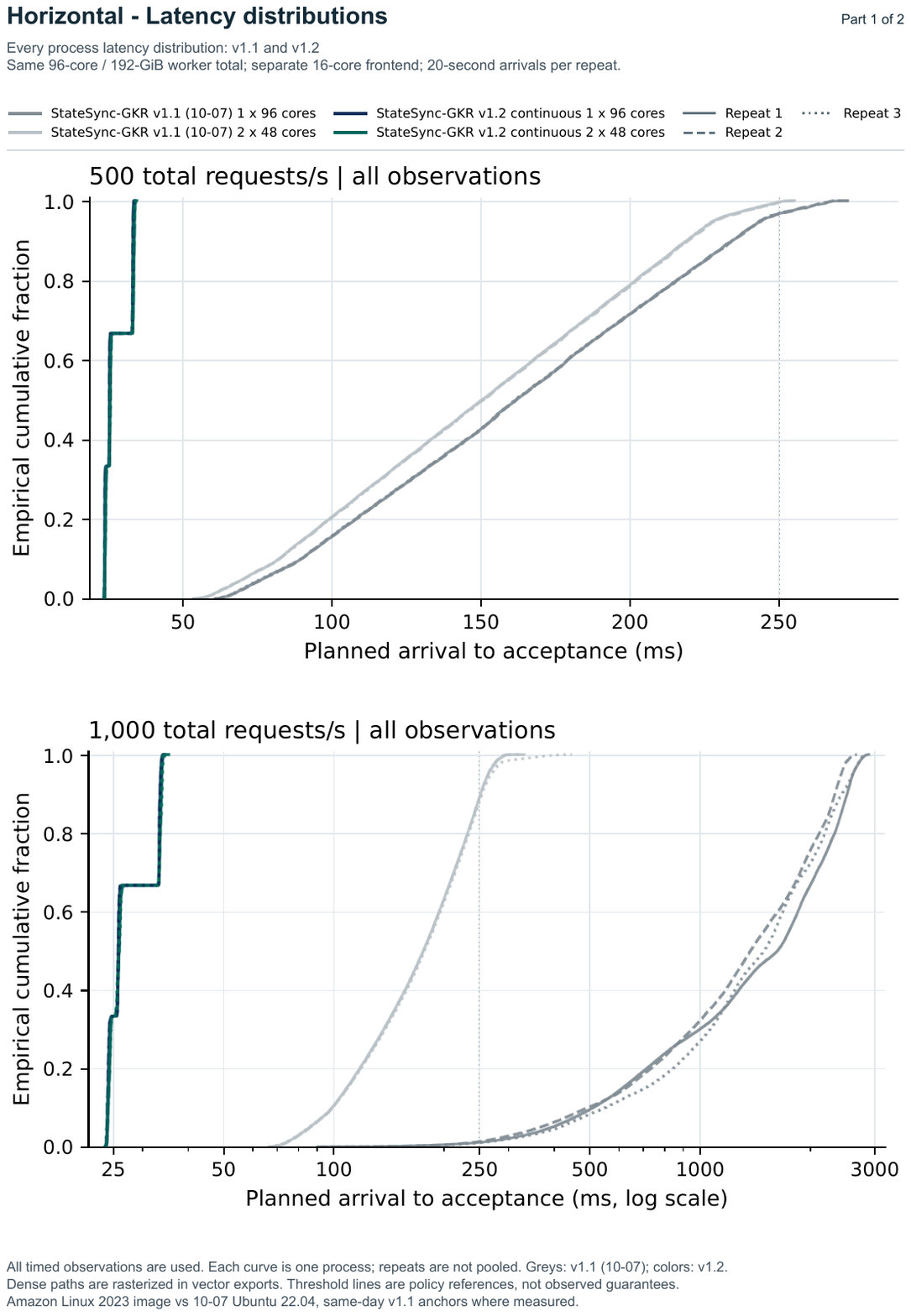}
\atlasfamily{A22}{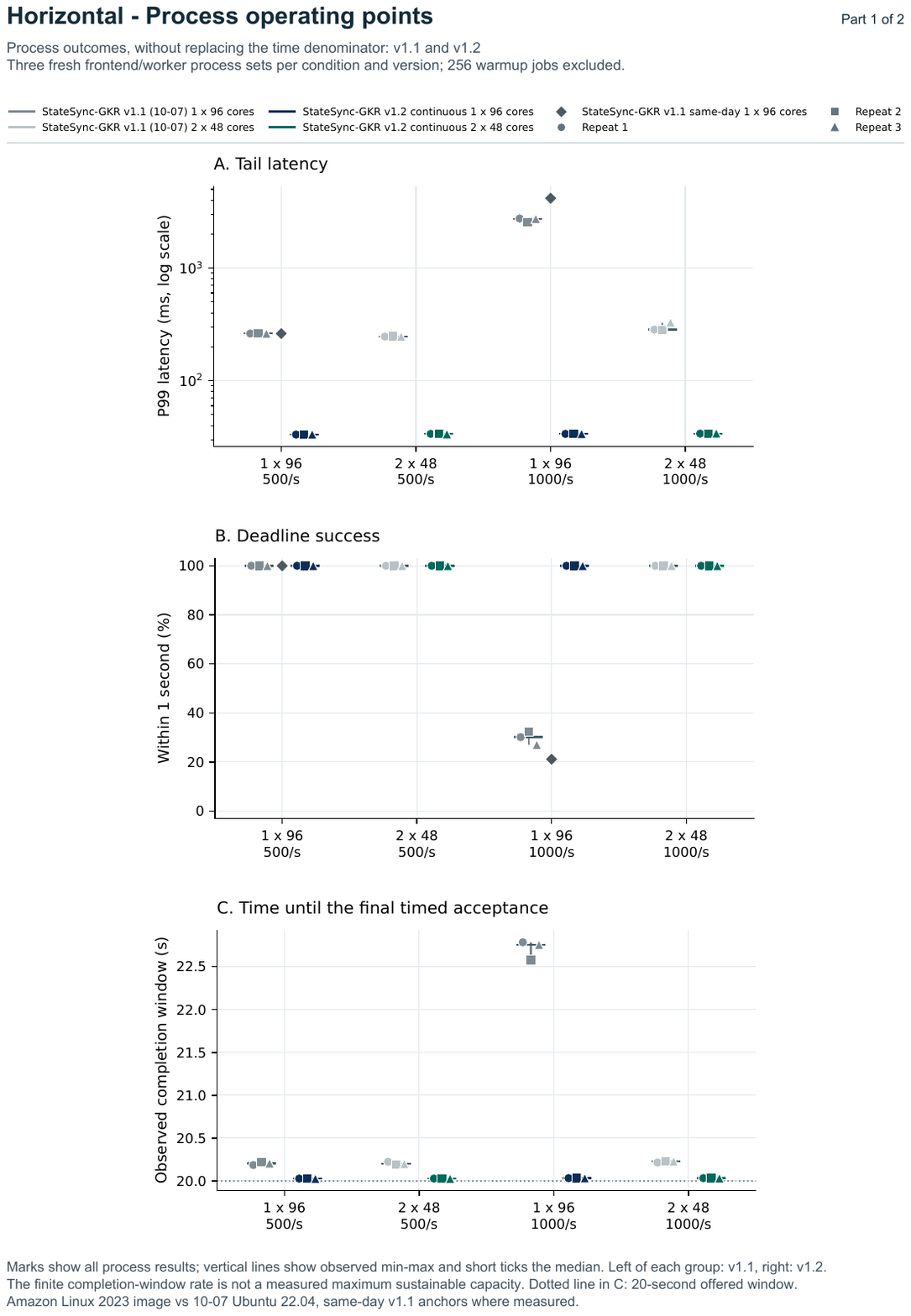}
\atlasfamily{A23}{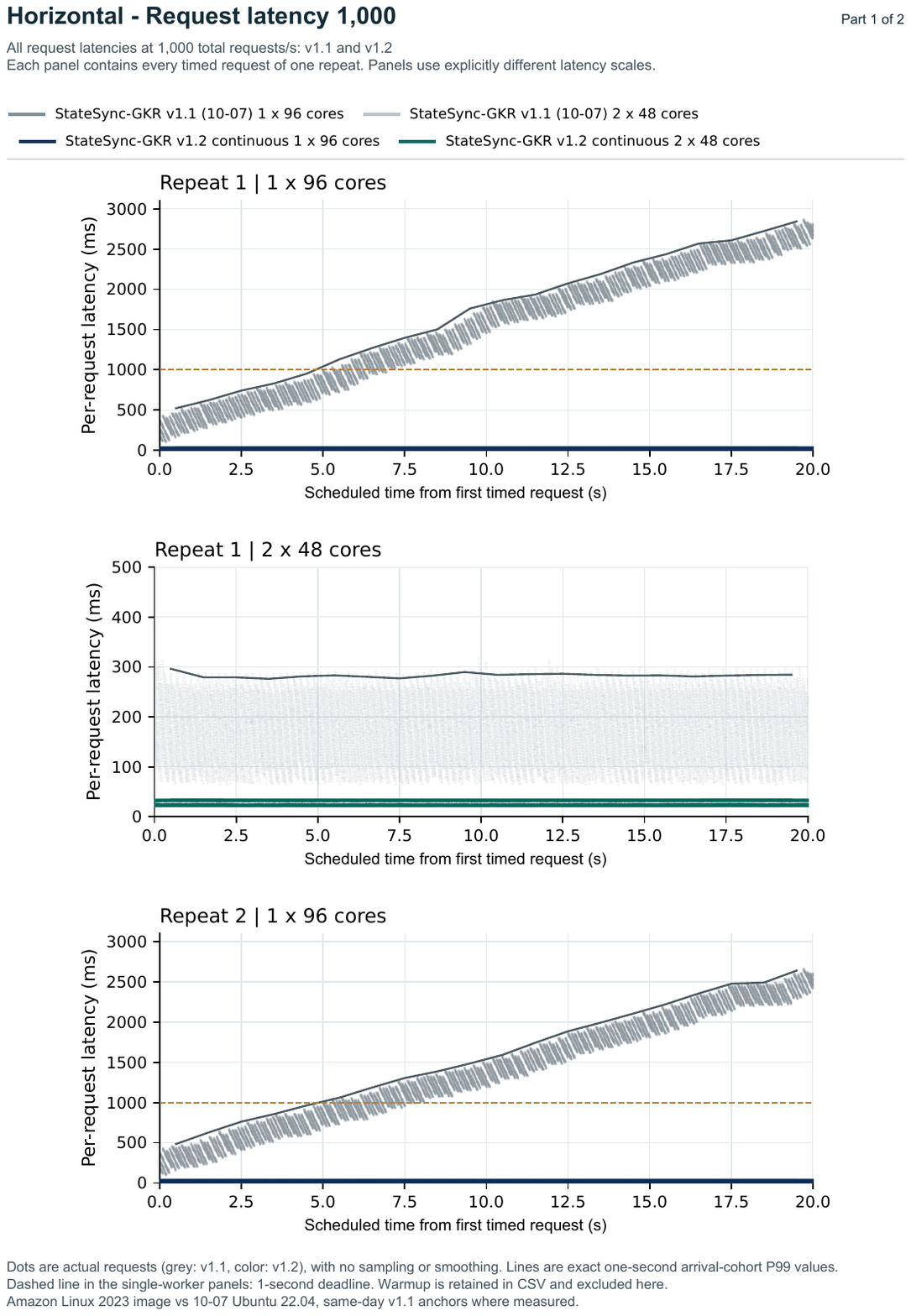}
\atlasfamily{A24}{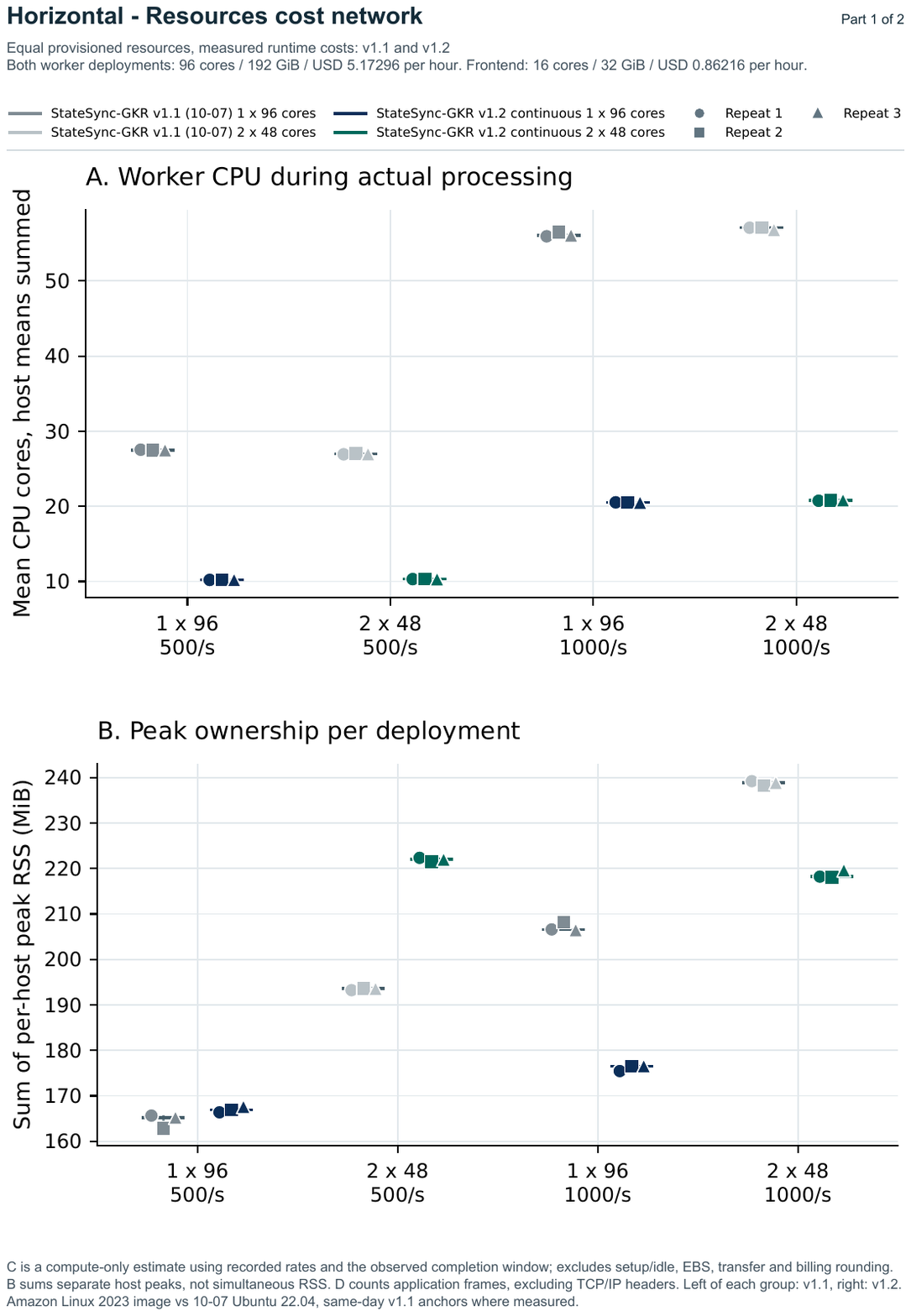}
\atlasfamily{A25}{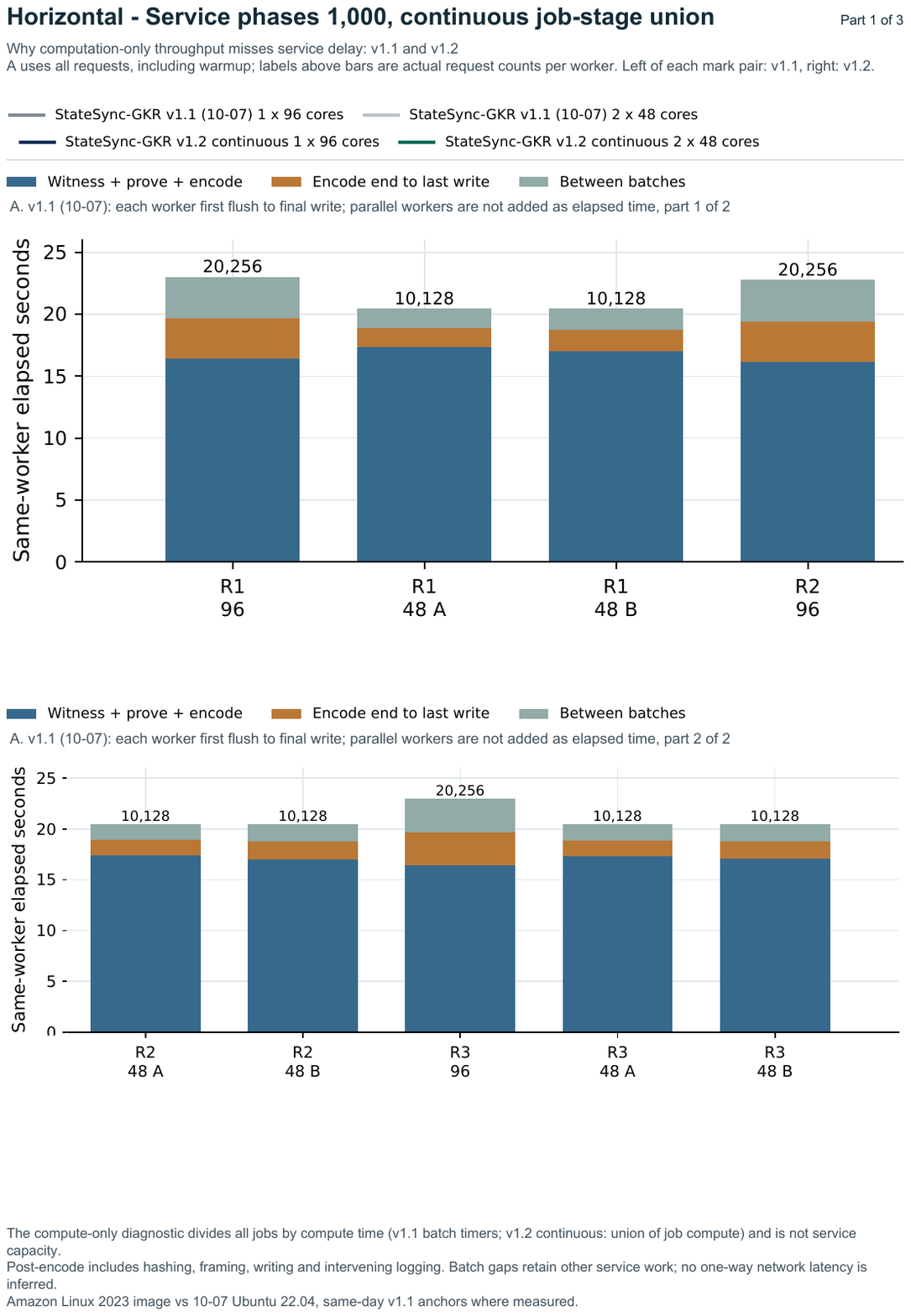}
\atlasfamily{A26}{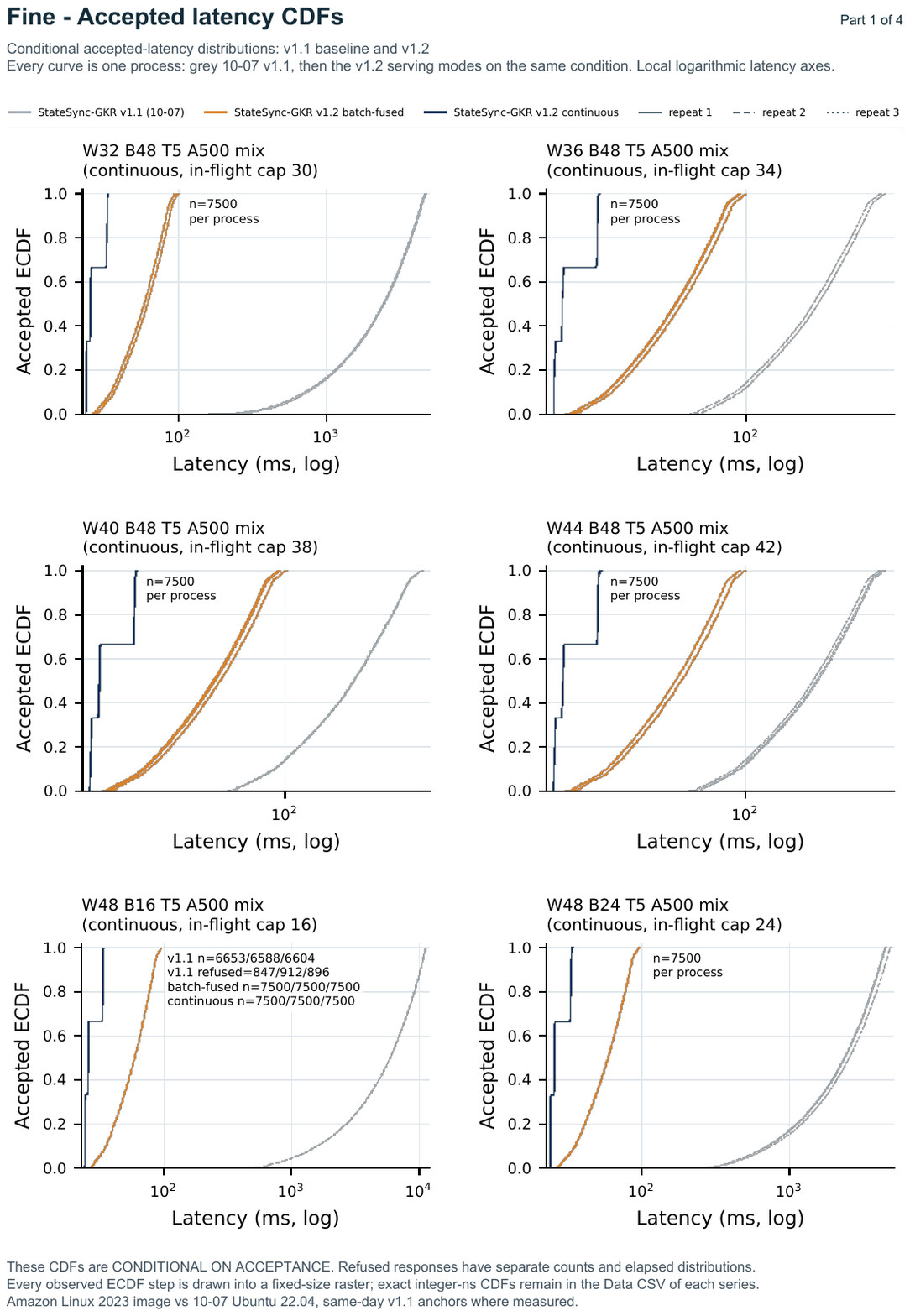}
\atlasfamily{A27}{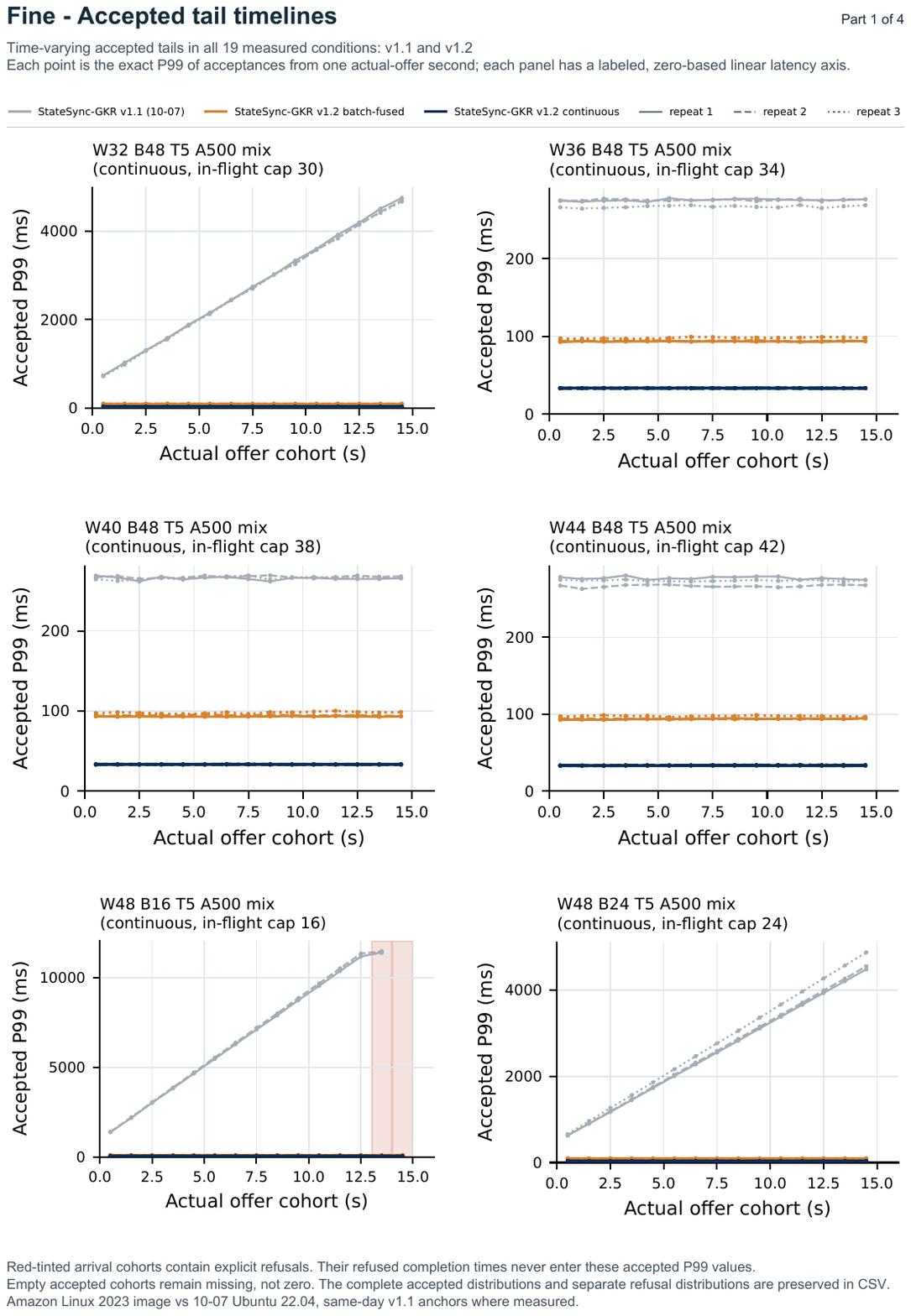}
\atlasfamily{A28}{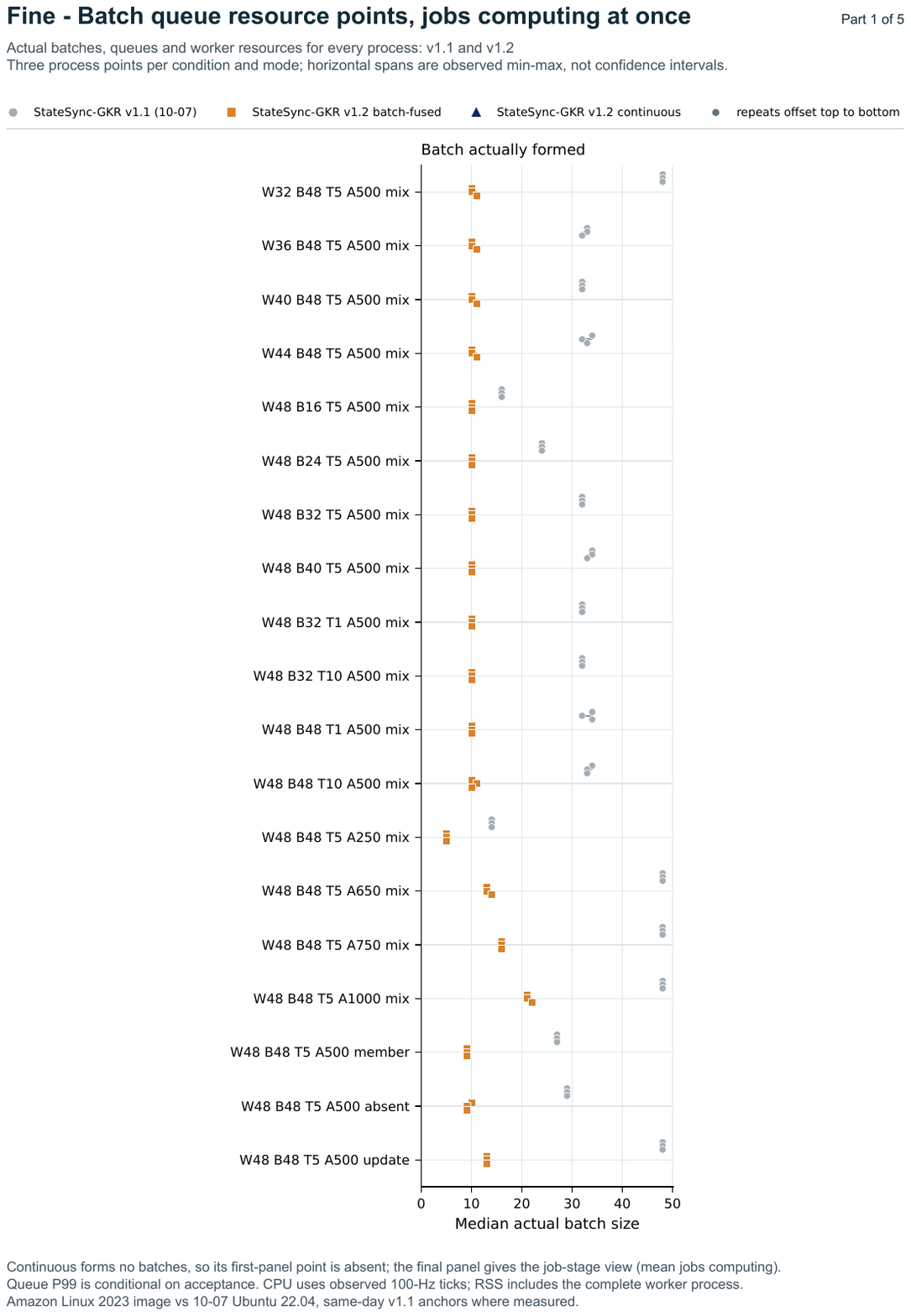}
\atlasfamily{A29}{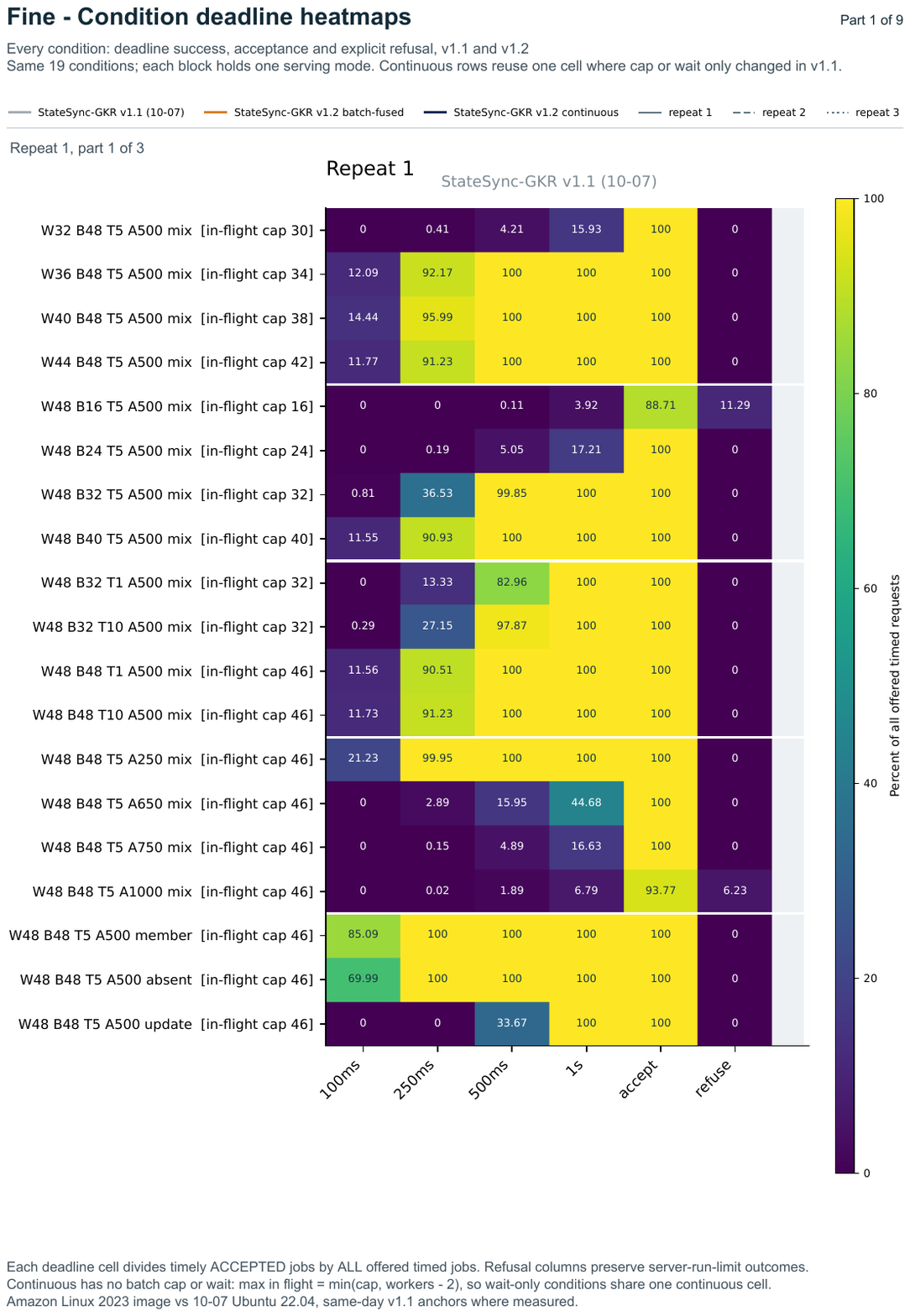}
\atlasfamily{A30}{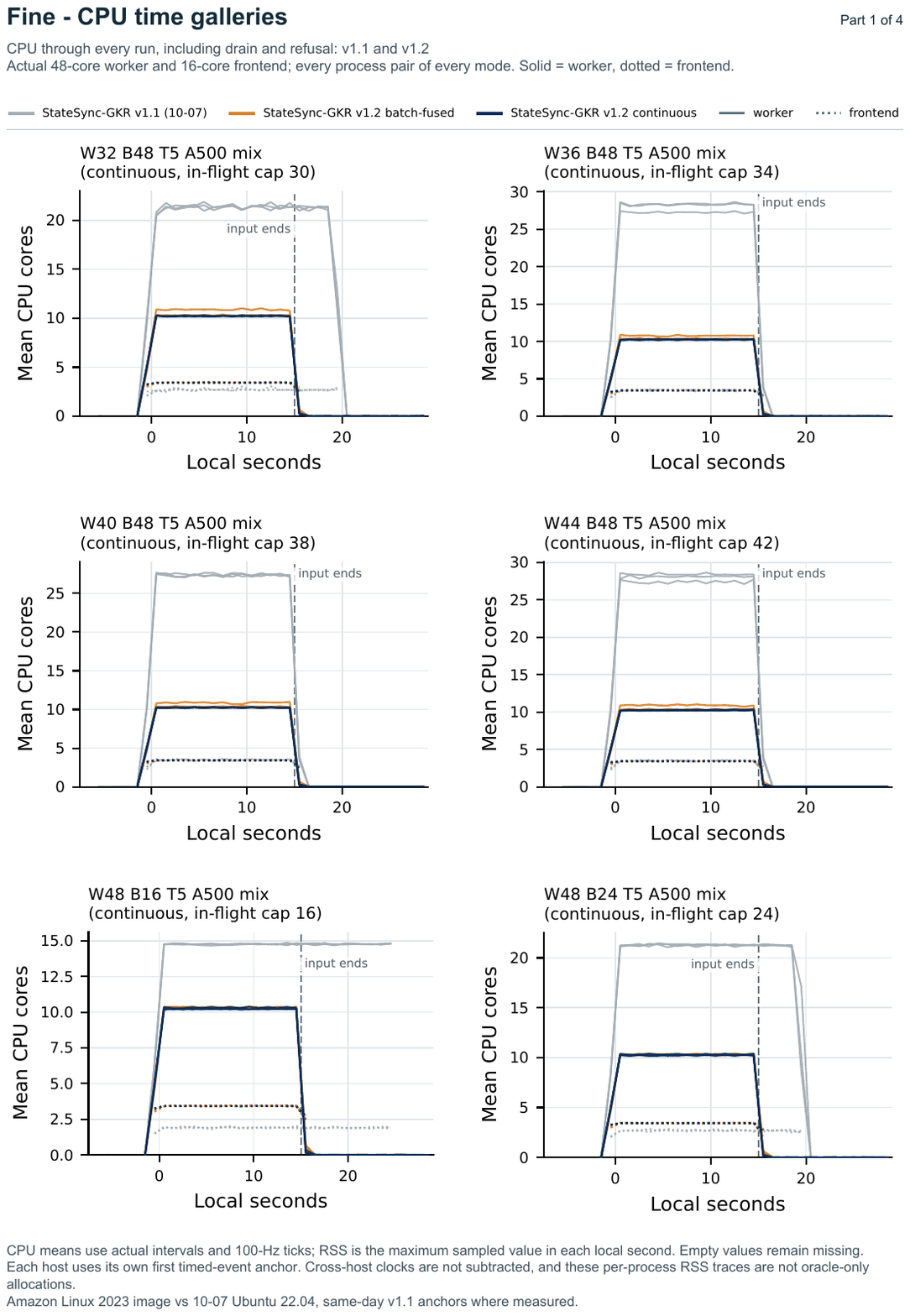}
\atlasfamily{A31}{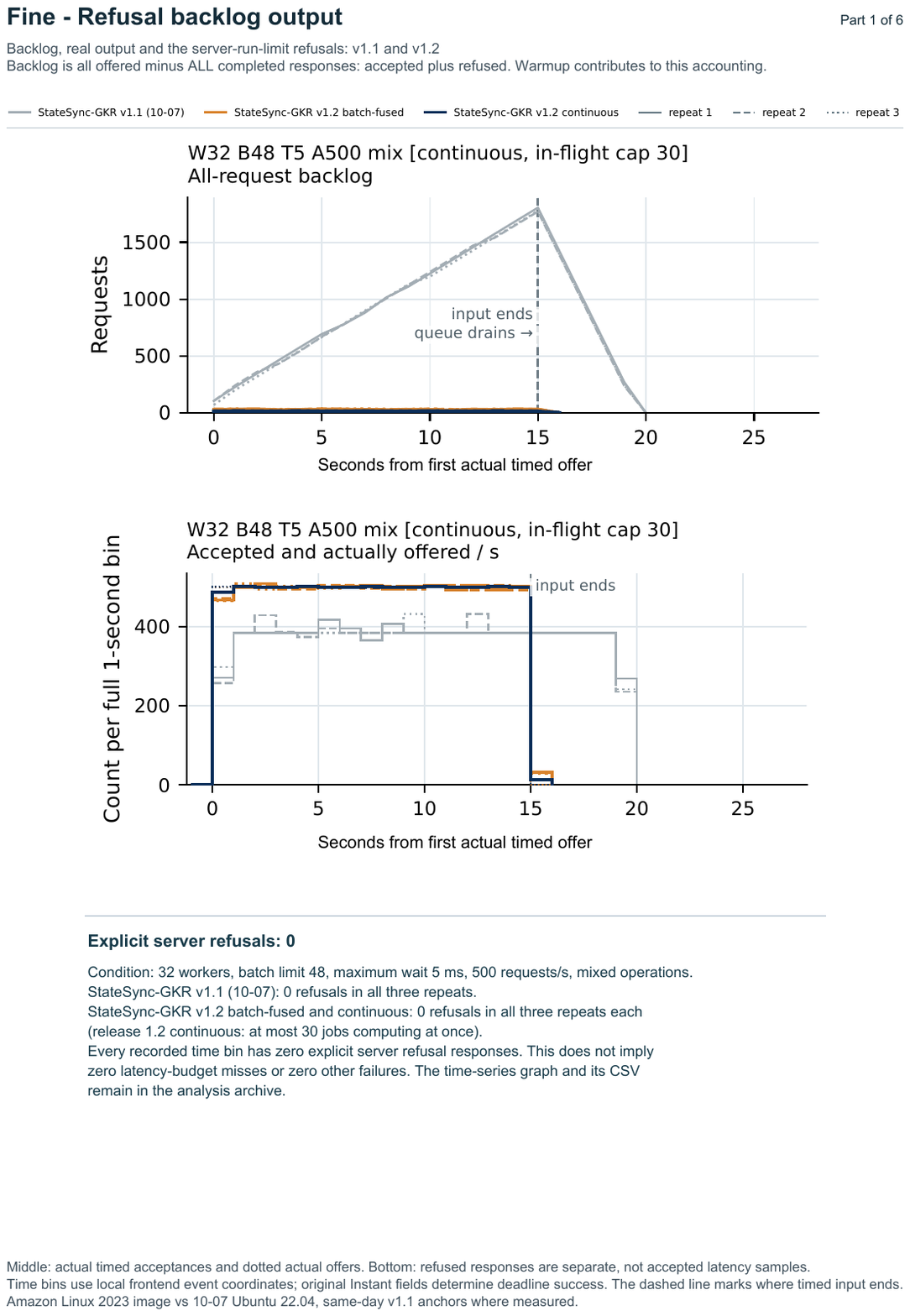}
\atlasfamily{A32}{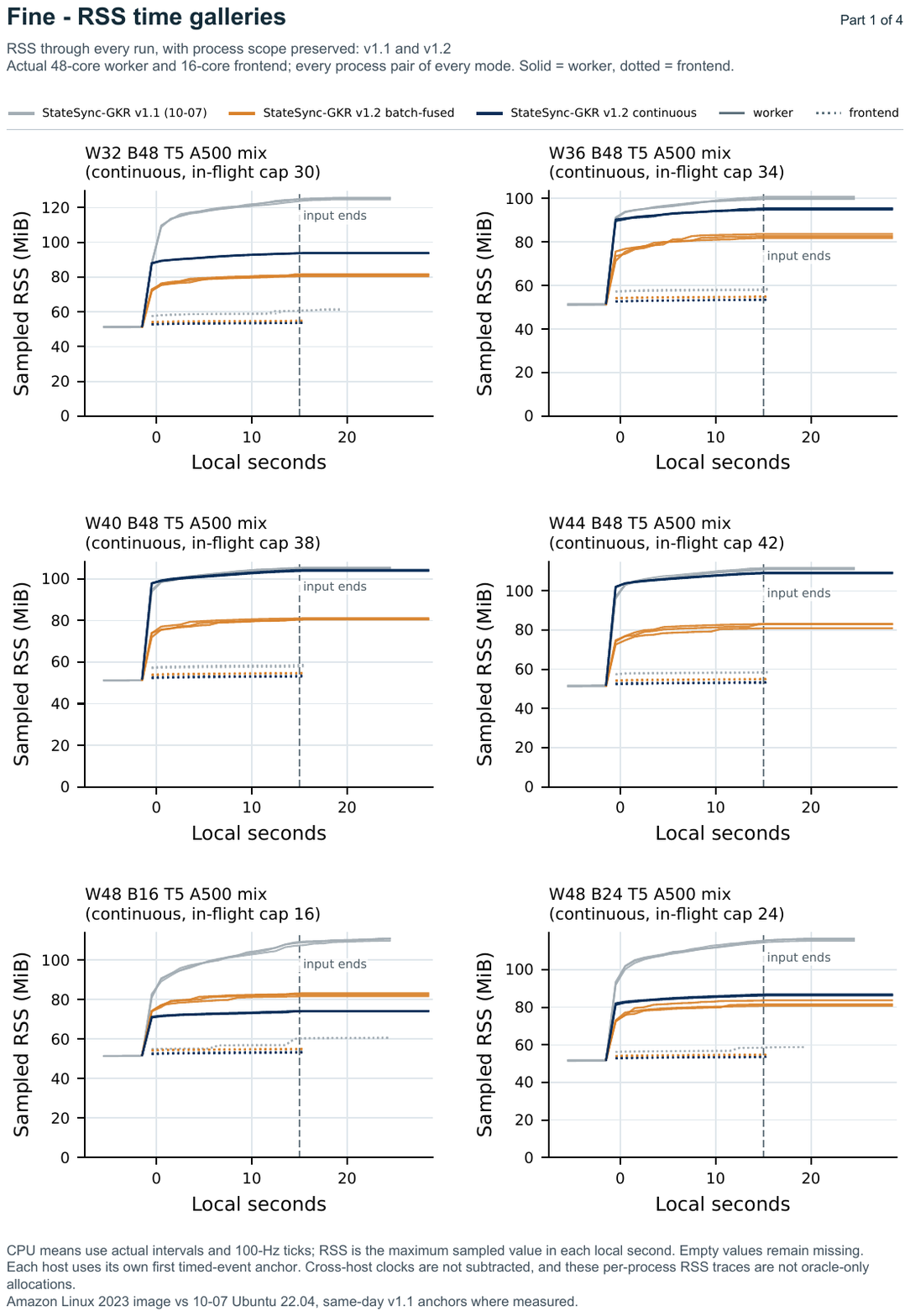}
\atlasfamily{A33}{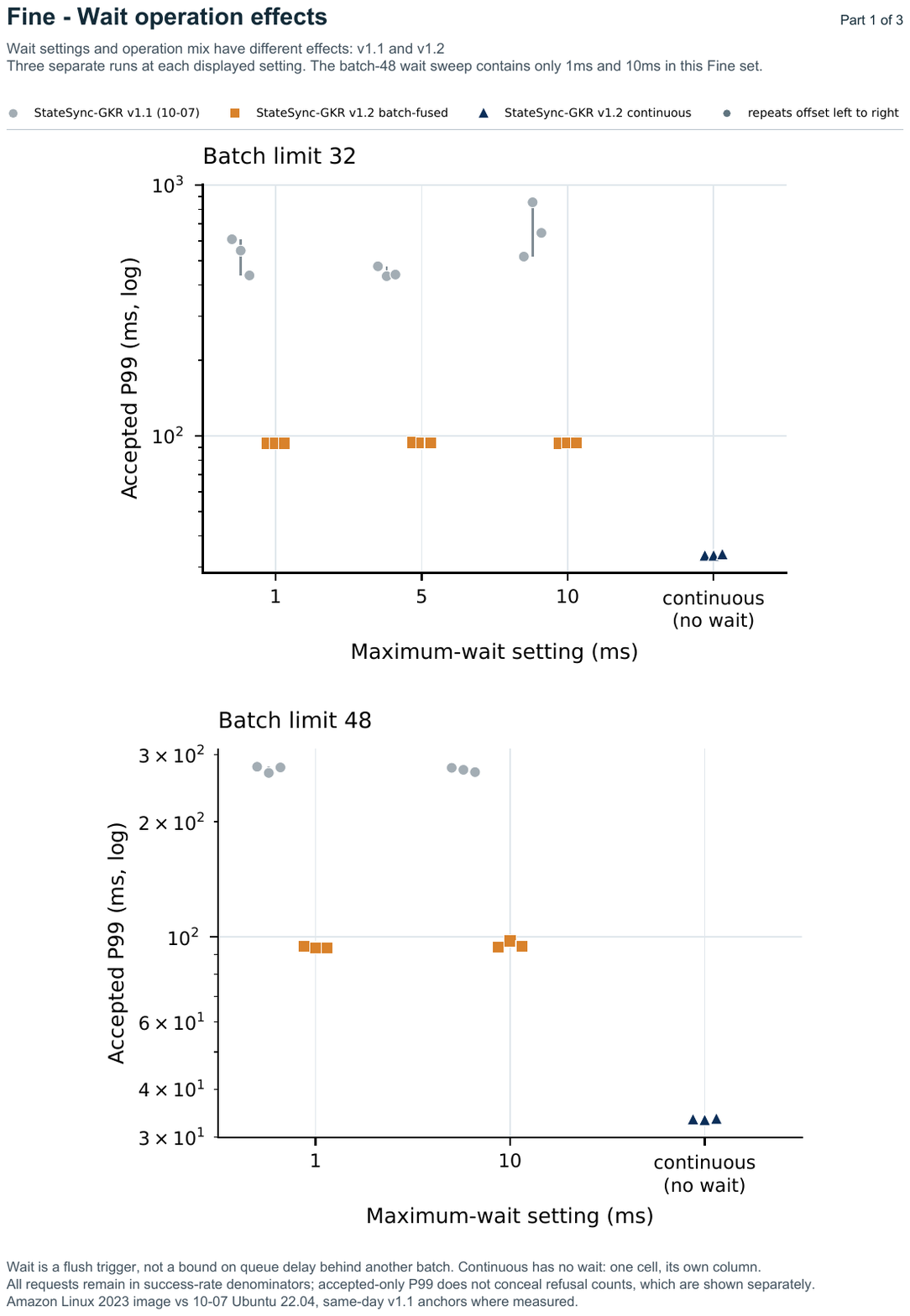}
\atlasfamily{A34}{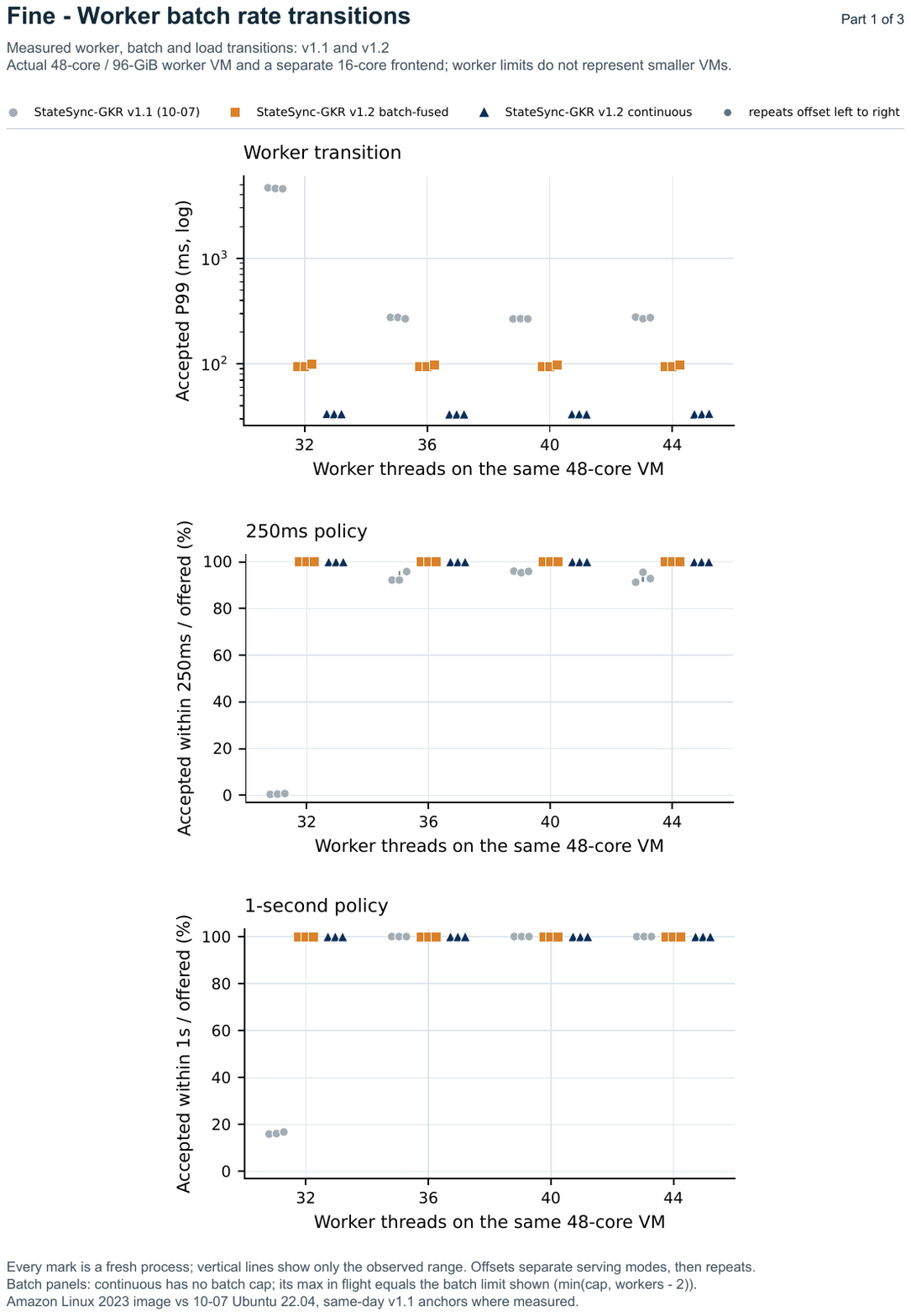}
\atlasfamily{A35}{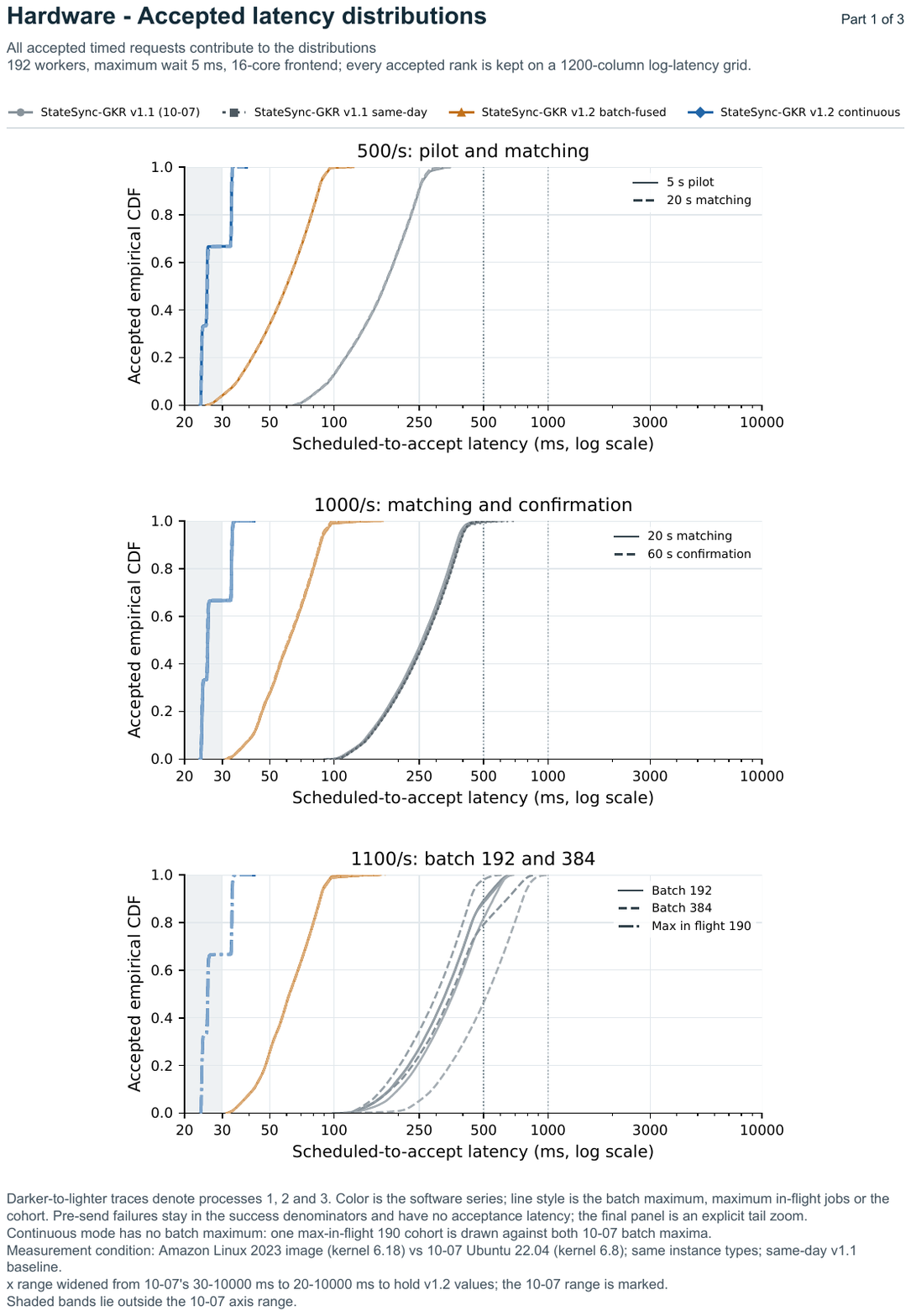}
\atlasfamily{A36}{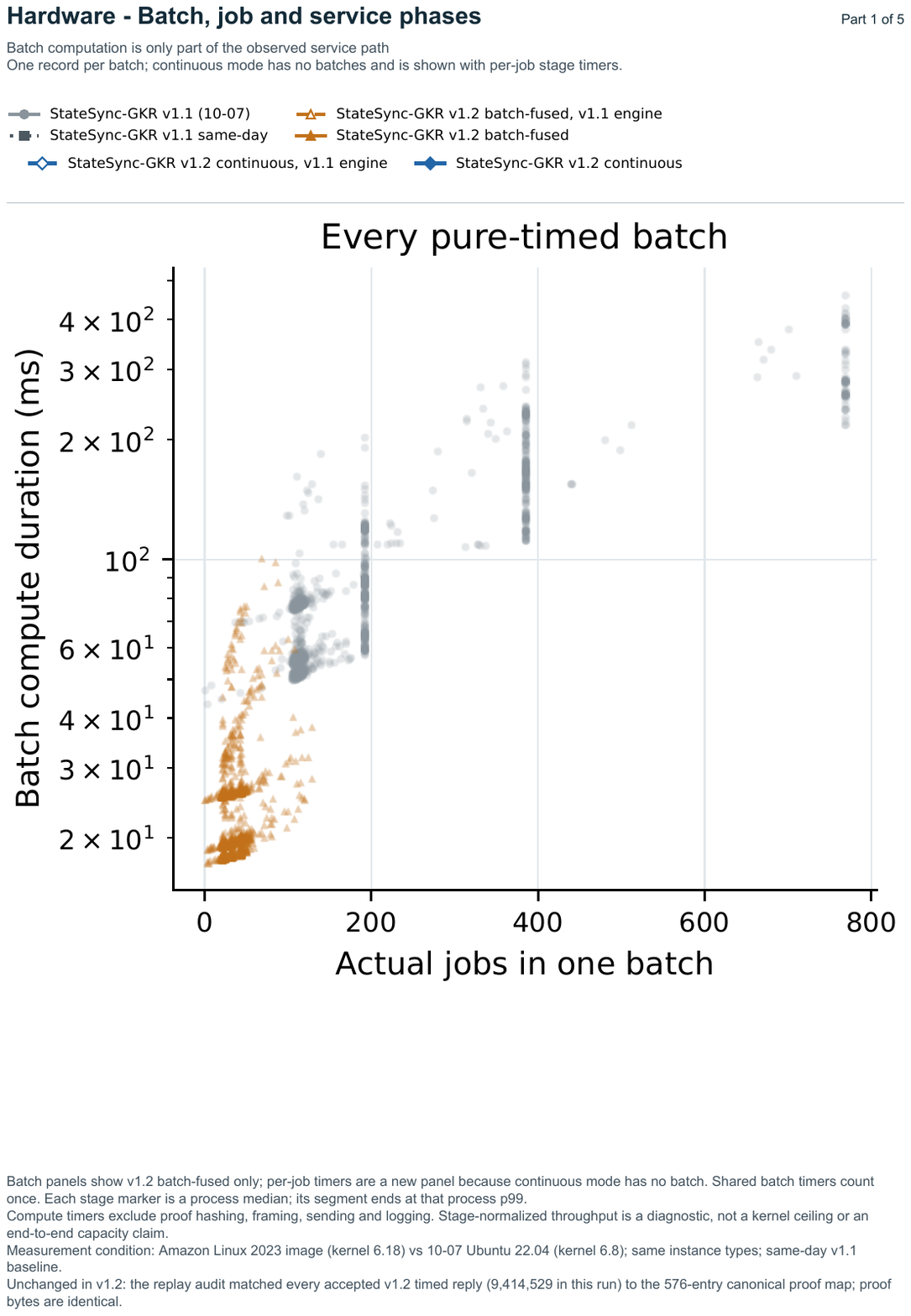}
\atlasfamily{A37}{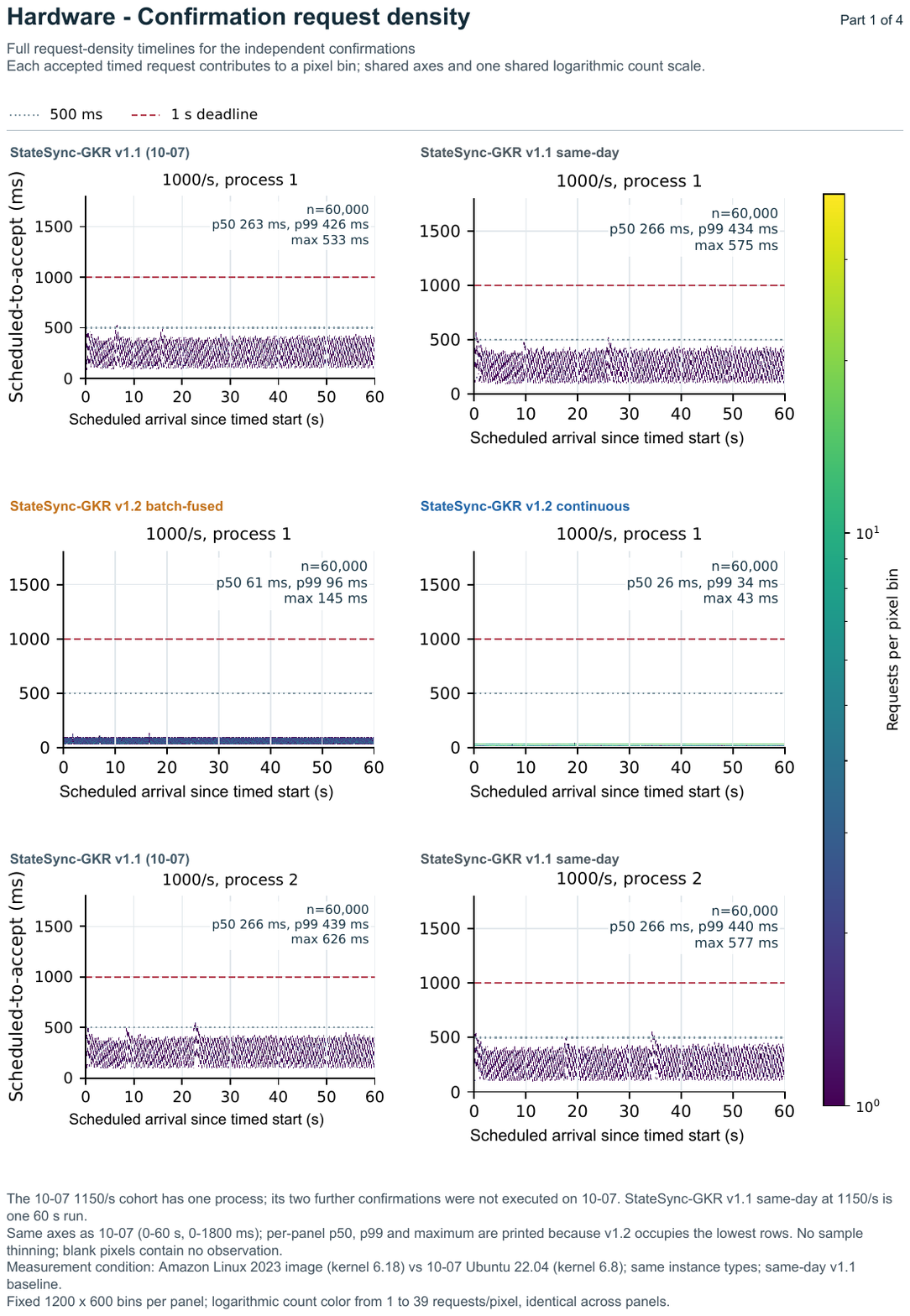}
\atlasfamily{A38}{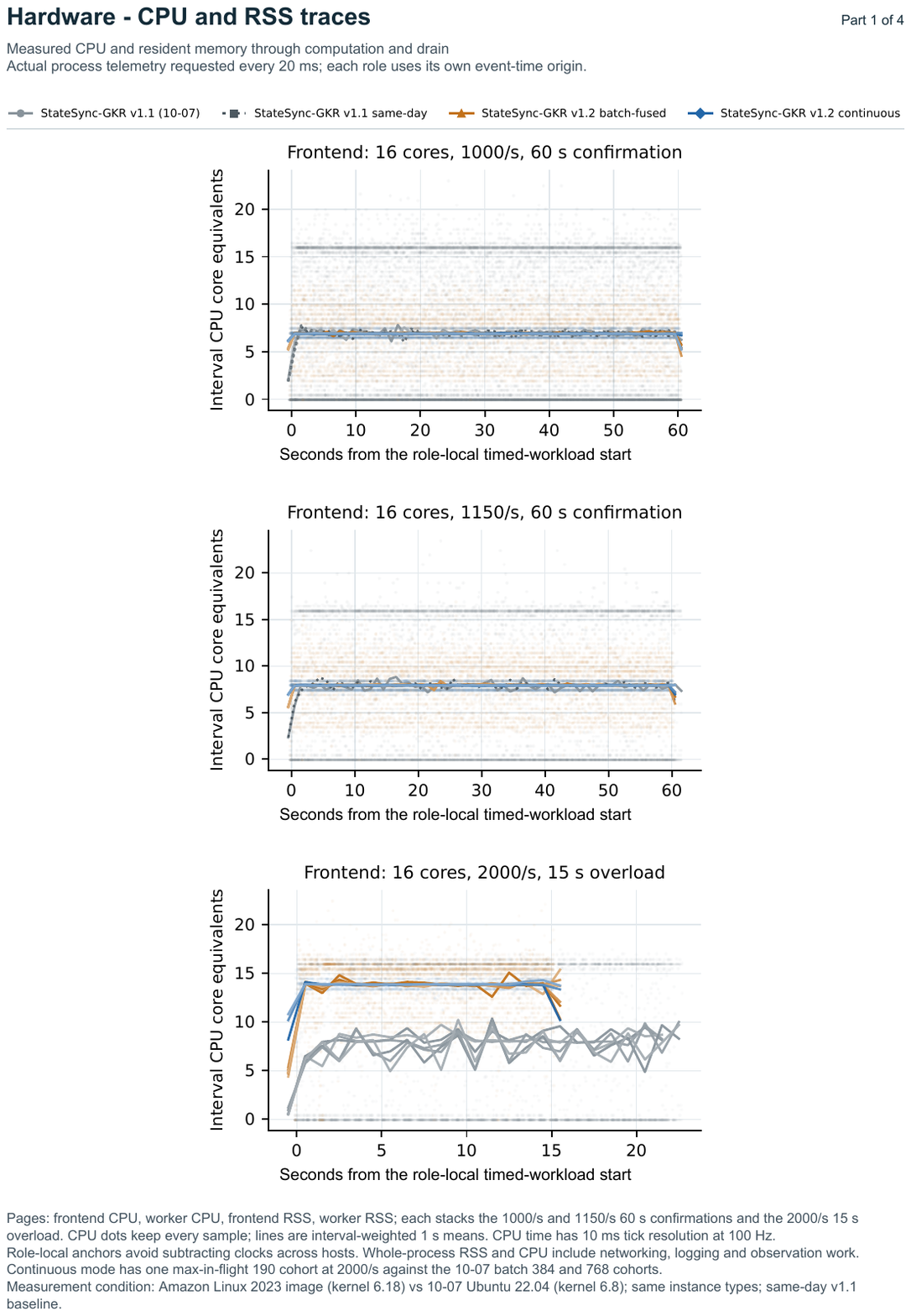}
\atlasfamily{A39}{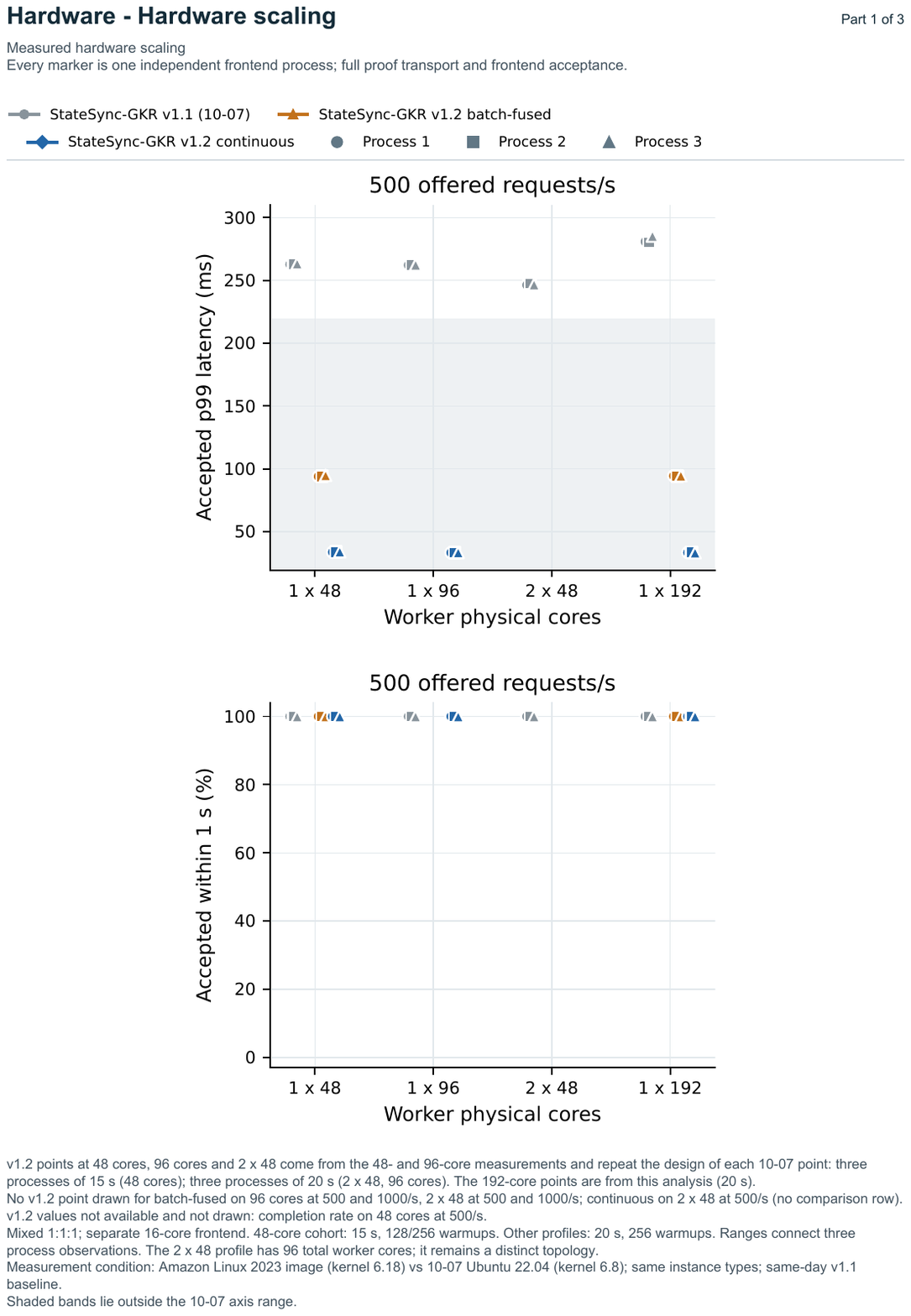}
\atlasfamily{A40}{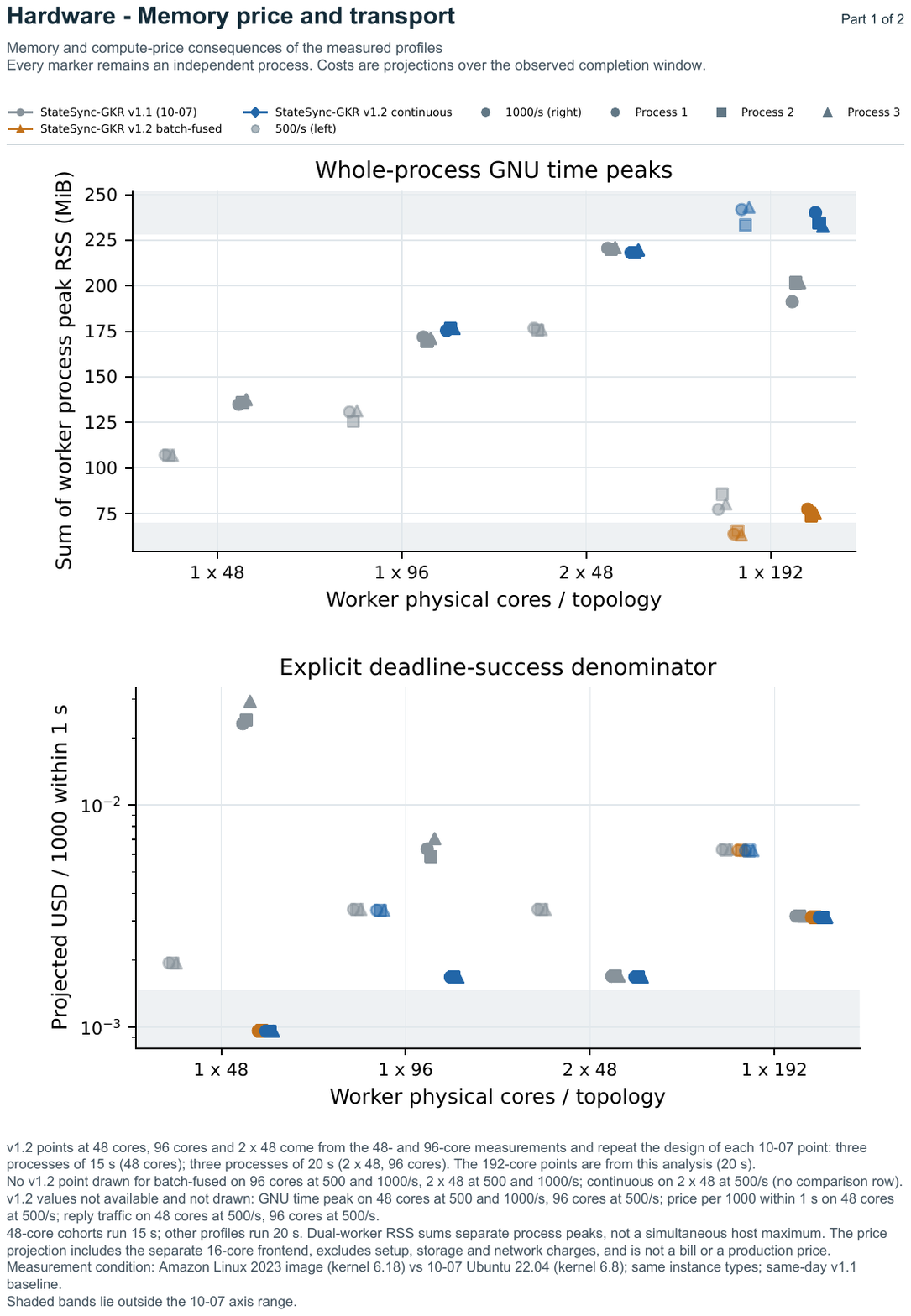}
\atlasfamily{A41}{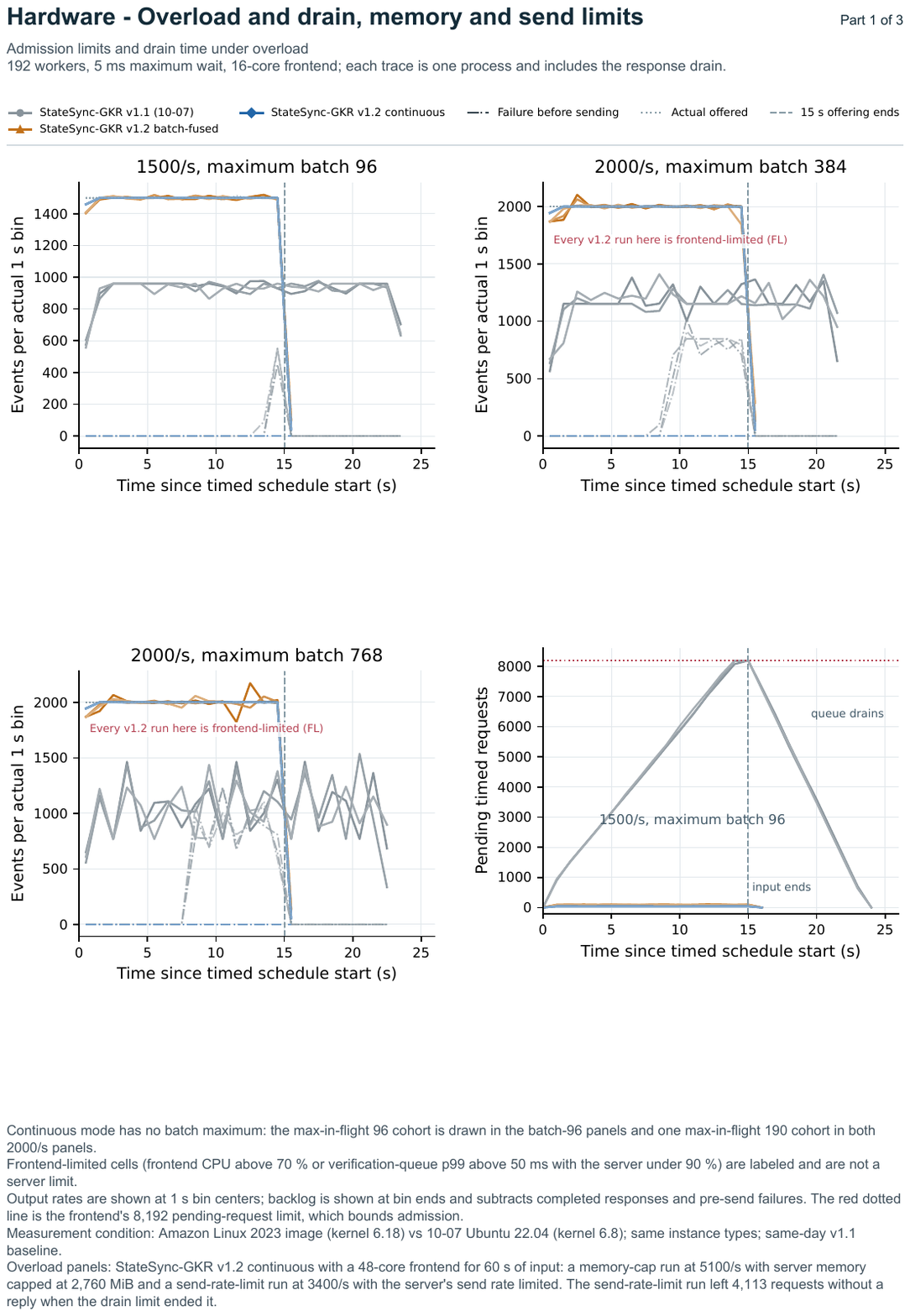}
\atlasfamily{A42}{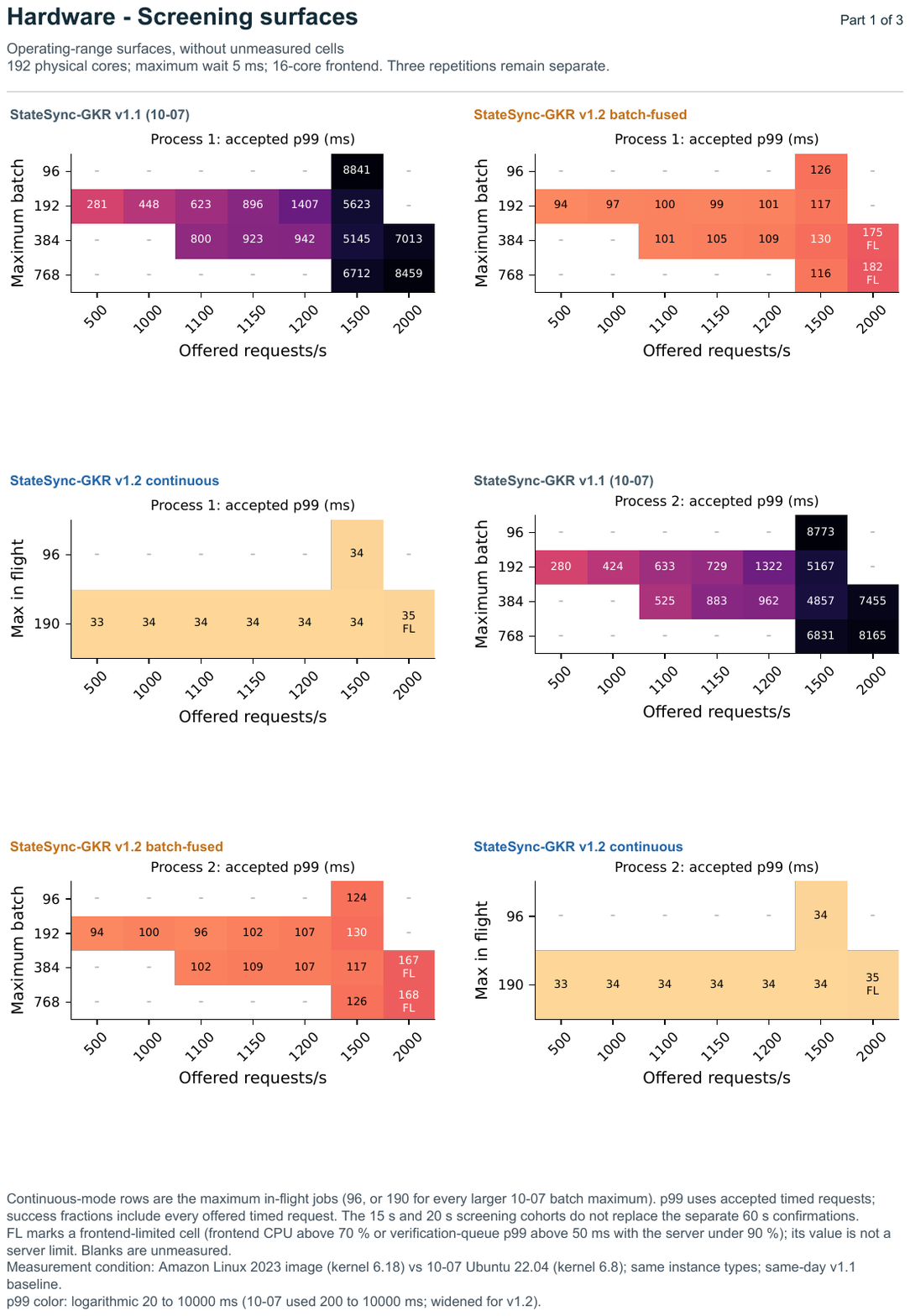}
\atlasfamily{A43}{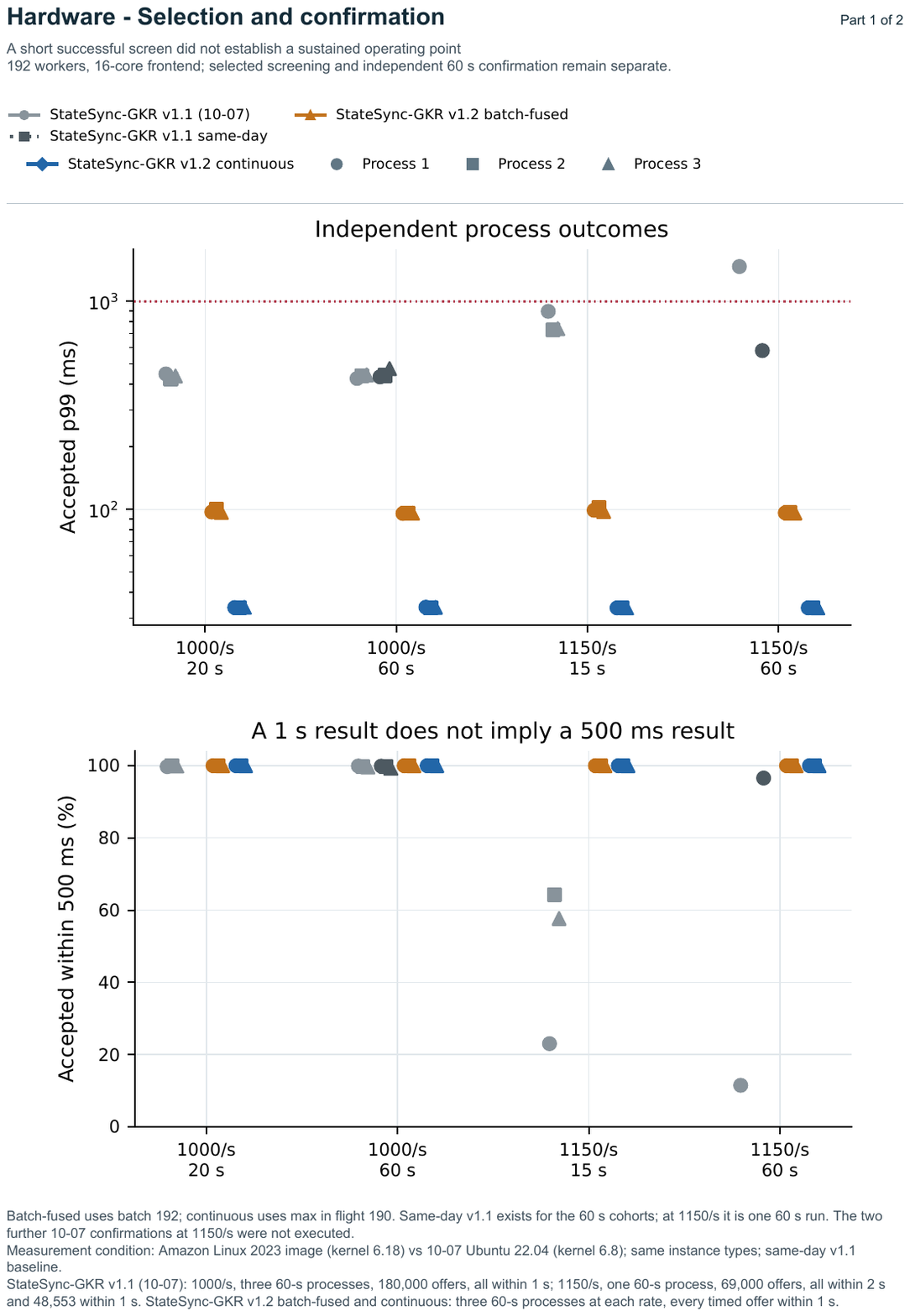}
\atlasfamily{A44}{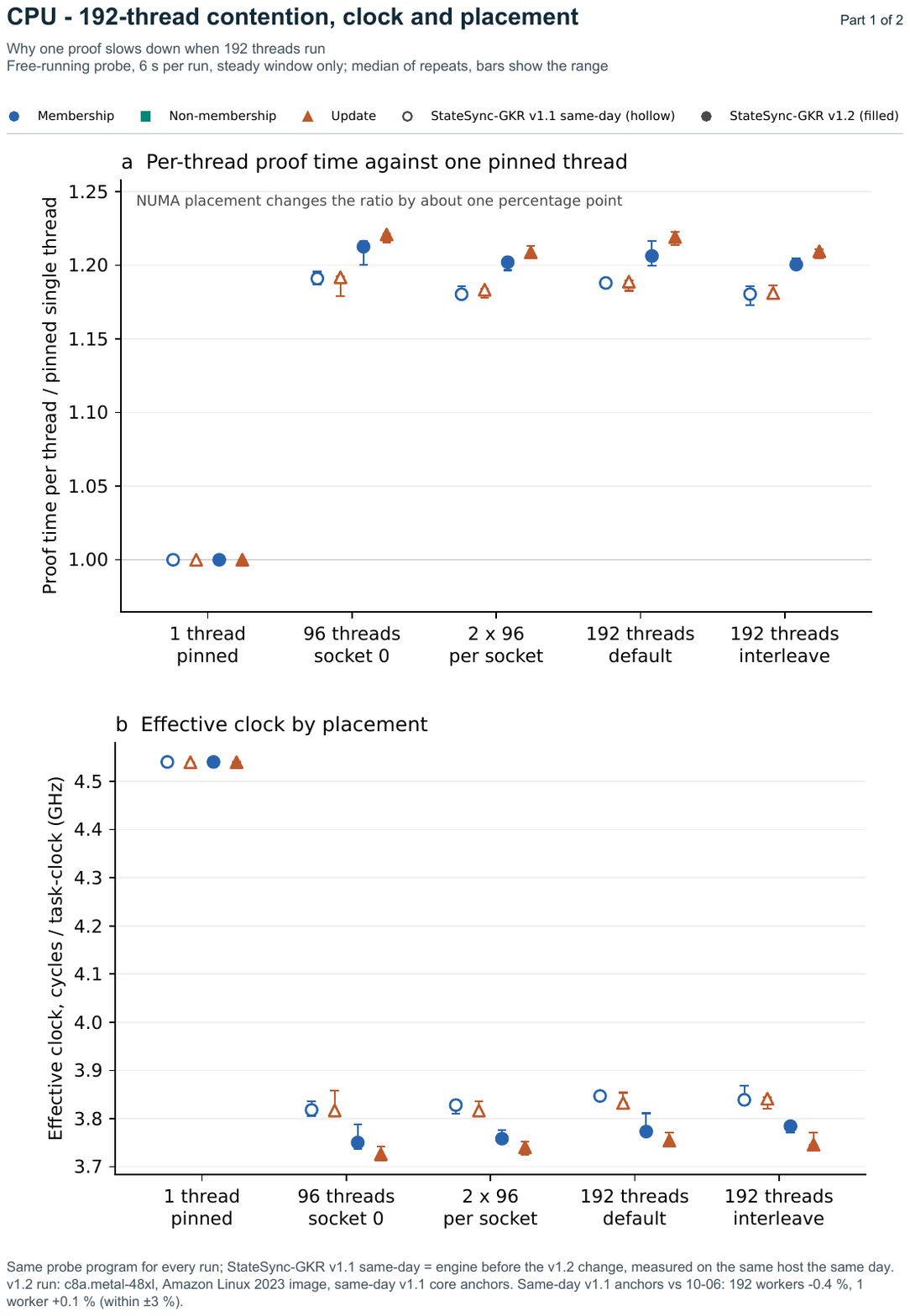}
\atlasfamily{A45}{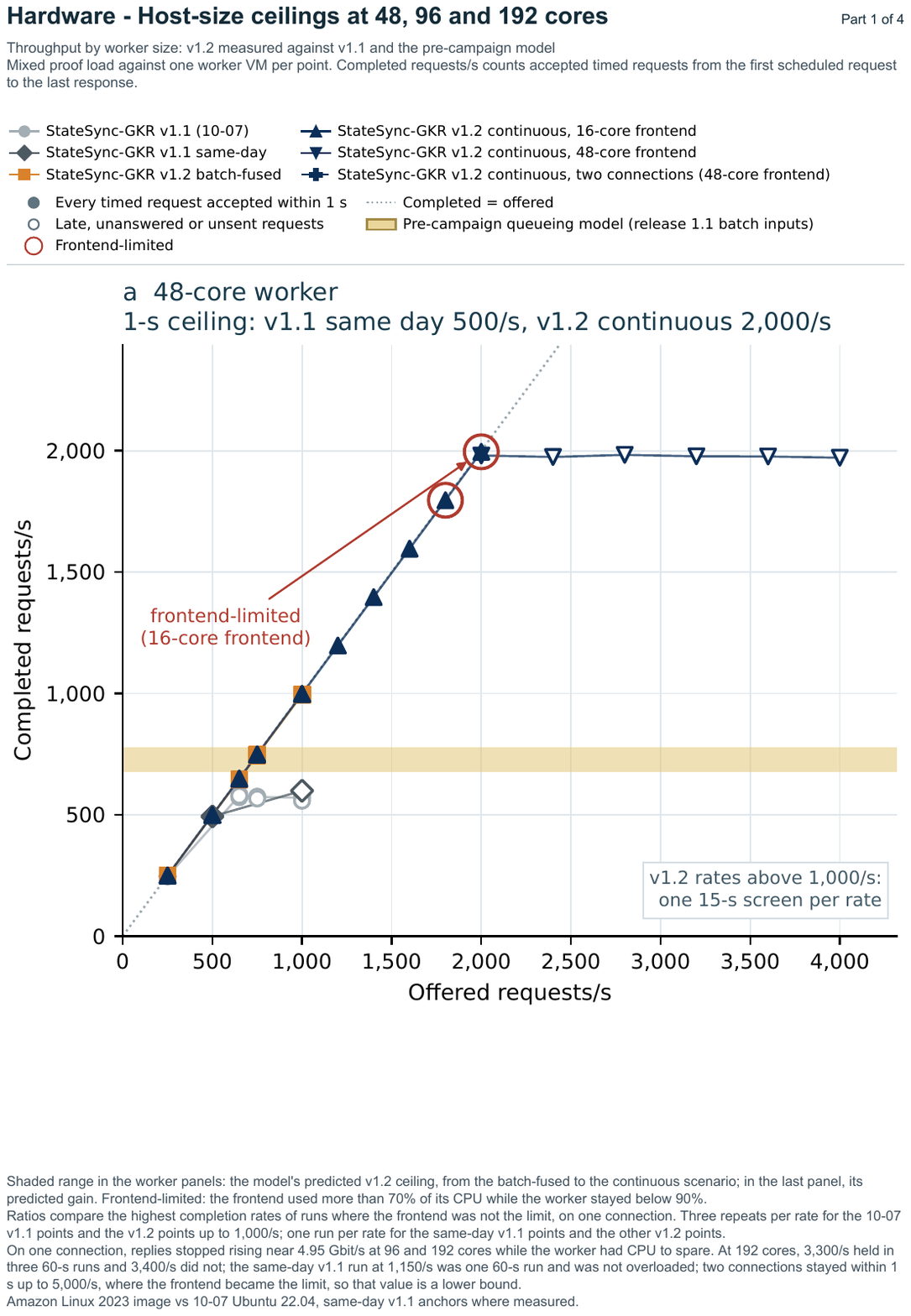}
\atlasfamily{A46}{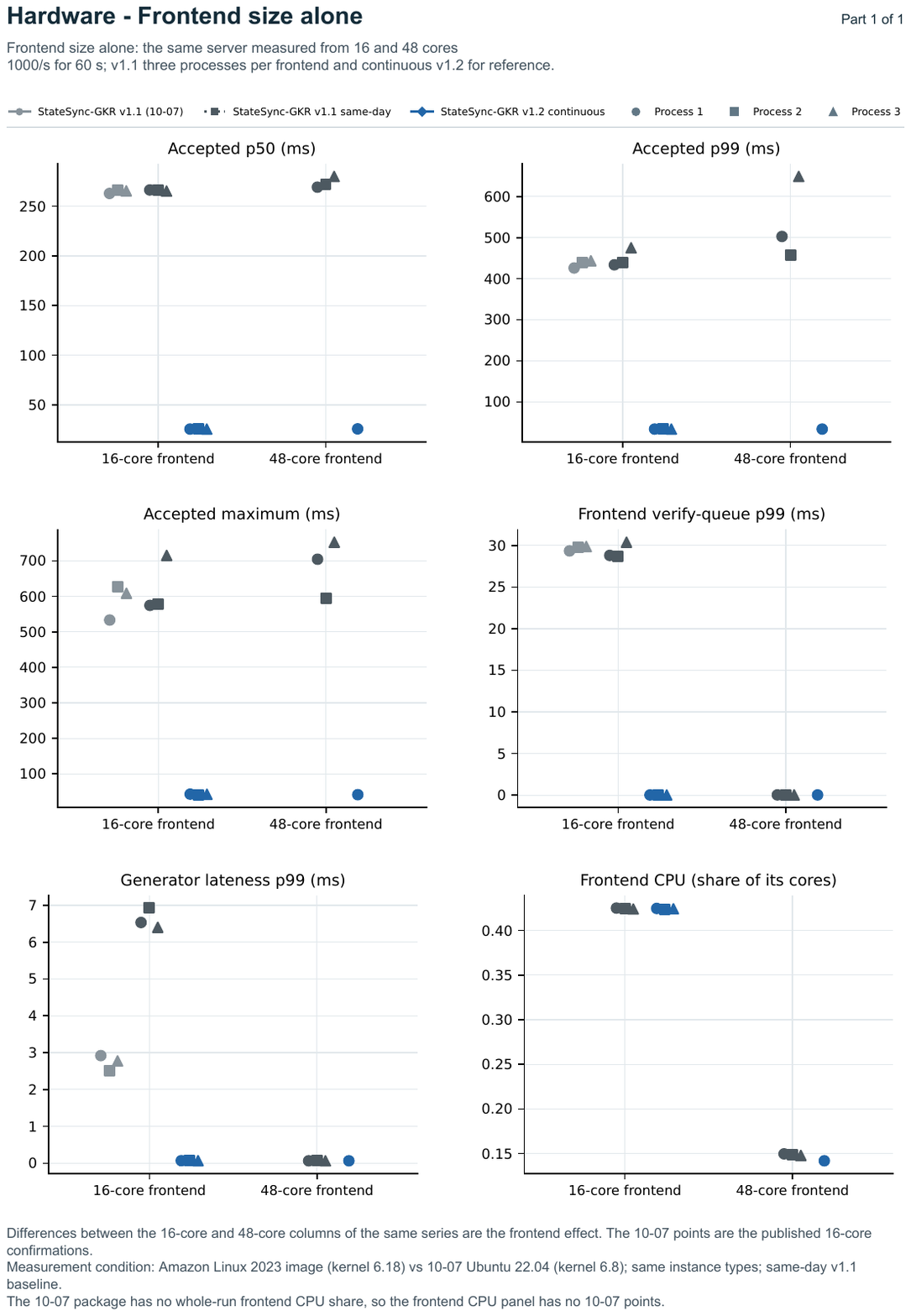}
\atlasfamily{A47}{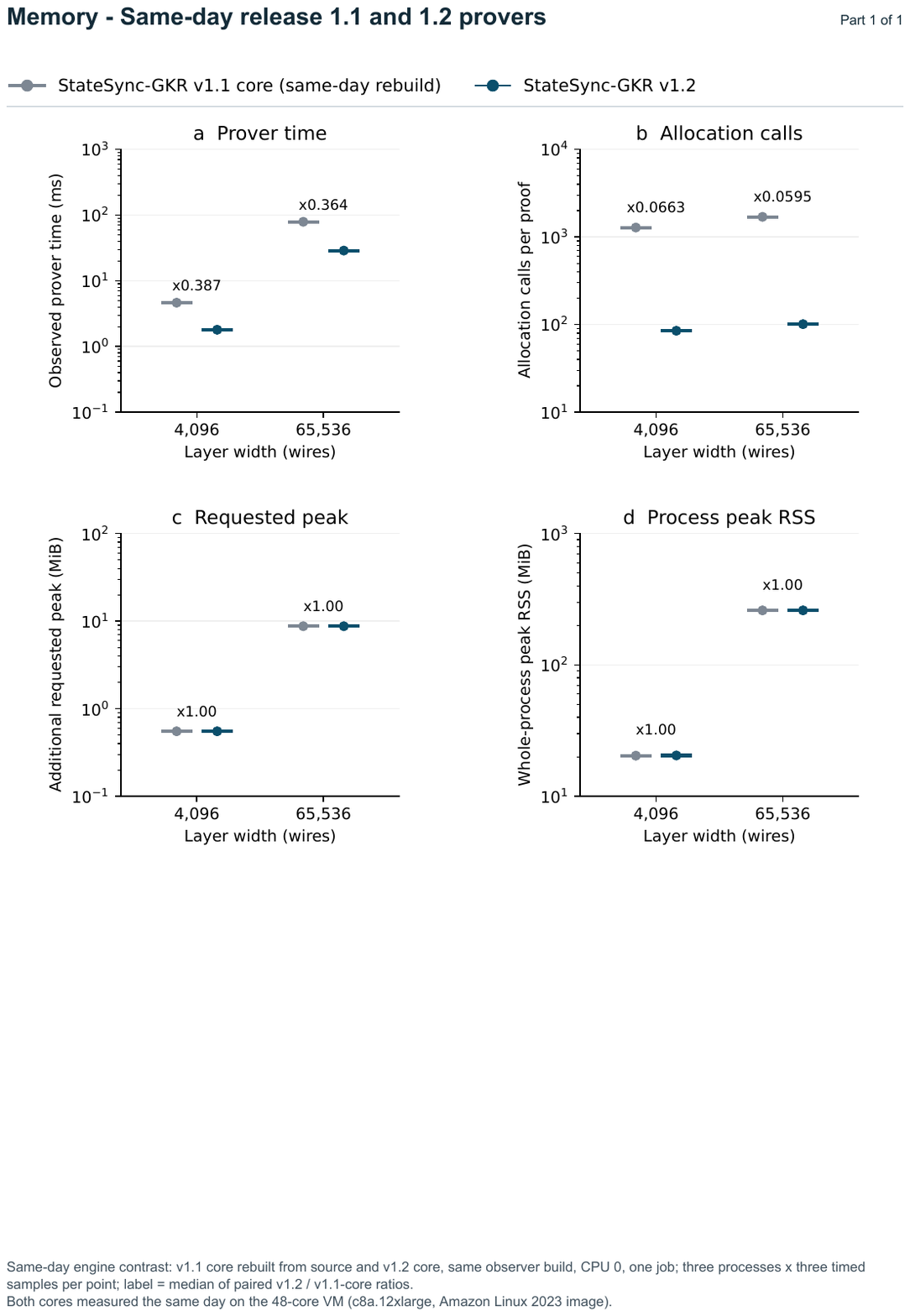}
\end{document}